\documentclass[10pt]{article}
\pdfoutput=1
\usepackage{jheppub,amsmath,amssymb}
\usepackage{booktabs}
\newcommand{\tabitem}{~~\llap{\textbullet}~~}
\usepackage{bm}
\usepackage{bbm}
\usepackage{enumitem}
\usepackage{mathtools}
\usepackage[usenames,dvipsnames]{xcolor}
\usepackage{adjustbox}
\usepackage{tikz}
\usetikzlibrary{decorations.pathmorphing}
\usetikzlibrary{arrows.meta}
\usepackage{graphicx}
\usepackage{hhline}
\usepackage{subfig}
\usepackage{hyperref}
\usepackage{color}
\usepackage{verbatim}
\usepackage{lscape}
\usepackage{float}
\usepackage{breqn}
\usepackage{multicol}
\usepackage{multirow}
\usepackage{makecell}
\usepackage{graphics}
\usepackage{tikz,todonotes}
\usepackage[utf8]{inputenc}
\usepackage{autobreak}
\usepackage{verbatim}

\usetikzlibrary{arrows,positioning,decorations.markings,decorations.pathmorphing,calc}
\pgfdeclarelayer{edgelayer}
\pgfdeclarelayer{nodelayer}
\usetikzlibrary{decorations.pathreplacing}
\pgfsetlayers{edgelayer,nodelayer,main}
\tikzset{none/.style={draw=none}}
\tikzset{new edge style 2/.style={black}}
\tikzset{new style 0/.style={black}}
\tikzset{rednode/.style={draw=none, scale=0.3pt,fill=red,circle, draw}}
\tikzset{redline/.style={line width=0.3mm,red}}
\tikzset{greyE/.style={line width=0.1mm,gray}}
\usepackage[utf8]{inputenc}
\usetikzlibrary{arrows,positioning,decorations.markings,decorations.pathmorphing,calc}

\usepackage{braket}

\newcommand{\beq}{\begin{equation}}
	\newcommand{\eeq}{\end{equation}}
\newcommand{\bal}{\begin{aligned}}
	\newcommand{\eal}{\end{aligned}}
\newcommand{\bea}{\begin{eqnarray}}
	\newcommand{\eea}{\end{eqnarray}}
\def\be{\begin{equation}}
	\def\ee{\end{equation}}

\def\beq{\begin{equation}}
	\def\eeq{\end{equation}}

\newcommand{\del}{\delta}

\newcommand{\eps}{\varepsilon}

\newcommand{\mpl}{M_{\rm Pl}}

\renewcommand{\eps}{\varepsilon}

\renewcommand{\L}{\mathcal L}

\usepackage{sujai}

\def\be{\begin{equation}}
	\def\ee{\end{equation}}
\def\ba{\begin{eqnarray}}
	\def\ea{\end{eqnarray}}

\usepackage[normalem]{ulem}

\def\ba{\begin{eqnarray}}
	\def\ea{\end{eqnarray}}

\def\L{\mathcal{L}}

\def\del{\nabla}

\def\mn{_{\mu \nu}}
\def\mnup{^{\mu \nu}}

\def\({\left(}
\def\){\right)}
\def\eps{\epsilon}

\def\mpl{M_{\rm Pl}}
\def\p{\partial}
\def\ie{{\em i.e. }}

\def\tilg8{\tilde{g}_8}

\newcommand{\ud}[2]{^{#1}_{\phantom{#1}#2}}
\newcommand{\du}[2]{_{#1}^{\phantom{#1}#2}}

\begin{document}
\preprint{Imperial/TP/2023/MC/02}	
	
	\title{Positivity-causality competition: a road to ultimate EFT consistency constraints}

	\author[a]{Mariana Carrillo Gonz\'alez, }
	\author[a]{Claudia de Rham, }
	\author[a]{Sumer Jaitly, }
	\author[a]{Victor Pozsgay, }
    \author[a,b,c]{Anna Tokareva}
	
	\affiliation[a]{Theoretical Physics, Blackett Laboratory, Imperial College, London, SW7 2AZ, U.K.}
\affiliation[b]{School of Fundamental Physics and Mathematical Sciences, Hangzhou Institute for Advanced Study, UCAS, Hangzhou 310024, China}
\affiliation[c]{International Centre for Theoretical Physics Asia-Pacific, Beijing/Hangzhou, China}
 
	\abstract{Effective field theories (EFT) are strongly constrained by fundamental principles such as unitarity, locality, causality, and Lorentz invariance. In this paper, we consider the EFT of photons (or other $U(1)$ gauge field) and compare different approaches to obtain bounds on its Wilson coefficients. We present an analytic derivation of the implications of unitarity (linear and non-linear positivity bounds) and compare these constraints with the requirement of causal propagation of the photon modes around non-trivial backgrounds generated by external sources. We find that the low energy causality condition can give complementary constraints to the positivity bounds. Applying both constraints together can significantly reduce the allowed region of the photon EFT parameters.}
	
	\maketitle
	
	\pagebreak

\section{Introduction}

Effective field theories (EFTs) allow us to describe low-energy physics without the knowledge of the fundamental theory in the ultraviolet (UV). These low-energy descriptions can include any interaction that is consistent with the symmetries of the system and thus can lead to a large number of free parameters. From a phenomenological perspective, bounding these parameters from observations without any theoretical guidance could result in an incorrect estimation of their actual value. Thus, it is desirable to understand how physical principles can reduce the freedom in this large parameter space. The assumption that the ultimate fundamental theory obeys the basic principles of quantum field theory and relativity puts constraints on its low-energy effective description. Namely, imposing unitarity, Lorentz invariance, analyticity, and locality of the scattering amplitudes leads to constraints on the couplings of the low-energy EFT \cite{Adams:2006sv, Pennington:1994kc, Pham:1985cr}. One such class of constraints is termed {\it positivity bounds}, as the simplest of them require strict positivity of linear combinations of EFT coefficients. The study of positivity bounds fits within the broader program of the S-matrix bootstrap, which aims to map the infinite-dimensional space of possible S-matrices of all consistent quantum field theories \cite{Eden:1966dnq, Martin:1969ina, Paulos:2016fap, Kruczenski:2022lot}. Techniques for obtaining positivity bounds have been fruitfully explored in the recent years following the revival due to \cite{Adams:2006sv}, leading to infinite sets of positivity bounds beyond the forward limit \cite{Vecchi:2007na,deRham:2017avq,Manohar:2008tc,Nicolis:2009qm}, for massive spinning particles (particularly spin-2) \cite{Bellazzini:2016xrt,deRham:2017zjm,Cheung:2016yqr, Bonifacio:2016wcb,Alberte:2019zhd,Alberte:2019xfh,Wang:2020xlt,Davighi:2021osh,Bellazzini:2019bzh}, massless gravity \cite{Haring:2022cyf, Bern:2021ppb, Chowdhury:2021ynh, Chiang:2022jep, Alberte:2020jsk, Alberte:2020bdz, Alberte:2021dnj, deRham:2022gfe, Hamada:2018dde, Noumi:2022wwf, Tokuda:2020mlf}, and non-linear bounds from crossing symmetry, Cauchy-Schwarz type inequalities and properties of Gegenbauer polynomials \cite{Chiang:2021ziz,Arkani-Hamed:2020blm, Bellazzini:2020cot, Tolley:2020gtv, Caron-Huot:2020cmc}, as well as applications to the SMEFT \cite{Zhang:2018shp, Bi:2019phv, Remmen:2019cyz,Zhang:2020jyn, Fuks:2020ujk,Remmen:2020vts, Yamashita:2020gtt}.

In this paper, we will analyze in detail the consequences of one of the physical conditions required to derive positivity bounds, namely causality. By imposing causality one reaches the conclusion that scattering amplitudes are analytic in the complex $s$ plane up to physical poles and branch cuts in the real axis \cite{Bogolyubov:1959bfo,PhysRev.104.1760,Bremermann:1958zz,Bogoliubov,osti_4650017,Eden:1966dnq}, which is used to derive positivity bounds. Causality is the requirement that no information can be received before it is generated. For a local field theory, this is encoded in the condition that the retarded Green's function does not have support outside the lightcone, which in turn is a consequence of micro-causality, that is, that commutators of local operators vanish outside the lightcone. Here, we assume that the spacetime in which the fields propagate has a chronology defined by a minimally coupled photon moving at the speed of light. Examples, where this is not assumed have been considered in \cite{Bruneton:2006gf,Bruneton:2007si,Babichev:2007dw} and allow for a large time advance. In such settings, an observable time advance does not imply that we can generate a closed timelike curve as is the case in the examples considered in \cite{Adams:2006sv}, which also assumes the chronology of the spacetime to be given by the photon as in this paper. 

Besides the implications of causality for the derivation of positivity bounds, one can also obtain constraints on the Wilson coefficients of an EFT without any assumptions of the UV by only requiring causal propagation at low energies around non-trivial backgrounds generated by external sources \cite{Camanho:2014apa,Camanho:2016opx,Bai:2016hui,Goon:2016une,Hinterbichler:2017qcl,Hinterbichler:2017qyt,AccettulliHuber:2020oou,Bellazzini:2021shn,Serra:2022pzl,Chen:2021bvg,deRham:2021bll,CarrilloGonzalez:2022fwg}. One should note that the strict requirement of a subluminal speed of sound is too strong. Modes can propagate superluminally in small regions of spacetime and not lead to any observable violation of causality. Instead of focusing on the speed of sound, we will consider a cleaner diagnostic by computing the time delay experienced by a propagating mode in the EFT relative to a free mode. The time delay is invariant under field redefinitions since it is directly related to the S-matrix as
\begin{equation}
\Delta T=-i\left\langle\operatorname{in}\left|\hat{S}^{\dagger} \frac{\partial}{\partial \omega} \hat{S}\right|\operatorname{in}\right\rangle  .
\end{equation}
By defining the scattering phase shift from the S-matrix eigenvalues as $\hat{S}|\operatorname{in}\rangle=e^{2 i \delta}|\operatorname{in}\rangle$, we can compute the time delay at fixed impact parameter 
\begin{equation}
    \Delta T_b=\left.2 \frac{\partial \delta_{\ell}}{\partial \omega}\right|_b  \ ,
\end{equation}
which corresponds to the time delay computed in the Eikonal approximation \cite{Wallace:1973iu,Wallace:1973ni}. In order to have a well-defined time delay, one requires a separation of scales between the background and the propagating modes. The scales of variation of the background should be much larger than the wavelength of the mode in order to observe a well-defined phase shift and hence time delay. This means that we will be working in the semi-classical regime, in other words, in the WKB regime. We can now diagnose violations of causality by checking whether a theory can give rise to closed-timelike curves, which is equivalent to obtaining a resolvable time-advance \cite{Adams:2006sv,Hollowood:2015elj}.  Additionally, this should be computed within the regime of validity of the EFT and the WKB approximation. The concept of resolvability arises from the uncertainty principle which tells us that, if we have a mode with frequency $\omega$, we cannot measure its time delay/advance if it is smaller than the uncertainty $\Delta t\sim \omega^{-1}$. Thus, if a mode experiences a time advance that is resolvable, that is, a time advance larger than the resolution scale of geometric optics, 
\begin{equation}
	\Delta T <- 1/\omega \ , 
\end{equation}
then the theory has a mode that violates causality. By imposing that our theory does not have any such modes, we can bound the value of Wilson coefficients. In this paper, we will analyze such bounds for the case of an EFT of a photon.

In principle, one would expect that the requirement of causal propagation in the infrared leads to weaker constraints than the requirement of a local, unitary, causal, and Lorentz invariant theory all the way to the UV. Nevertheless, the actual constraints obtained from our requirement of causal propagation and the actual constraints obtained by deriving positivity bounds are not guaranteed to encode the full power of these assumptions. As we will see, it is possible that these bounds test both similar and different regions of the parameter space. Thus, in some cases, causality and positivity can lead to similar bounds, while in other cases they can probe different regions of the parameter space and lead to complementary bounds that can be combined to obtain a much stronger constraint on the Wilson coefficients of the EFT.

The main purpose of this paper is to compare the implications arising from the requirement of causal propagation with the constraints from positivity bounds by considering the example of an EFT of photons. The positivity bounds on the EFT coefficients were studied previously in \cite{Adams:2006sv,Arkani-Hamed:2020blm,Henriksson:2021ymi,Henriksson:2022oeu,Haring:2022sdp}. In this paper, we formulate new constraints on the operators up to dimension-12 (we consider 7 independent couplings) in the form of analytic expressions. We develop a new approach based on scattering amplitudes for indefinite polarization states of photons parameterized by four angles. This allows us to make the bounds stronger than those derived from the scattering of given polarization states, as the obtained inequalities must be valid for all possible indefinite polarization states. We formulate a set of linear positivity bounds derived in a way similar to \cite{deRham:2017avq},\footnote{To the best of our knowledge, the bounds of this type were never formulated before for amplitudes without $s-u$ crossing symmetry.} and analytically perform the optimisation of the inequalities. In addition, we formulate the two non-linear bounds constraining dimension-10 operators between dimension-8 and dimension-12 couplings. We also discuss how one-loop corrections to the amplitudes would affect the bounds. 

To compare the strength of the causality and positivity requirements, we consider different slices in the 6-dimensional space of couplings. We show that, in some directions of the parameter space, positivity bounds provide stronger constraints while causality conditions do not lead to compact bounds in all directions. The later is due to  technical challenged related to the implementation of the WKB approximation, which prevent us from including non-sign definite contributions to the time delay for certain Wilson coefficients. This does not imply that causality requirements cannot lead to compact bounds in these specific directions, but rather that in some situations, our current setup does not lead to a compact bound in all directions. On the other hand, for dimension-10 operators we obtain compact causality bounds which are so far stronger than the analytic positivity bounds formulated in this work. We emphasize that this does not represent a contradiction but simply signals the fact that the ultimate unitarity constraints must be stronger than those which were formulated so far. Thus, the causality conditions can provide a hint for the formulation of better constraints on the EFT parameter space. 

The paper is organised as follows. In Section 2 we introduce the photon EFT and review the connection between the Lagrangian and scattering amplitudes parametrisations. In Section 3 we describe the novel methods of obtaining positivity bounds for the scattering of indefinite polarization states of photons. Section 4 is dedicated to deriving the causality constraints from the absence of a resolvable time advance in photon propagation on top of a spherically-symmetric background generated by an external source. In Section 5, we compare positivity and causality bounds in different slices of the parameter space. After that, in Section 6 we present the conclusions and discussion on the complementarity of positivity and causality bounds.
Throughout the manuscript, we work in four spacetime dimensions, with signature $(-+++)$.

\section{Photon Effective Field Theory}

\subsection{EFT of $U(1)$ Gauge Field}

The most generic EFT of a massless vector field $A_{\mu}$ enjoying a $U(1)$ gauge symmetry can be formulated in the following way. We define the field-strength (or Faraday) tensor $F\mn$ and its Hodge dual $\tilde{F}\mn$ in the following way
\begin{equation}
	F\mn = \p_{\mu} A_{\nu} - \p_{\nu} A_{\mu} \,, \qquad \tilde{F}\mn = \frac12 \epsilon_{\mu \nu \rho \sigma} F^{\rho \sigma} \,.
	\label{eq:defF}
\end{equation}
We make the hypothesis that the $U(1)$-symmetric Lagrangian only depends on gauge-invariant quantities, \ie
\begin{equation}
	\L_{F} = \L(F,\tilde{F},\p F, \p \tilde{F}, \dots) \,.
	\label{eq:gaugeinvariantquantities}
\end{equation}
Note that we will forbid any operator with an odd number of $\tilde{F}$ to avoid any parity breaking. We will assume that our EFT breaks down at a scale $\Lambda$ and we will consider operators of at most $4$ fields and dimension-$12$. We assume that operators with more than $4$ fields are suppressed at the EFT order that we work at, which allows us to compare the causality bounds to positivity bounds. After neglecting redundant terms that can be removed through field redefinitions, we write a set of independent operators up to dimension-$12$ which reads
\begin{align}
	\label{eq:Lagrangian}
	\L =& - \frac14 F\mn F\mnup \nonumber \\
	&+\frac{c_1}{\Lambda^4} F\mnup F\mn F^{\alpha \beta} F_{\alpha \beta} + \frac{c_2}{\Lambda^4} F\mnup F^{\alpha \beta} F_{\mu \alpha} F_{\nu \beta} \nonumber \\
	&+ \frac{c_3}{\Lambda^6} F^{\alpha \mu} F^{\nu \beta} \p_{\mu} F_{\beta\gamma} \p_{\nu} F_{\alpha}^{\phantom{\alpha}\gamma} + \frac{c_4}{\Lambda^6} F^{\alpha \mu} F^{\nu \beta} \p_{\beta} F_{\mu\gamma} \p^{\gamma} F_{\alpha \nu} + \frac{c_5}{\Lambda^6} F^{\alpha \mu} F^{\nu \beta} \p_{\beta} F_{\nu\gamma} \p^{\gamma} F_{\alpha \mu} \nonumber \\
	&+ \frac{c_6}{\Lambda^8} F\mnup \p_{\mu} F_{\nu \rho} \p^{\rho} \p^{\alpha} F^{\beta\gamma} \p_{\alpha} F_{\beta\gamma} + \frac{c_7}{\Lambda^8} F\ud{\mu}{\gamma} \p_{\mu} F_{\nu \rho} \p^{\nu} F_{\alpha\beta} \p^{\rho} \p^{\gamma} F^{\alpha\beta} \nonumber \\
	&+ \frac{c_8}{\Lambda^8} F^{\mu\gamma} \p_{\mu} F_{\nu\rho} \p^{\rho} \p^{\beta} F_{\alpha\gamma} \p^{\alpha} F\ud{\nu}{\beta} \,.
\end{align}
The $4$-point tree-level scattering amplitude arising from this theory can be parametrized as 
\begin{subequations}
	\begin{align}
		\mathcal{A}_{++++}&=\frac{f_2}{\Lambda^4}\left(s^2+t^2+u^2\right)+\frac{f_3}{\Lambda^6} s t u+\frac{f_8}{\Lambda^4}\left(s^2+t^2+u^2\right)^2 \,, \\
		\mathcal{A}_{++--}&=\frac{g_2}{\Lambda^4} s^2+\frac{g_3}{\Lambda^6} s^3+\frac{g_4}{\Lambda^8}s^4+\frac{g_4^{\prime}}{\Lambda^4} s^2 t u \,, \\
		\mathcal{A}_{+++-}&=\frac{h_3}{\Lambda^6} s t u \,, 
	\end{align}
\label{eq:AmplParam}%
\end{subequations}
where all other helicity configurations can be obtained by symmetry considerations (parity, time-reversal, boson exchange, crossing symmetry) and we consider a set up where all particles are incoming. The scattering amplitude parameters above are related to the Wilson coefficients in Eq.~\eqref{eq:Lagrangian} via
\begin{align}
f_2=2\left(4 c_1+c_2\right)	\ , \quad g_2=2\left(4 c_1+3 c_2\right) \, \nonumber \\
f_3=-3\left(c_3+c_4+c_5\right) \ , \quad	g_3=-c_5 \ ,  \quad h_3=-\frac{3}{2} c_3 \ , \nonumber \\
f_4=\frac{1}{4}c_6 \ , \quad g_4 =\frac{1}{2} (c_6 - c_8)+c_7 \ , \quad  g_4'=-\frac{1}{2} (c_7+ c_8) \ . \label{eq:conversion}
\end{align}
Note that throughout the text we occasionally refer to the amplitude parameters themselves as `Wilson coefficients'.
\subsection{Partial UV completions}
 Together with the positivity and causality bounds, we will show explicitly the values for the coefficients of known partial UV completions. We will focus on tree level, partial UV completions given by the interactions of the photon with a scalar and an axion. In some cases, we will also show the values for the partial UV completions involving a graviton that were analyzed in \cite{Henriksson:2021ymi,Henriksson:2022oeu,Haring:2022sdp}. We refer to the spin-2 partial UV completion from \cite{Henriksson:2021ymi} as minimally-coupled spin-2. Lastly, we will also analyze some of these bounds that can be compared to one-loop, partial UV completions from QED-like theories. We will consider the standard (spinor) QED \cite{Euler:1935qgl,Euler:1935zz,Costantini:1971cj,Karplus:1950zz}, scalar QED \cite{Weisskopf:1936hya,Yang:1994nu}, and vector QED \cite{Yang:1994nu,Vanyashin:1965ple}. Table \ref{tab:UVcomp} shows the coefficients for all these partial UV completions.

\begin{table}[h!]
	\centering
	\begin{tabular}{ c || c | c | c | c | c | c | c | c }
	UV completion & $g_2$ & $f_2$ & $f_3$ & $g_3$ & $h_3$ & $f_4$ & $g_4$ & $g_4'$ \\ \hline
	scalar & $1$ & $1$ & $3$ & $1$ & $0$ & $\frac12$ & $1$ & $0$   \\
	axion & $1$ & $-1$ & $-3$ & $1$ & $0$ & $-\frac12$ & $1$ & $0$  \\ \hline \hline
	scalar QED & $1$ & $\frac34$ & $\frac{5}{14}$ & $\frac{3}{28}$ & $\frac{1}{28}$ & $\frac{1}{84}$ & $\frac{41}{420}$ & $-\frac{1}{168}$  \\
	spinor QED & $1$ & $-\frac{3}{11}$ & $-\frac{10}{77}$ & $\frac{4}{77}$ & $-\frac{1}{77}$ & $-\frac{1}{231}$ & $\frac{13}{660}$ & $-\frac{5}{462}$ \\
	vector QED & $1$ & $\frac{1}{28}$ & $\frac{5}{294}$ & $-\frac{47}{1764}$ & $\frac{1}{588}$ & $\frac{1}{1764}$ & $\frac{131}{8820}$ & $-\frac{23}{1176}$  \\ \hline \hline
	spin-2 even I* & $1$ & $1$ & $0$ & $1$ & $0$ & $\frac{1}{2}$ & $1$ & $-6$  \\
	spin-2 even II & $1$ & $0$ & $0$ & $-1$ & $0$ & $0$ & $1$ & $-2$  \\
	spin-2 odd* & $1$ & $-1$ & $0$ & $1$ & $0$ & $-\frac12$ & $1$ & $-6$  \\
	min.-coupled spin-2 & $1$ & $0$ & $0$ & $-\frac12$ & $0$ & $0$ & $\frac12$ & $-1$ 
\end{tabular}
	\caption{Values of the Wilson coefficients for known partial UV completions, where $g_2$ is normalized to unity. The partial UV completions with a $*$ superscript do not satisfy the causality and positivity bounds. This simply indicates that the non-minimal couplings that they contain are ruled out; see Appendix \ref{app:spin2} for more details.}
	\label{tab:UVcomp}
\end{table}
One should note that the time delay for the even sector of the axion vanishes as it should. Similarly, the time delay for the odd sector vanishes for the scalar partial UV completions. On the other hand, the so-called odd and even spin-2 partial UV completions do not satisfy this feature. See Appendix \ref{app:spin2} for more details. 
\section{Positivity bounds}
In this section, we will explore constraints on the Wilson coefficients of the photon EFT that arise from requiring a so-called `standard' or `consistent' UV completion \cite{Adams:2006sv}. By `standard' we mean that the UV theory has an S-matrix satisfying certain properties that encapsulate unitarity, causality, locality, and Lorentz invariance:
\begin{itemize}
    \item \textbf{Unitarity}: The S-matrix being unitary follows from the completeness of the asymptotic in/out Hilbert space and implies the conservation of probability in scattering processes. This leads to the optical theorem, which roughly states that,
    \begin{equation}
        \text{Im}\cA_{i\xr i}(s,0) = \frac12\sum_n\, |\cA_{i\xr n}(s)|^2\,,
    \end{equation}
    i.e. that the imaginary part of the elastic scattering amplitude from state $i$ to $i$ is related to the absolute values of the amplitudes for all $i$ to $n$ processes, for any intermediate state $n$. An immediate consequence of this is that the imaginary part of the forward limit elastic amplitude is positive. Since the imaginary part of the amplitude arises from non-analytic structures such as poles and branch cuts, unitarity also informs us of the analytic structure of the amplitude.
    
    One slightly more technical statement that we utilise is positivity of $t$-derivatives of the discontinuity (or imaginary part as phrased above) at $t=0$. This can be derived from properties of the partial wave expansion, e.g. see \cite{deRham:2017avq, deRham:2017zjm,deRham:2022gfe}. Non-analyticity of the amplitude generated by massless loops can undermine this, however for spin-1 loops it is possible to circumvent this with an IR regulating mass.
    \item \textbf{Causality}: The statement of microcausality in the quantum field theory, i.e. that space-like separated operators commute, is connected to the analytic structure of the amplitude \cite{Bogolyubov:1959bfo}. The amplitude is assumed to be analytic in the upper half complex $s$ plane for fixed negative $t$. The assumption of the Schwarz reflection property extends this to the lower half plane, with the final result that the amplitude is analytic up to the poles and branch cuts required by unitarity.
    \item \textbf{Locality}: We assume that the fixed $t$ amplitude is polynomially bounded in the complex $s$ plane as $|s|\xr\infty$. This assumption is implicit in the derivation of scattering amplitudes via the LSZ prescription as it allows one to Fourier transform between position and momentum space correlation functions. This has been linked in quantum field theory with the notion of locality and is required for the existence of a dispersion relation representation \cite{Bremermann:1958zz, Epstein:1969bg}.
    \item \textbf{Boundedness}: For theories with a mass gap, the Froissart-Martin bound demands that the elastic amplitude grows slower than $s^2$ as $|s|\xr \infty$, allowing for a twice subtracted dispersion relation to be written. Whilst this boundedness is not proven for theories without a mass gap, we will assume that the exact amplitude obeys this property \cite{Froissart:1961ux,Martin:1965jj,Jin:1964zz}. 
\end{itemize}

 In this section we assume that there are no poles due to the exchange of states with spin $\geq2$ (such as the infamous $t$-channel pole arising from graviton exchange) and so we can safely assume any poles have been already subtracted, leaving us with the pole-subtracted amplitude, without violating the boundedness properties of the amplitude. 
 
\subsection{Indefinite helicity amplitudes and manifest crossing symmetry}\label{ihamp}

Positivity bounds can be derived for any scattering amplitude that is elastic, meaning that the initial and final states are the same. In these cases, unitarity implies positivity of the discontinuities of the amplitude. Thus, we construct a general in-going state of two photons, $\ket{\rm in}=\sum_{h_1,h_2} a_{h_1 h_2}\ket{h_1,h_2}$ where the helicity labels are summed over the two polarization states, $h_i=\pm$, and study the elastic scattering of this state. Bounds on such indefinite helicity amplitudes have been fruitfully explored in previous literature \cite{Bellazzini:2015cra,Cheung:2016yqr,deRham:2017imi,deRham:2018qqo,Alberte:2019xfh,Alberte:2019zhd,Alberte:2021dnj, deRham:2022gfe}.

As the initial state is a sum of helicity eigenstates, the scattering amplitude will decompose into a sum of amplitudes between helicity eigenstates which notably contains amplitudes for processes that are themselves not elastic. The helicity amplitudes for the process with in-state of helicity $|h_1 , h_2 \rangle$ and out-state of helicity $|-h_3,-h_4 \rangle$ are denoted,
\be
(\textbf{in-out})\qquad\cA_{h_1 h_2 \rightarrow -h_3 -h_4}(s,t,u)  \equiv \cA_{h_1 h_2  h_3 h_4}(s,t,u)\qquad(\textbf{all in})\, .
\ee
Generally crossing relations/symmetries between helicity amplitudes are extremely complicated (see for example \cite{deRham:2017zjm}) however for the massless, bosonic and equal spin scattering we are interested in, the relations are a lot simpler and are given by,
\begin{equation}
\begin{aligned}
    \cA_{h_1 h_2  h_3 h_4}(s,t,u) &= \cA_{h_1 h_4  h_3 h_2}(u,t,s) \,, \\
    \cA_{h_1 h_2  h_3 h_4}(s,t,u) &=\cA_{h_1 h_3  h_2 h_4}(t,s,u) \,, \\
    \cA_{h_1 h_2  h_3 h_4}(s,t,u) &=\cA_{h_1 h_2  h_4 h_3}(s,u,t)\,.
\end{aligned}
\end{equation}
If we denote the elastic scattering of our generic initial state as the $s$--channel process, with amplitude $\cA_s(s,t,u)$, the corresponding crossed $u$--channel process will have the amplitude $\cA_u(s,t,u)$ given by,
\ba
\cA_u(s,t,u) = \cA_s(u,t,s) &=&\sum_{h_i} a_{h_1h_2} a^*_{-h_3-h_4} \cA_{h_1 h_2  h_3 h_4}(u,t,s)  \\
&=&\sum_{h_i} a_{h_1h_2} a^*_{-h_3-h_4} \cA_{h_1 h_4  h_3 h_2}(s,t,u)\,.
\ea
Note that the final expression cannot arise as the amplitude for an elastic scattering process and so we cannot use positivity of the $u$--channel discontinuity or apply positivity bounds in the usual manner. This can be simply remedied by requiring that the coefficients in our general initial state are factorisable (making it a separable state) such that $a_{h_1 h_2 }=\alpha_{h_1} \beta_{h_2}$ in which case the above expression becomes,
\ba
\cA_u(s,t,u) &=& \sum_{h_i} \alpha_{h_1} \beta_{h_2}\alpha_{-h_3}^* \beta_{-h_4}^*\cA_{h_1 h_4  h_3 h_2}(s,t,u)  \\
&=& \sum_{h_i} \alpha_{h_1} \beta_{-h_2}^* \alpha_{-h_3}^* \beta_{h_4} \cA_{h_1 h_2  h_3 h_4}(s,t,u) \, ,
\ea
which identifies the amplitude $\cA_u$ as the elastic amplitude for the initial state, $|{\rm in}  \rangle = \sum \alpha_{h_1} \beta_{-h_2}^* |h_1 h_2 \rangle$.

By comparing the above two lines we can deduce that the amplitude will obey manifest $s-u$ crossing symmetry if the $\beta_h$ coefficients satisfy,
\be\label{crossing1}
\beta_{h}=\beta_{-h}^* e^{i \gamma}\implies\cA_s(s,t,u) = \cA_u(s,t,u)\,,\quad\quad\gamma\in\mathbb{R}\,,
\ee
and similarly manifest $s-t$ crossing symmetry if,
\be
\beta_{h}=\alpha_{-h}^* e^{i \psi}\implies\cA_s(s,t,u) = \cA_t(s,t,u)\,,\quad\quad\psi\in\mathbb{R}\,.
\ee
If both the above conditions are satisfied the amplitude will be triple crossing symmetric, i.e. unchanged under any permutation of $s,t,u$ and $\cA_s = \cA_t = \cA_u$. A useful parametrisation of the $\alpha,\beta$ coefficients that guarantees correct normalisation of states is given by (for real values of the angles and phases),
\begin{equation}\label{para}
    \alpha_+ = \cos\theta\,,\quad\alpha_-=\sin\theta \,e^{\ri \phi}\,,\quad\beta_+ = \cos\chi\,,\quad\beta_- = \sin\chi\,e^{\ri\psi}\,.
\end{equation}
The motivation for imposing manifest crossing symmetries is that different, a priori stronger positivity bounds can be derived for such amplitudes \cite{Tolley:2020gtv,Caron-Huot:2020cmc}. Imposing the above conditions amounts to restrictions on the angles introduced in Eq.~\eqref{para} and leaves one with families of crossing symmetric amplitudes parameterised by the remaining angles. For example, the one-parameter family of manifestly triple crossing symmetric (which we denote by $\stuu$) amplitudes is,
\begin{dmath}
    \cA_\stuu
=\frac 12 \left(\cA_{++--}+\cA_{+--+}+\cA_{+-+-}\right)+2\cA_{+---}\cos \phi
+\frac 12 \cA_{++++}\cos 2\phi \,,
\end{dmath}
whereas the indefinite helicity amplitude with no manifest crossing symmetry, denoted $\ih$, is given by,
\begin{dmath}
            \cA_{\ih}=\frac{1}{2} (\cos (2 \theta ) (\cA_{++--}-\cA_{+--+}) \cos (2 \chi )+\cA_{++--}+4 \cA_{+---} \sin (\chi ) \cos (\chi ) \cos (\psi )+\cA_{+--+}+\sin (2 \theta ) \sin (2 \chi ) (\cA_{++++} \cos (\psi +\phi )+\cA_{+-+-} \cos (\phi -\psi ))+4 \cA_{+---} \sin (\theta ) \cos (\theta ) \cos (\phi ))\,.
\end{dmath}
Finally the manifestly $s-u$ symmetric amplitudes, $\cA_\su$ are obtained by evaluating $\cA_\ih$ on any value of $\theta$ or $\chi$ such that $\cos(2\theta)\cos(2\chi)=0$.

\subsection{Dispersion relations and positivity bounds}
\begin{figure}
    \centering
    \begin{tikzpicture}[scale=2.4,>={Stealth[inset=0pt,length=5pt,width=5pt]}]
  \clip (-2.5,-1.3) rectangle (2.5,1.3);
  \draw [->] (-2,0) -- (2,0) node [right] {};
  \draw [->] (0,-1.3) -- (0,1.3) node [above] {};
  \draw [decorate, decoration={zigzag}] (-2,0) -- (2,0);  
  \draw [red,thick,postaction={decorate,decoration={markings,mark=at position 0.5 with {\arrow{>}}}}] (-2,0.1) -- (-0.5,0.1);
  \draw [red,thick,postaction={decorate,decoration={markings,mark=at position 0.5 with {\arrow{<}}}}] (-2,-0.1) -- (-0.5,-0.1);
  \draw [red,thick,postaction={decorate,decoration={markings,mark=at position 0.5 with {\arrow{>}}}}] (0.5,0.1) -- (2,0.1);
  \draw [red,thick,postaction={decorate,decoration={markings,mark=at position 0.5 with {\arrow{<}}}}] (0.5,-0.1) -- (2,-0.1);
  \draw [red,thick, postaction={decorate,decoration={markings,mark=at position 0.5 with {\arrow{>}}}}] (-0.5,0.1) arc [start angle=180, end angle=0, radius=0.5];
  \draw [red,thick, postaction={decorate,decoration={markings,mark=at position 0.5 with {\arrow{<}}}}] (-0.5,-0.1) arc [start angle=-180, end angle=0, radius=0.5];
  \node [above right] at (0.5,0.1) {$\els$};
  \node [below left] at (2.1,1.3) {$\boxed{\mu}$};
  \node [circle,fill,inner sep=0pt,minimum size=2pt] at (0.15,0.15) {};
\node [above right] at (0.13,0.13) {$s$};
\node [fill,inner sep=0pt,minimum size=4pt] at (0,0) {};
\node [fill,inner sep=0pt,minimum size=4pt] at (0.3,0) {};
\node [below] at (0.3,0) {$-t$};
\draw [dashed,red,thick, postaction={decorate,decoration={markings,mark=at position 0.5 with {\arrow{<}}}}] (-2,0.1) arc [start angle=180, end angle=0, radius=2];
\draw [dashed,red,thick, postaction={decorate,decoration={markings,mark=at position 0.5 with {\arrow{>}}}}] (-2,-0.1) arc [start angle=180, end angle=360, radius=2];
\node [above,left] at (2.3,0.2) {$\infty$};
\end{tikzpicture}
    \caption{Structure of the pole subtracted amplitude, and the integration contour used for the dispersion relation. The dashed line corresponds to the two arcs at infinity which close the two disjoint contours in the upper and lower half plane. The value of $s$ in the dispersion relation is taken inside the semi-circular region near the origin. The boxes denote the start of the two branch cuts, the $s$-channel going from $\mu=0$ to the right and the $u$-channel from $\mu=-t$ to the left.}
    \label{fig:contour}
\end{figure}
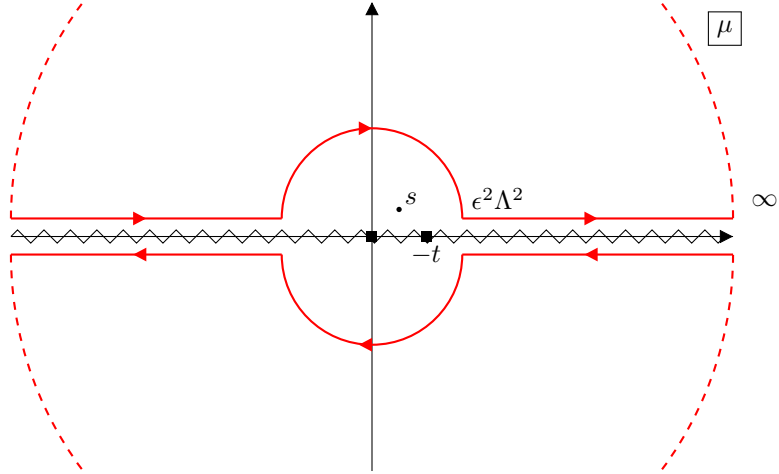
From our assumptions on the UV theory, the analytic structure of the amplitude in the complex $s$ plane for fixed $t<0$ is known, and depicted in Fig.~\ref{fig:contour}. Dispersion relations are an essential tool in connecting these UV assumptions to the amplitudes in the EFT; here, we give a brief summary of their derivation. Consider the integral,
\begin{equation}
    \frac{1}{2\pi\ri}\left(\oint_+ +\oint_-\right)\frac{\cA_s(\mu,t)}{(\mu-s)^3}\,\rd\mu=0\,,
\end{equation}
where the two closed contours, labelled $+$ in the upper half plane and $-$ in the lower, are shown in red in Fig.~\ref{fig:contour}. The branch cuts completely cover the real axis and overlap, so we are forced to use this combination of disjoint contours, rather than the one typically used in situations where there is an analytic region between the cuts. From the residue theorem, we know that both integrals are zero as they contain no poles. Additionally, the assumption that the amplitude grows more slowly than $\mu^2$ as $|\mu|\xr\infty$, or in other words that there exists a twice subtracted dispersion relation, implies that the integral over the arcs at infinity vanishes due to the $\mu^3$ suppression in the denominator of the integrand. Separating the remainder of the contour into the path along the branch cuts and the two arcs at the intermediate scale $|\mu|=\epsilon^2\Lambda^2$ one arrives at the dispersion relation,
\begin{dmath}\label{mdr}
    \frac{1}{2 \pi \ri} \int_{\text{arcs}}\rd\mu\, \frac{\cA_s(\mu,t)}{(\mu-s)^3}= \int^{\infty}_{\epsilon^2\Lambda^2}\frac{\rd\mu}{\pi}\frac{\disc_s\cA_s(\mu,t)}{(\mu-s)^3}+\int^{\infty}_{\epsilon^2\Lambda^2-t}\frac{\rd\mu}{\pi}\frac{\disc_s\cA_u(\mu,t)}{(\mu-u)^3}\,,
\end{dmath}
where the integral on the left-hand side is taken \textit{anticlockwise} on the arcs at $|\mu|=\els$, i.e. opposite to how is shown in the above figure. The discontinuity is defined as,
\begin{equation}
    \disc_s\cA(s,t) \equiv \lim_{\epsilon\xr0^+} \frac{1}{2\ri}\left(\cA(s+\ri \epsilon,t)-\cA(s-\ri\epsilon,t)\right)\, ,
\end{equation}
and under the assumption of the Schwarz reflection principle, $\cA(s^*,t)=\cA(s,t)^*$, this is equal to the imaginary part of $\cA(s,t)$. Note that we have also used crossing symmetry to relate the left-hand branch cut discontinuity of $\cA_s$ to the physical right-hand cut of $\cA_u$, which also obeys positivity properties as both processes are elastic.\\

The power of the dispersion relation representation comes from the fact that the left-hand side is a low-energy quantity, computable in an effective field theory, whilst the right-hand side includes information about the amplitude at arbitrarily large energy scales in the UV. Therefore it connects the Wilson coefficients of the EFT to a high-energy quantity that can be constrained by our requirements of a consistent UV completion, and thereby, directly constrain the low-energy theory itself. From this point, positivity bounds are a direct consequence of unitarity, which implies positivity of the discontinuities in the integrals of the right-hand side, for all energies. Taking the limit $|s|\xr0$ and $t\xr0^-$ in the dispersion relation gives positive quantities on both sides,
\begin{equation}
    \frac{1}{2 \pi \ri} \int_{\text{arcs}}\rd\mu\,\frac{\cA_s(\mu,0)}{\mu^3} = \int^{\infty}_{\epsilon^2\Lambda^2}\frac{\rd\mu}{\pi}\frac{\disc_s\cA_s(\mu,0)}{\mu^3}+\int^{\infty}_{\epsilon^2\Lambda^2}\frac{\rd\mu}{\pi}\frac{\disc_s\cA_u(\mu,0)}{\mu^3}>0\, .
\end{equation}
    Inserting the tree-level EFT amplitude $\cA_{++--}(s,0)$ into the left-hand side and computing the integral simply yields the coefficient $g_2$, thus we arrive at the first positivity bound, $g_2>0$.\\

More generally by taking $s$ and $t$ derivatives of the dispersion relation given by Eq.~\eqref{mdr} and then taking the limits $|s|\xr0$ and $t\xr0^-$ we can derive further positivity statements. There are an infinite number of dispersion relations one could obtain in this way, but as the number of derivatives increases so does the number of Wilson coefficients in the bounds. In order to bound the coefficients $g_2,f_2,g_3,f_3,h_3$ we shall first examine the leading three dispersion relations. To condense expressions we introduce the notation,
\begin{equation}
    C(n,m)\equiv \frac{n!/2}{2\pi \ri}\int_{\rm arcs} \rd\mu\,\frac{\dell_t^m \cA(\mu,0)}{\mu^{n+1}}\quad\text{and}\quad I_{s,u}(n,m)\equiv\int^{\infty}_{\epsilon^2\Lambda^2}\frac{\rd\mu}{\pi}\frac{\dell_t^m\disc_s\cA_{s,u}(\mu,0)}{\mu^{n+1}}>0\,,
\end{equation}
which gives the first three dispersion relations as,
\begin{equation}\label{disps}
\begin{aligned}
        \fl(2,0)&=\DS(2,0)+\DU(2,0)>0\,,\\
        \fl(3,0)&=3\DS(3,0)-3\DU(3,0)\,,\\    \fl(2,1)&=\frac{\disc_s\cA_u(\epsilon^2\Lambda^2,0)}{\pi(\epsilon^2\Lambda^2)^3}+\DS(2,1)-3\DU(3,0)+\DS(2,1)\,.\\
\end{aligned}
\end{equation}
Due to positivity and the $\mu$ suppression in the integrands $I(n,m)$, we have the inequality, $$I_{s,u}(n+1,m)<\frac{1}{\els}I_{s,u}(n,m)\,,$$ allowing us to construct positive quantities even when negative factors of the discontinuity integrals appear in the above expressions. From doing this we obtain the following linear positivity bounds,
\begin{equation}\label{bnd1}
    \fl(2,0)>0\,,\quad \fl(3,0)+\frac{3}{\els}\fl(2,0)>0\,,\quad \fl(2,1)+\frac{3}{\els}\fl(2,0)-\frac{\disc_s\cA_u(\epsilon^2\Lambda^2,0)}{\pi(\epsilon^2\Lambda^2)^3}>0\,.
\end{equation}
In the final line, the explicit factor of the discontinuity appears as a consequence of the $t$ dependence of the lower limit of the $u$-channel integral in the dispersion relation. The discontinuity is positive and so serves to improve the strength of the positivity bounds if we are able to compute it in the EFT. 

If we take arbitrary linear combinations of the first three dispersion relations in Eq.~\eqref{disps}, we can derive other positivity statements. In such a combination the coefficient of $\fl(2,1)$ must be positive for the combination to be positive, as it contains $I_s(2,1)$ which, if multiplied by a negative number cannot be compensated by anything in $C(2,0)$ or $C(3,0)$. Then we have a continuum of positive quantities parameterised by $\Omega,\Theta\in\mathbb R$,
\begin{equation}
    C(2,1)-\frac{\disc_s\cA_u(\epsilon^2\Lambda^2,0)}{\pi(\epsilon^2\Lambda^2)^3} +\Theta C(3,0)+\frac{\Omega}{\els}C(2,0) > 0\,,\quad\forall \,\Omega \geq 3\left|\Theta+\frac12\right|+\frac32 \,.
\end{equation}
Since $\fl(2,0)$ is positive, the strongest bound will be obtained when $\Omega$ takes its minimum value. The bound should be then satisfied for any $\Theta$ and so by varying over this parameter we find the equivalent statement,
\begin{equation}\label{bnd2}
\boxed{
    |\fl(3,0)|\,<\frac{3}{\epsilon^2\Lambda^2}\fl(2,0) \quad\cup\quad C(2,1)-\frac12 C(3,0)+\frac{3}{2\els}C(2,0)-\frac{\disc_s\cA_u(\epsilon^2\Lambda^2,0)}{\pi(\epsilon^2\Lambda^2)^3}  > 0\,.}
\end{equation}

Finally, two further $C(n,m)$ coefficients we examine are,
\begin{equation}\label{disps2}
\begin{aligned}
        \fl(4,0)&=12\DS(4,0)+12\DU(4,0)>0\,,\\
        \fl(3,1)&=3\DS(3,1)-3\DU(3,1)+12\DU(4,0)-\frac{3\disc_s\cA_u(\els,0)}{\pi(\els)^4}\,.\\  
\end{aligned}
\end{equation}
We do not include $C(2,2)$ as it receives 1-loop corrections which are logarithmically divergent as $t\xr0^-$. From these expressions we can derive, in a similar manner to before, the linear positivity bounds,
\begin{equation}
\boxed{
        0<C(4,0)<\frac{12}{(\els)^2}C(2,0)\quad\cup\quad C(3,1)+\frac{3}{\els}C(2,1)-\frac{3}{2\els}C(3,0)+\frac{9\,C(2,0)}{2(\els)^2 }>0\,.}
\end{equation}

\subsection{Weak coupling assumption}

All the above positivity bound statements can be applied to the various elastic photon amplitudes that were derived in Section \ref{ihamp}. These positivity bounds accounted for the presence of loops in the EFT, which would generate non-analyticities at arbitrarily low energy, due to the masslessness of the photon. In the three leading quantities $C(2,0)$, $C(3,0)$, and $C(2,1)$ the 1-loop corrections appear as polynomials in the Wilson coefficients and in particular do not introduce logarithms. At 2-loop or higher, it is possible for this to change and for logarithms to arise. Running loop corrections in quantities such as $C(4,0)$ and $C(3,1)$ (provided that they themselves do not diverge as $t\xr0^-$), which would take the form of logarithms \cite{Haring:2022sdp}, imply that the Wilson coefficients are dependent on the arbitrary dim-reg scale $\mu$ which can be chosen to be $\mu=\els$. With this choice of dim-reg scale the logarithmic corrections which depend on ratios of $\mu$ to $\els$, would vanish leaving some finite corrections.

The assumption that loop corrections to the amplitude in the EFT are suppressed to the point of being negligible, either by a weak coupling in the UV (as for example in perturbative string theory) or a large hierarchy of scales (e.g. for gravitational interactions at low energies) is often termed \textit{weak-coupling}. If we assume that the UV completion involves interactions that are controlled by a dimensionless coupling $g_*^2$ such that the EFT Lagrangian can be organised into a form exhibiting a single mass scale and the UV coupling (see for example \cite{Giudice:2007fh, Liu:2016idz,deRham:2017xox,deRham:2018qqo}):
\begin{equation}
    \cL_{\rm EFT} = \frac{1}{g_*^2}\left(\Lambda^4 L_0(A/\Lambda,\dell/\Lambda)+g_*^2L_1+\ldots\,\right) \,,
\end{equation}
then the loop counting parameter is effectively $ g_*^2$. In this above expression the $L_{\mathfrak{L}}$ functions descend into the EFT from $\mathfrak{L}$-loop corrections in the UV theory. In fact from analysis of Feynman diagrams, one can determine that $\mathfrak{L}$--loop corrections are suppressed by a factor of $(g_*^2/16\pi^2)^{\mathfrak L}$, implying that the theory becomes strongly coupled (which in this context means a breakdown of the loop expansion) at $g_*=4\pi$. As a consequence, if one assumes that $g_*\ll 1$, loop contributions and thus non-analyticities are suppressed throughout the regime of validity of the EFT and positivity bounds can be applied to the tree-level EFT amplitudes.

Finally, we note that even if $g_*>4\pi$, there can be situations in which loop corrections in the EFT become suppressed by a large hierarchy of scales. One notable example of this is in the EFT of gravity, where power counting arguments reveal that loop corrections are ever-more suppressed by factors of $E/\mpl$ where $E$ denotes a typical low-energy scale of the physical system, be it a scattering center of mass energy, or the curvature scale of the gravitational background \cite{Burgess:2003jk}. This suppression however becomes weaker as $E$ increases and so is less robust than the weakly-coupled UV scenario which remains accurate for all energies up to the cutoff.

In the weak-coupling approximation, non-analyticities in the amplitude start at the cutoff $\Lambda_c$ which is generically higher than the scale explicit in the EFT action\footnote{The scale explicitly appearing in the action $\Lambda$ is where classically the effects of irrelevant interactions become comparable in strength to that of relevant and marginal interactions. This typically indicates the scale, at which tree-level perturbative unitarity breaks down at which point either loop corrections or new degrees of freedom must contribute to restore unitarity. If loop corrections do not restore unitarity, i.e. the unitarity breaking is non-perturbative in nature, then new degrees of freedom must arise at this scale, which we then identify as the \textit{cutoff} of the EFT \cite{deRham:2014wfa}.} $\Lambda_c^2\gtrsim\Lambda^2$ and below this scale, the EFT amplitude is polynomial in Mandelstam variables. The corresponding dispersion relation exhibits branch cut integrals $I_{s,u}$ having lower limits starting at $\Lambda^2$. The upshot is that the assumption of weak coupling practically means we can set all loop contributions to zero as well as freely take $\epsilon\xr1$ from below. This also means that the explicit term involving the discontinuity evaluated at $\epsilon^2\Lambda^2$ is not present in the bounds. Whilst this assumption is not strictly necessary in our setup as we could in principle compute the 1-loop corrections, for the sake of comparison to causality bounds it is natural to neglect 1-loop corrections as the scattering time advance/delay is computed in the semi-classical regime.

\subsection{Applying positivity bounds to the EFT: linear bounds}
Ignoring loop corrections in the EFT then we find the above $C$ coefficients to be,
\begin{equation}
\label{LBs}
    \begin{aligned}
        \Lambda^4 C(2,0)&=g_2+f_2\sin (2 \theta ) \sin (2 \chi ) \cos (\psi +\phi )\,,\\
        \Lambda^6 C(3,0)&=3g_3 \cos (2 \theta ) \cos (2 \chi ) \,,\\
        \Lambda^6 C(2,1)&=\frac{1}{2} (-f_3 \sin (2 \theta ) \sin (2 \chi ) \cos (\psi +\phi )+3 g_{3} \cos (2 \theta ) \cos (2 \chi )-3 g_{3}\\&-4 h_{3} (\sin (\theta ) \cos (\theta ) \cos (\phi )+\sin (\chi ) \cos (\chi ) \cos (\psi )))\,,\\
        \Lambda^8 C(4,0)&=12 (2 f_{4} \sin (2 \theta ) \sin (2 \chi ) \cos (\psi +\phi )+g_{4})\,,\\
        \Lambda^8 C(3,1)&=6 (2 f_{4} \sin (2 \theta ) \sin (2 \chi ) \cos (\psi +\phi )+g_{4})-3 (2 g_{4}+g_{4}') \cos (2 \theta ) \cos (2 \chi )\,,
    \end{aligned}
\end{equation}
and the linear positivity bound statements allowing $\epsilon\xr1^-$ are,
\begin{equation}
\label{L1}
\begin{aligned}
    g_{2}+ f_{2} \sin (2 \theta ) \sin (2 \chi )\cos (\psi +\phi )>  \left| g_{3} \cos (2 \theta ) \cos (2 \chi )\right| \,,
    \end{aligned}
\end{equation}
\begin{equation}
\label{L2}
\begin{aligned}
    6 g_{2}&> 6 g_{3}+8 h_{3} \sin (\chi ) \cos (\chi ) \cos (\psi )+8 h_{3}  \sin (\theta ) \cos (\theta ) \cos (\phi )\\&\phantom{<}+\sin (2 \theta ) \sin (2 \chi ) \cos (\psi +\phi ) \left(-6 f_{2}+2 f_{3} \right)\,
\end{aligned}
\end{equation}
\begin{equation}\label{L3}
    \begin{aligned}
        f_{2} \sin (2 \theta ) \sin (2 \chi ) \cos (\psi +\phi )+g_{2}>2 f_{4} \sin (2 \theta ) \sin (2 \chi ) \cos (\psi +\phi )+g_{4}>0\,,
    \end{aligned}
\end{equation}
\begin{equation}\label{g4pb}
    \begin{aligned}
                  \sin (2 \chi ) &(2 \sin (2 \theta ) (3 f_{2}-f_{3}+8 f_{4}) \cos (\psi +\phi )-4 h_{3} \cos (\psi ))+6 g_{2}+8 g_{4}\\&>6 g_{3}+4 (2 g_{4}+g_{4}') \cos (2 \theta ) \cos (2 \chi )+4 h_{3} \sin (2 \theta ) \cos (\phi )\,. 
    \end{aligned}
\end{equation}
From the first inequality one can obtain the bounds on $g_2$ and $f_2$, namely
\begin{equation}
    g_2>0\,,\quad g_2>|f_2|\,.
\end{equation}
Note that it is not possible to have $g_2=0$ in any unitary interacting theory, as $g_2$ can be expressed as an integral over a positive discontinuity which is non-zero if there are interactions. In fact it can be seen from the above bounds that if $g_2$ were to go to zero, all other Wilson coefficients under consideration would be forced to also vanish.

The dependence on angles makes it difficult in general to immediately read off the strongest constraints on the Wilson coefficients; however, two simple statements are
\begin{equation}
    g_4>2|f_4|\,,\quad g_2>|f_2-2f_4|+g_4\,.
\end{equation}
Notice that the last inequality provides the upper bound on $g_4$, as can also be seen in the plots below.

Finally, whilst we find the bound \eqref{g4pb}, for the purposes of comparison to causality constraints we shall not impose it on the Wilson coefficients in the plots shown in this paper as the latter constraints are independent of $g_4'$ as is explained below. 

\subsection{Non-linear bounds}
In the spirit of \cite{Tolley:2020gtv} we may use the Cauchy-Schwarz inequality to derive non-linear bound statements. One can define a normalised positive distribution from the $s$ and $u$-channel discontinuities, then Cauchy-Schwarz inequality implies,
\begin{equation}
    I_{s,u} (3,0)^2<I_{s,u} (2,0)I_{s,u} (4,0) \,.
\end{equation}
The latter term can be related to $C(2,0)C(4,0)$ resulting in the inequality
\begin{equation}
\label{NL1}
    \frac{4}{3} C(3,0)^2<C(2,0)C(4,0)\,.
\end{equation}
In a similar way, one can derive the bound for $C(2,1)$,
\begin{equation}
\begin{split}
    C(2,1)-\frac{1}{2}C(3,0)=I_s(2,1)+I_u(2,1)- \frac32( I_s(3,0)+I_u(3,0)) \, .
    \end{split}
\end{equation}
The negative term on the right-hand side can be made positive by adding a larger quantity, again given by the Cauchy-Schwarz inequality,
\begin{equation}
    \left(I_s(3,0)+I_u(3,0)\right)^2<\left(I_s(2,0)+I_u(2,0)\right)\left(I_s(4,0)+I_u(4,0)\right) \, .
\end{equation}
The right-hand side of this inequality can be written in terms of $C(2,0)C(4,0)$ which leads us to the bound,
\begin{equation}
    C(2,1)-\frac{1}{2}C(3,0)+\frac{\sqrt{3}}{4}\sqrt{C(2,0)C(4,0)}>0 \,.
\end{equation}
In summary, we have the two non-linear bounds involving $C(4,0)$,
\begin{equation}
\label{NL12}
    \boxed{
    \frac{4}{3} C(3,0)^2<C(2,0)C(4,0)\,,\quad\cup\quad C(2,1)-\frac{1}{2}C(3,0)+\frac{\sqrt{3}}{4}\sqrt{C(2,0)C(4,0)}>0\,.
    }
\end{equation}
Given the inequality $0<C(4,0)<12C(2,0)/(\els)^2$, these two non-linear bounds are strictly stronger than the two linear bounds given in Eq.~\eqref{bnd2}.

In terms of EFT parameters, the two non-linear bounds are,
\begin{equation}
\label{NL2}
\left(g_2+f_2\cos{(\phi+\psi)\sin{(2\theta)}\sin{(2\chi)}}\right)\left(g_4+2 f_4\cos{(\phi+\psi)\sin{(2\theta)}\sin{(2\chi)}}\right)>g_3^2\cos^2{(2\theta)}\cos^2{(2\chi)}\,,
\end{equation}
\begin{equation}
\begin{split}
    &3\sqrt{\left(g_2+f_2\cos{(\phi+\psi)\sin{(2\theta)}\sin{(2\chi)}}\right)\left(g_4+2 f_4\cos{(\phi+\psi)\sin{(2\theta)}\sin{(2\chi)}}\right)}\\&>
    3 g_3+2 h_3 (\sin{(2\theta)}\cos{\phi}+\sin{(2\chi)}\cos{\psi})+f_3 \cos{(\phi+\psi)}\sin{(2\theta)}\sin{(2\chi)}\,.
    \end{split}
\end{equation}
Notice that the last bound becomes the same as $D^{\rm stu}$ lower bound \cite{Tolley:2020gtv} for the full crossing symmetric choice of angles ($\phi=\psi,~\chi=\theta=\pi/4$). The latter bound reads,
\begin{equation}
   -\frac{3}{2}\sqrt{\left(g_2+f_2\cos{(2\phi)}\right)\left(g_4+2 f_4\cos{(2\phi)}\right)} <-\frac12\left(3 g_3+4 h_3 (\cos{\phi})+f_3 \cos{(2\phi)}\right) \,.
\end{equation}
The upper $D^{\rm stu}$ derived in \cite{Tolley:2020gtv} can be formally applied to the photon EFT, 
\begin{equation}
    -\frac12\left(3 g_3+4 h_3 (\cos{\phi})+f_3 \cos{(2\phi)}\right)<8\sqrt{\left(g_2+f_2\cos{(2\phi)}\right)\left(g_4+2 f_4\cos{(2\phi)}\right)} \,.
\end{equation}
However, this bound was consistently formulated only for gapped theories, as it relies on the full crossing symmetry for polynomial expansion of the amplitude. For gapless theories, the right-hand side of this bound will also include logarithmic terms which make it impossible to express the bound through $C(4,0)$.

Remarkably, the bounds \eqref{NL12} already result in restricting the values of $g_3,~f_3, ~h_3$ to be in the compact ranges, regardless of the value of $\epsilon$. Thus, these bounds do not rely on the assumption of weak coupling. Figure \ref{fig:g3h3f3}-a shows that in the case $g_4=1$ these non-linear bounds coincide with the linear bound \eqref{bnd2} for $\epsilon=1$. In general, these bounds provide stronger restrictions on the parameter space of the EFT, especially for smaller values of $g_4$, see Figure \ref{fig:g3h3f3}-b.

\subsection{Visualising positivity bounds}
To visualise the regions of parameter space allowed by positivity bounds, we plot ratios of the Wilson coefficients with $g_2>0$. In other words, in what follows, if a Wilson coefficient $g_n$ appears in the axis of a plot it is understood to represent the ratio $g_n/g_2$. The positivity bounds we use to generate plots are unaffected by $g_4'$ and so we effectively have a six-dimensional space of ratios, in which we select two-dimensional slices by fixing a number of Wilson coefficients. This is effectively equivalent to absorbing the value of $g_2$ into the scale appearing in the effective action, i.e. setting $g_2=1$.

As mentioned earlier, the positivity bounds above depend on several angular variables arising from the indefinite helicity of the incoming state, and in particular the bound holds for any value of these angles. Then, to find the strongest constraint on the Wilson coefficients, we must either vary over these angles or find a way to eliminate angles analytically and obtain the strongest bound as is described in Appendix \ref{app:angle}. In the plots below we use a combination of these approaches. 

\begin{figure}%
    \centering
    \subfloat[\centering ]{\includegraphics[width=0.4\textwidth]{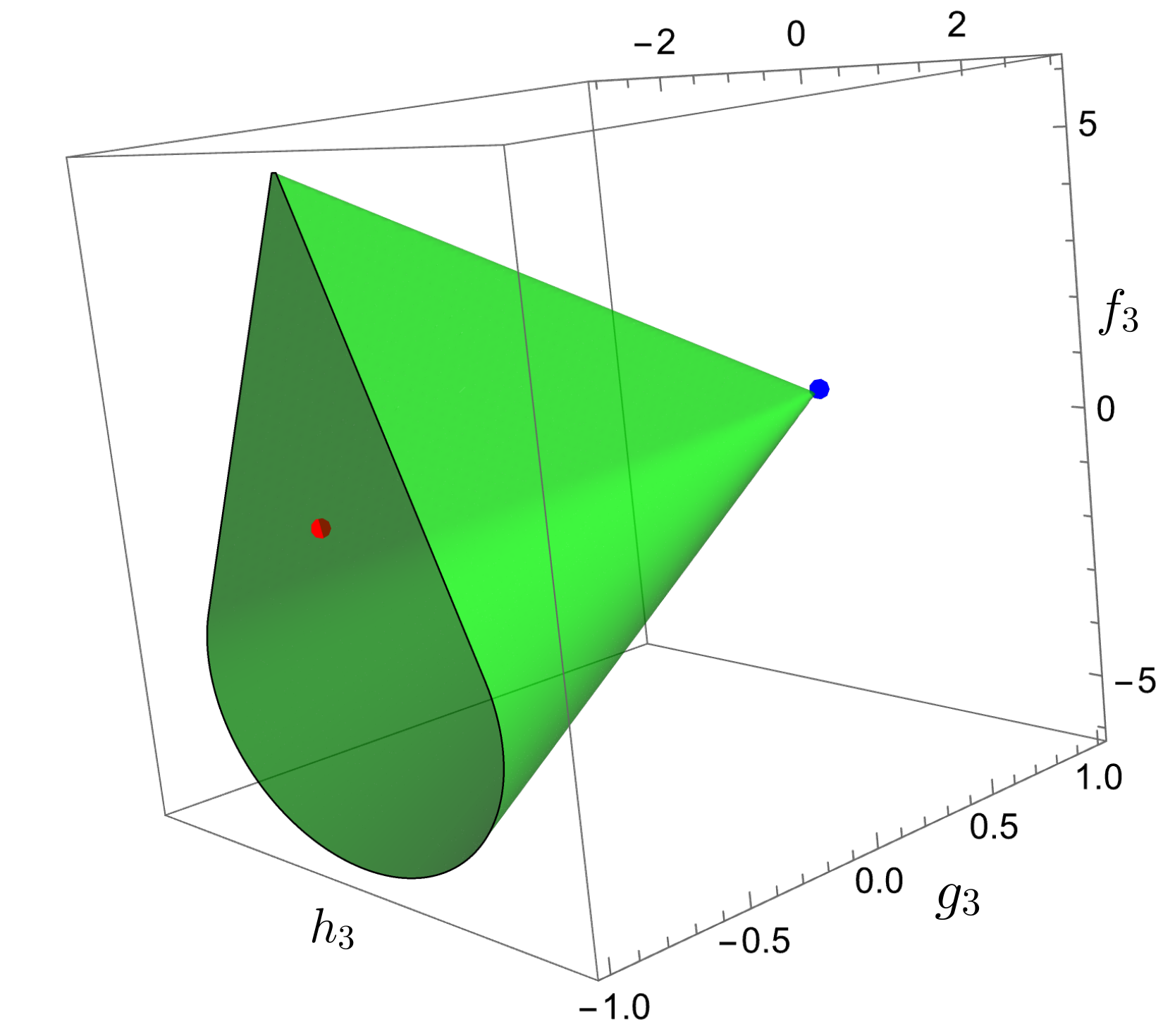}}%
    \qquad
    \subfloat[\centering ]{\includegraphics[width=0.4\textwidth]{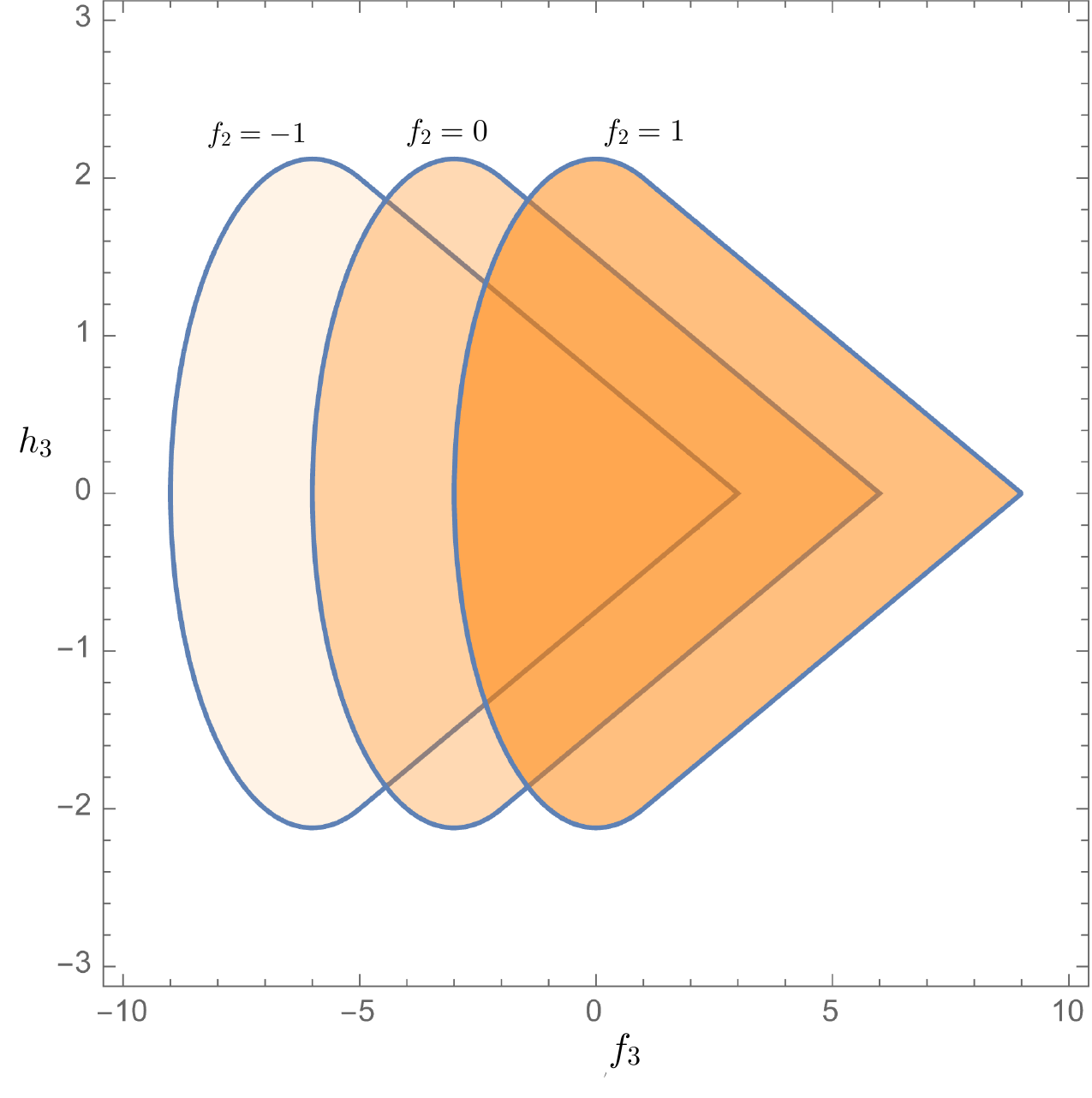}}%
    \caption{(a) Allowed region for $g_3,~f_3,~h_3$ for $f_2=0$, $g_2=1$, $g_4=1$, $f_4=0$. The blue point corresponds to the scalar-axion UV completion (partial UV completion where both scalar and axion have the same coupling strength). The red point represents massive spin-2 (even II) UV completion. (b) $f_3-h_3$ plots for $g_2=1$, $g_3=-1$ and $f_2=-1,~0,~1$. This bound does not depend on the values of $g_4$, $f_4$ and all other couplings in the EFT.} 
    \label{fig:g3h3f3slice}%
\end{figure}

In Figure \ref{fig:g3h3f3slice}-a we take the slice of the bounds corresponding to $g_2=g_4=1$, $f_2=f_4=0$. We present the constraints on $g_3,~f_3,~h_3$ obtained analytically from linear inequalities after optimization with respect to angles. Even though the original inequalities \eqref{L1} and 
\eqref{L2}\footnote{The other linear bounds are weaker than \eqref{L1} and 
\eqref{L2}.} are linear, the outcome of the optimization procedure described in Appendix~\ref{app:angle} leads to non-linear relations. The mixed axion-scalar UV completion, as well as massive spin-2 (even), lie precisely on the boundary of the allowed region.  Figure \ref{fig:g3h3f3slice}-b represents the $f_3-h_3$ slices for different values of $f_2$.

\begin{figure}%
    \centering
    \subfloat[\centering ]{\includegraphics[width=0.4\textwidth]{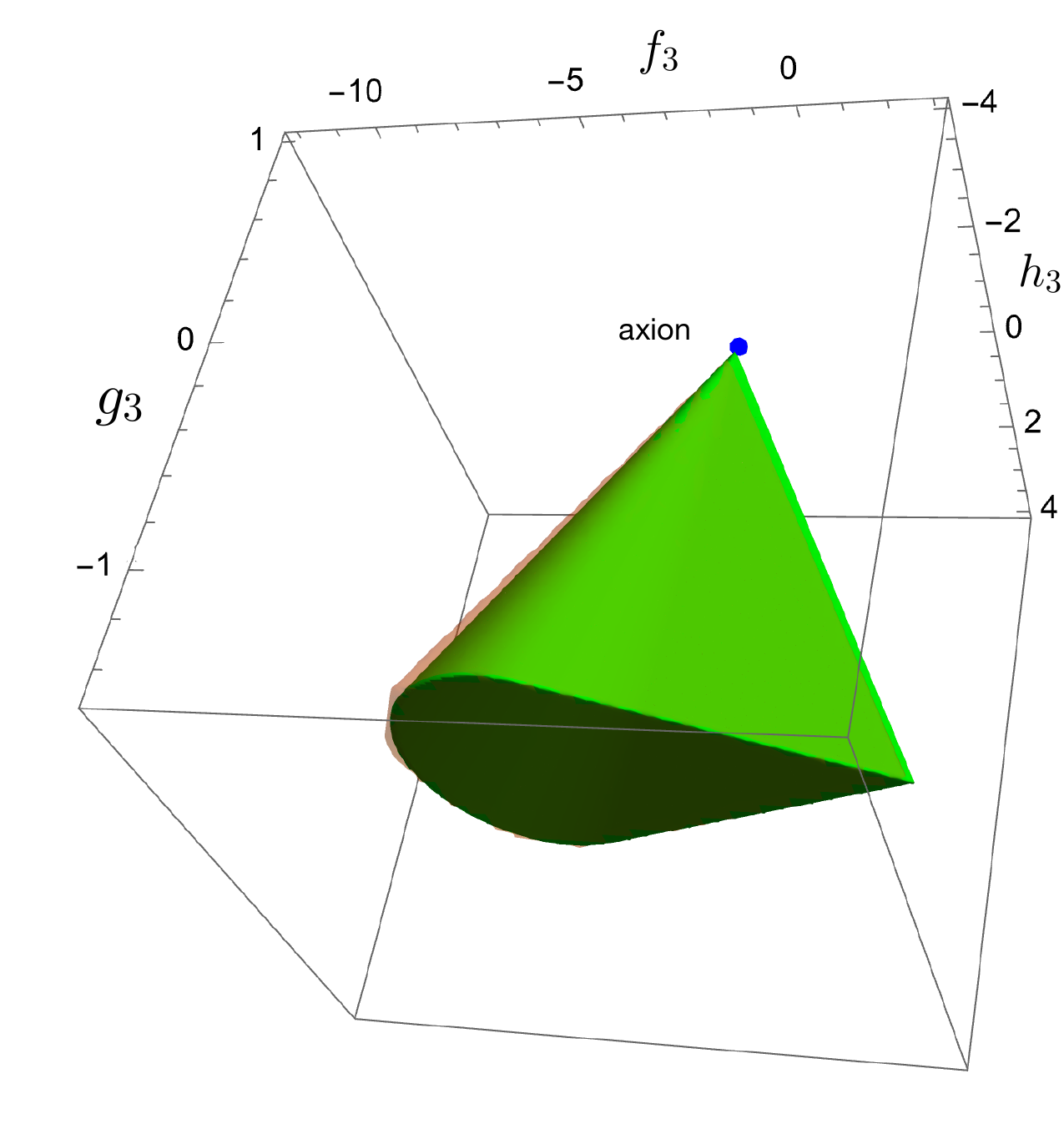}}%
    \qquad
    \subfloat[\centering ]{\includegraphics[width=0.4\textwidth]{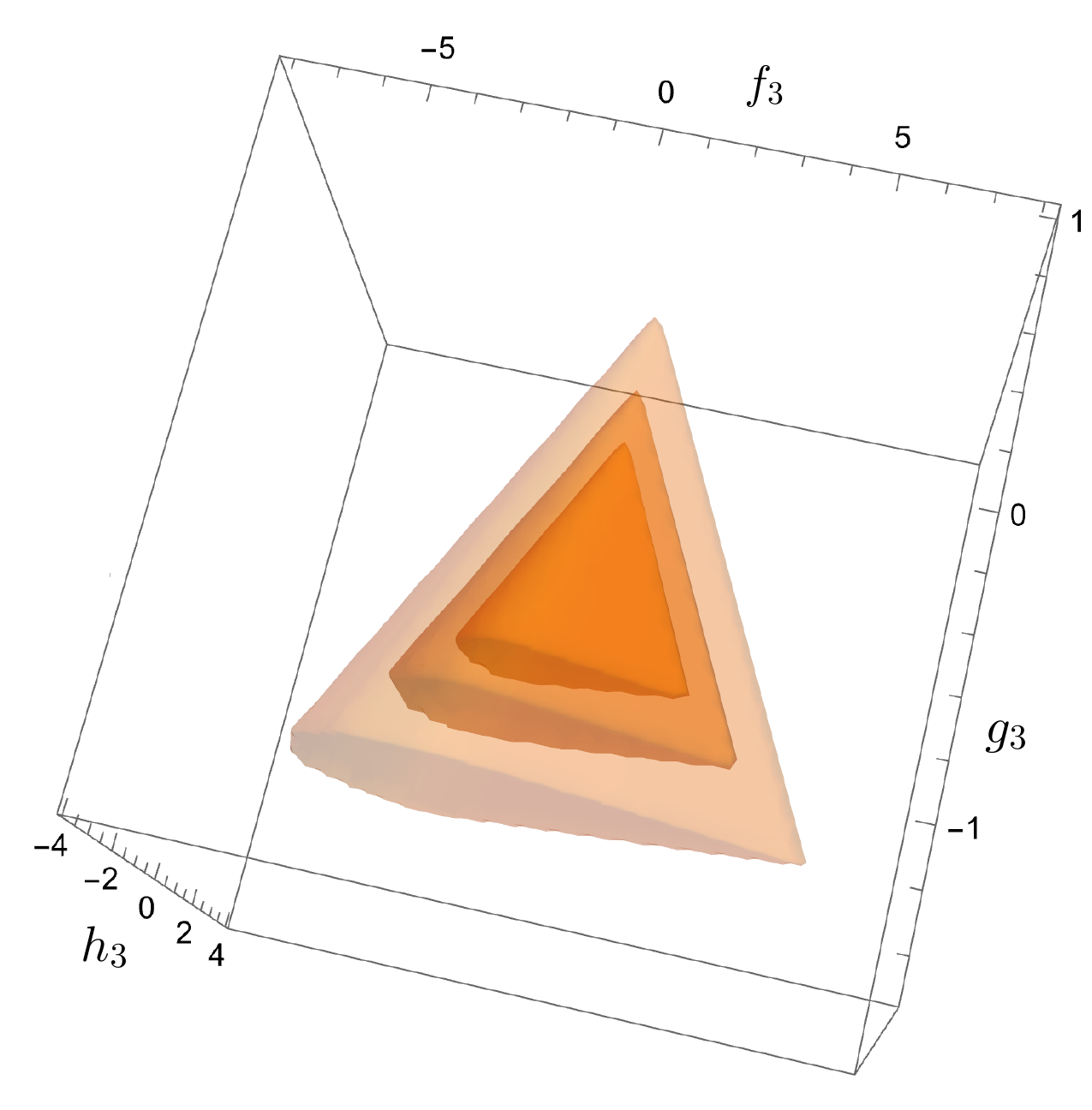}}%
    \caption{(a) Allowed regions from analytically obtained linear ($\epsilon=1$, green) bounds and numerical non-linear bounds of $g_3,~f_3,~h_3$ for $f_2=-1$, $g_2=1$, $g_4=1$, $f_4=-1/2$ (orange-shaded region). These two regions coincide upon a small computational error related to the non-optimal choice of angles in the numerical optimization of inequalities. The blue point corresponds to the axion partial UV completion. (b) Non-linear bounds of $g_3,~f_3,~h_3$ for $g_2=1$, $f_2=0$, $f_4=0$ and for different values of $g_4=1,~1/2,~1/4$. The inner region corresponds to the lowest value of $g_4$. }
    \label{fig:g3h3f3}%
\end{figure}

In Figure~\ref{fig:g3h3f3} we explore the correspondence between the two strongest linear bounds \eqref{L1} and \eqref{L2} derived under the assumption of weak coupling and the non-linear bounds \eqref{NL12} which do not rely on that assumption. We take $g_2=1,~f_2=f_4=0$. The left-hand panel shows the linear bounds after analytical optimization with respect to the angles, taking $\epsilon=1$ (solid green cone) as well as the non-linear bounds \eqref{NL12} after numerical scanning over the angles (orange-shaded cone). The latter only deviates from the linear bounds by a small region which can be entirely attributed to the numerical error. As a non-trivial conclusion, non-linear bounds coincide with linear ones for $g_4=1$. If $g_4<1$, the non-linear bounds are stronger (recall that the linear bounds \eqref{L1} and \eqref{L2} do not depend on $g_4$). We plot the bounds \eqref{NL12} for different values of $g_4$. One can see that the allowed region is shrinking while keeping the same shape when $g_4$ is decreasing.

\begin{figure}
    \centering
    \includegraphics[width=0.6\textwidth]{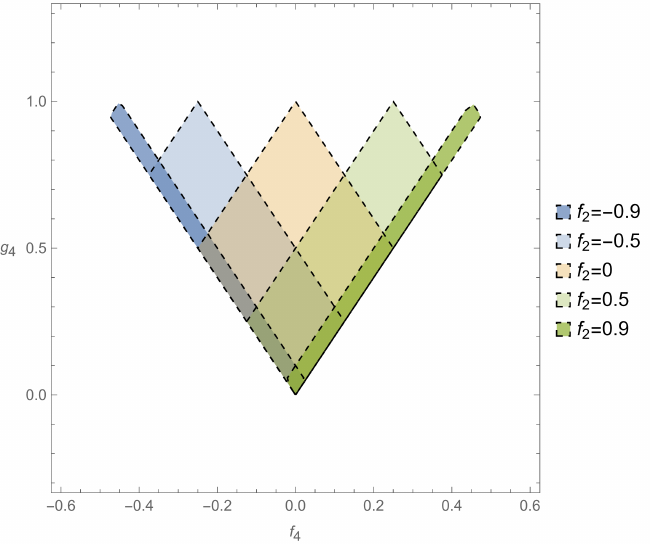}
    \caption{Positivity bounds in the $f_4-g_4$ plane with Wilson coefficients set to $g_3=f_3=h_3=0$, varying $f_2$. As the value of $f_2\xr1$ the allowed region approaches the line segment going from $(f_4,g_4)=(0,0)$ to $(0.5,1)$ and similarly as $f_2\xr-1$ approaches the line segment going from $(f_4,g_4)=(0,0)$ to $(-0.5,1)$.}
    \label{fig:f4g4vf2}
\end{figure}

Last, Figure \ref{fig:f4g4vf2} shows the bounds on the $f_4 - g_4$ plane following from \eqref{L3} for $g_2=1$ at different values of $f_2$. Remarkably, the allowed region shrinks towards a line when the value of $f_2$ becomes close to $\pm 1$. 
We save further analysis of positivity bounds for future work after the comparison with causality bounds.

\newpage
\section{Causality bounds}
In this section, we analyze bounds arising on the Wilson coefficients of the photon EFT from a related, but different perspective than in the previous sections. Here, we will obtain bounds by imposing causal propagation of the two physical photon modes around a non-trivial electromagnetic background. This is a purely low-energy calculation and does not rely any specific assumptions on the UV completion of the theory. We will follow the analysis performed in \cite{CarrilloGonzalez:2022fwg} for a scalar field theory. More specifically, we will consider the propagation of a linearized mode around a spherically-symmetric electromagnetic background in the regime where the scale measuring the variations of the background is much larger than the scale at which the perturbative mode varies. This will allow us to compute the time delay experienced by the mode traveling on a non-trivial background compared to that of a mode traveling in a background with $F_{\mu\nu}=0$. Causal propagation dictates that the time delay is bounded as
\begin{equation}
	\Delta T > -1/\omega \ , \label{eq:CausalBound}
\end{equation}
where $\omega$ is the frequency of the mode. This simply indicates that one should not have a measurable time advance. 

\subsection{Spherically-symmetric backgrounds} \label{sec:spherical}
To proceed we compute the equations of motion (explicitly shown in Appendix \ref{ap:eom}) and consider small fluctuations $\mathcal{A}_{\mu}$ on top of a background $\bar{A}_{\mu}(r)$ such that, in spherical coordinates,
\begin{equation}
	A_{\mu}(t,r,\theta,\varphi) = \bar{A}_{\mu}(r) + \mathcal{A}_{\mu}(t,r,\theta,\varphi) \,,
	\label{eq:Aspherical}
\end{equation} 
where the spherically-symmetric background is given by\footnote{Note that a radial component that depends only on $r$ could be included but this is just a gauge mode that does not contribute to the field strength.}
\begin{equation}
	\bar{A}_{\mu}(r) \mathrm{d} x^{\mu} = \bar{A}_0(r) \mathrm{d}t \,.
\end{equation}
We assume that this background is sourced by an arbitrary, spherically-symmetric, external current $J^\nu$. One is then left with the task to expand the vector perturbations in spherical harmonics. In order to do so, we will assume azimuthal symmetry and thus neglect any $\varphi$ dependence, meaning that we effectively take the quantum number $m$ to be vanishing and then the spherical harmonics reduce to Legendre polynomials instead. We will also assume that the time dependence factorises out in the form $e^{i \omega t}$ where $\omega$ is the frequency of the wave. We parameterise the gauge field as

\begin{equation}
	\mathcal{A}_{\mu} (t,r,\theta) = \frac{1}{r} \sum_{I=1}^4 \sum_{\ell \geq 0} D_I^{\ell}(r) Z^{(I)\ell}_{\mu}(\theta) e^{i \omega t} \,,
\end{equation}
where we take the basis

\begin{align}
	Z^{(1)\ell}_{\mu} &= \delta_{\mu}^t Y_{\ell}(\theta) \,, \nonumber \\
	Z^{(2)\ell}_{\mu} &= \delta_{\mu}^r Y_{\ell}(\theta) \,, \nonumber \\
	Z^{(3)\ell}_{\mu} &= \frac{r}{\sqrt{\ell(\ell+1)}} \delta_{\mu}^{\theta} \p_{\theta} Y_{\ell}(\theta) \,, \nonumber \\
	Z^{(4)\ell}_{\mu} &= \frac{r}{\sqrt{\ell(\ell+1)}} \sin\theta \delta_{\mu}^{\varphi} \p_{\theta} Y_{\ell}(\theta) \,.
\end{align}
The functions $Y_{\ell}(\theta)$ are the usual Legendre polynomials and $\ell$ denotes the order of a given partial wave. The functions $Z^{(I)\ell}_{\mu}$ with $I=1,2,3$ have polar or even parity, whereas $I=4$ has axial or odd parity\footnote{The parity of each mode can be understood by their behavior under a space inversion $\theta \rightarrow \pi - \theta$. Under this transformation we have $\left(Z^{(1,2,3)\ell}_{\mu}, Z^{(4)\ell}_{\mu} \right) \rightarrow \left((-1)^{\ell} Z^{(1,2,3)\ell}_{\mu}, (-1)^{\ell + 1} Z^{(4)\ell}_{\mu} \right)$.}. It follows that the perturbation modes $D_{I}^{\ell}(r)$ are also real and inherit polar parity for $I=1,2,3$ and axial parity for $I=4$, meaning that those two sectors decouple at leading order. One of the $D_I^\ell$ modes can be removed straightforwardly via a gauge transformation of the form $A_{\mu} \rightarrow A_{\mu} + \p_{\mu} \chi$ with 
\begin{equation}
	\chi(t,r,\theta) = - \frac{D_3^{\ell}(r)}{\sqrt{\ell(\ell+1)}} Y_l(\theta) e^{i \omega t} \,. 
\end{equation}
We can now consider the redefinition 
\begin{align}
	u_1^{\ell} &\equiv D_1^{\ell} - \frac{i \omega r}{\sqrt{\ell (\ell + 1)}} D_3^{\ell} \,, \nonumber \\
	u_2^{\ell} &\equiv \frac{1}{r} \left( D_2^{\ell} - \frac{r}{\sqrt{\ell (\ell + 1)}} (D_3^{\ell})' \right) \,, \nonumber \\
	u_4^{\ell} &\equiv D_4^{\ell} \,,
\end{align}
so that that  the vector perturbations take the following form
\begin{equation}
	\mathcal{A}_{\mu} = \left( \frac{u_1^{\ell}}{r} Y_{\ell}, u_2^{\ell} Y_{\ell}, 0, \frac{u_4^{\ell}}{\sqrt{\ell( \ell +1)}} \sin \theta  Y_l' \right) e^{i \omega t} \,.
\end{equation}
Note that we have introduced a $1/r$ factor in the definition of $u_2^{\ell}$ which makes this function dimensionful, however, this will simplify further analysis. By analyzing the equations of motion, one can further notice that one of them corresponds to a constraint that removes the remaining unphysical degree of freedom and we are left only with two propagating modes (given by $u_2^{\ell}$ and $u_4^{\ell}$) as expected for a massless vector field in $4$ dimensions. This constraint equation sets a linear combination of $u_1^{\ell}$, $u_2^{\ell}$, and derivatives of $u_2^{\ell}$ to zero and can be found at any desired order in the EFT expansion. The explicit expression is given in Eq.~\eqref{eq:constraint}. 

We can solve the equations of motion for the 2 remaining physical degrees of freedom, $u_2^{\ell}$ and $u_4^{\ell}$, by removing higher-order radial derivatives iteratively in order to obtain a second-order differential equation for each mode and using the WKB approximation. We work with $\ell>0$ since a vector field does not propagate a monopole mode\footnote{While this is a well known fact, it can be seen explicitly in our setting by setting $\ell=0$ in the equations of motion in which case we are left with only two modes, namely $D^{0}_1$ and $D^0_2$. One of this modes can be removed by a gauge transformation $\chi(t,r,\theta) = -D_1^{0}(r)Y_0 e^{i \omega t}/(i \omega r) \,$. Meanwhile, the second one is set to zero by Maxwell's equations. Thus there are no propagating modes.
}. In this case, it is well-known that we need to take $r \rightarrow e^{\rho}$ to better describe the problem at low $\ell$ \cite{Langer:1937qr}. Once the equations of motion are expressed in this variable, we can remove the friction term and then change variables back to $r$. For more detail on how to perform this calculation see \cite{CarrilloGonzalez:2022fwg,deRham:2020zyh}. We now proceed to solve the equations of motion for the two physical degrees of freedom by using the WKB approximation to find the phase shift experienced by these modes when propagating around non-trivial backgrounds.

\paragraph{Regime of validity of the EFT and WKB approximation}
Following the same  notation as in \cite{CarrilloGonzalez:2022fwg}, \ie we  denote by $r_0$ the typical oscillation length of the background $\bar{A}_0(r)$ and by $\bar{\Phi}_0$ its typical amplitude, which carries mass dimensions. We also introduce the reduced (dimensionless) radial coordinate $R$ defined as
\begin{equation}
	\bar{A}_0(r) = \bar{\Phi}_0 f(R) \, , \quad R\equiv\frac{r}{r_0} \ ,
\end{equation}
The impact parameter of the free theory is designated by $b$, and its reduced partner $B$
\begin{equation}
	\omega b = \ell + 1/2 \,, \qquad B \equiv \frac{b}{r_0} \,.
\end{equation}
In order to stay within the regime of validity of the EFT, we following parameters need to remain small 
\begin{equation}
	\epsilon_1 \equiv \frac{\bar{\Phi}_0}{r_0 \Lambda^2} \ll 1 \,, \qquad \epsilon_2 \equiv \frac{1}{r_0 \Lambda} \ll 1 \,, \qquad \Omega \ \epsilon_2 \equiv \frac{\omega}{r_0 \Lambda^2} \ll 1 \,.
\end{equation}
Furthermore, we will take $\epsilon_1 \ll \epsilon_2$ to neglect the $\epsilon_1^4$ contributions and $\omega r_0\gg1$ as well as $\Omega\gg1$ to ensure the consistency of our expansion within the WKB approximation. The former is simply the requirement for the WKB approximation to hold and tells us that the the typical scale of variation of the perturbation is shorter than that of the background. The latter ($\Omega\gg1$) is required so that higher WKB corrections which cannot be included consistently in the phase shift calculation can be ignored at the order we work at. Overall, this requires the following hierarchy for our parameters
\begin{equation}
	\epsilon_1 \ll \epsilon_2  \ll 1 \ll \Omega \ll \frac{1}{\epsilon_2}  \ . \label{eq:HierarchyParam}
\end{equation}

\paragraph{Time Delay}
Now that we have all the ingredients, we can apply the machinery to get the phase shift and then the time delay. The phase shift experienced by the propagating modes can be found by using the WKB approximation to solve their equations of motion, which are given by
\begin{equation}
	u_I^{\ell}(R)'' = - W_{I,\ell}(R) u_I^{\ell} \,, \qquad \text{for } I=2,4 \,, \label{eq:ModesEOM}
\end{equation}
where the explicit expressions for $W_{I,\ell}(R)$ can be found in Eq.~\eqref{eq:W2} and Eq.~\eqref{eq:W4}. Note that we slightly abuse the notation in the above expression by including the field $u_I^{\ell}$ when in fact  we are now describing the evolution of the field-redefined $u_I^{\ell}$ such that their respective equations of motion are free of any friction terms. The phase shift is found by looking at the behavior of the solution at infinity, that is,  $\lim\limits_{r\rightarrow\infty} u_I^{\ell} \propto\left(e^{2 i \delta_{I,\ell}} e^{i \omega r}- e^{i \pi \ell} e^{-i \omega r}\right)$. Thus, the phase-shift takes the following form
\begin{equation}
	\delta_{I,\ell}(\omega) = (\omega r_0) \left[ \int_{R^{t}_{I,\ell}}^{\infty} \frac{U_{I,\ell}(R)}{\sqrt{1- \left(\frac{R^{t}_{I,\ell}}{R}\right)^2}} \mathrm{d}R + \frac{\pi}{2} \left( B - R^{t}_{I,\ell} \right) \right] \,,
\end{equation}
where $R^{t}_{I,\ell}$ is the turning point for the degree of freedom $u_I^{\ell}$ defined by $	W_{I,\ell}(R^{t}_{I,\ell}) = 0$ and $U_{I,\ell}$ is such that
\begin{equation}
	\sqrt{W_{I,\ell}(R)} = \sqrt{1- \left(\frac{R^{t}_{I,\ell}}{R}\right)^2} + \frac{U_{I,\ell}(R)}{\sqrt{1- \left(\frac{R^{t}_{I,\ell}}{R}\right)^2}} \,.
\end{equation}
The explicit expression for all the functions appearing in this section can be found in Appendix~\ref{ap:SphericalExpr}. Note that we perform an expansion around the turning point in a way as to ensure the convergence of the integral. In particular, we have $U_{I,\ell}(R^{t}_{I,\ell}) = 0$. Furthermore, it is easy to show that $U_{I,\ell} = \mathcal{O}(\epsilon_1^2)$ and $R^{t}_{I,\ell}=B+\mathcal{O}(\epsilon_1^2)$, hence the turning point can safely be replaced by its leading-order value $B$ since any corrections would contribute to $\mathcal{O}(\epsilon_1^4)$ which can be neglected in the expansion scheme we have chosen. The phase shift then reads

\begin{equation}
	\delta_{I,\ell}(\omega) = (\omega r_0) \left[ \int_{B}^{\infty} \frac{U_{I,\ell}(R)}{\sqrt{1- \frac{B^2}{R^2}}} \mathrm{d}R + \frac{\pi}{2} \left( B - R^{t}_{I,\ell} \right) \right] \,.
\end{equation}
From there, obtaining the time delay from the phase-shift is then straightforward,
\begin{equation}
	(\omega \Delta T_{b,I,\ell}(\omega)) = 2 \left. \frac{\p \delta_{I,\ell}(\omega)}{\p \omega} \right|_b = 2(\omega r_0) \left[ \int_{B}^{\infty} \frac{\p_{\omega} \left( \omega U_{I,\ell}(R) \right)}{\sqrt{1- \frac{B^2}{R^2}}}  \mathrm{d}R + \frac{\pi}{2} \left( B - \p_{\omega} \left( \omega R^{t}_{I,\ell} \right) \right) \right] \, , \label{eq:TimeDelay}
\end{equation}
where $|_b$ means that we perform the derivative at fixed impact parameter $b$.

\subsection{Causality Bounds}
In this section, we will work with the scattering amplitudes parameters in Eqs.~\eqref{eq:AmplParam}, instead of the Wilson coefficients of the Lagrangian in Eq.~\eqref{eq:Lagrangian}. This will allow for a straightforward comparison with the positivity bounds in the previous section.

In order to find the causality bounds, we impose the requirement that we cannot get a resolvable time advance, that is, that Eq.~\eqref{eq:CausalBound} is satisfied for both even and odd sectors with the time delay given by Eq.~\eqref{eq:TimeDelay}. The precise method has been described in previously in \cite{CarrilloGonzalez:2022fwg} and is summarized for completion in Appendix~\ref{ap:method}.

\paragraph{Sign-definite contributions}
We will now investigate the contribution to the time delay from the different scattering amplitude parameters, in both the even and odd sector. In particular, we identify the ones that are sign-definite as this will lead us to predict whether the causality bounds will be one-sided or compact. The results can be found in Table \ref{tab:signdef}. First, we define the following positive integrals

\begin{align}
	\mathcal{A}^+ &= 2(\omega r_0) B^2 \int_B^{\infty} \frac{(f'(R)/R)^2}{\sqrt{1-B^2/R^2}} > 0 \,, \\
	\mathcal{B}^+ &= 2(\omega r_0) \int_B^{\infty} \frac{B^2}{R^2} \frac{(f'(R)/R - f''(R))^2}{\sqrt{1-B^2/R^2}} > 0 \,, \\
	\mathcal{C}^+ &= 2(\omega r_0) \int_B^{\infty} \frac{B^2}{R^2} \sqrt{1-B^2/R^2} \left( \frac{f'(R)}{R} - f''(R) \right)^2 > 0 \,.
	\label{eq:positiveInt}
\end{align}
We also introduce the following (non sign-definite) integral for convenience,
\begin{equation}
	(\omega \Delta T_{b,4,\ell}^{(f_3)}(\omega)) = - \frac23 (\omega r_0) \epsilon_1^2 \epsilon_2^2 \int_B^{\infty} \frac{B^2}{R^2} \frac{\left[ \frac{f'(R)^2}{R^2} -(f''(R)^2 + f'(R) f^{(3)}(R)) \right]}{\sqrt{1-B^2/R^2}} \,.
\end{equation}
We simplicity and concreteness, we work with a localized background of the form
\begin{equation}
    f(R)=\left(\sum_{n=0}^p a_{2 n} R^{2 n}\right) e^{-R^2} \ ,
\end{equation}
where the coefficients $a_{2n}$ are taken to be at most of order $1$. This ensures that the background and all its derivatives are under control. For simplicity, we will restrict ourselves to the case where the power of the polynomial is $p=3$, which provides sufficient freedom to derive relatively strong bounds. Furthermore, note that the expansion scheme we have chosen is such that the final time delay is linear in \textit{all} scattering amplitude parameters and hence we have
\begin{equation}
	(\omega \Delta T_{b,I,\ell}(\omega)) = \sum_{J} \mathcal{W}_J (\omega \Delta T_{b,I,\ell}^{(J)}(\omega)) \,,
	\label{eq:WJ}
\end{equation}
where the Wilson coefficients are denoted by $\mathcal{W}_J$ and the index $J$ runs from $1$ to $8$ such that $\mathcal{W}_J = \left\lbrace f_2 , g_2 , f_3 ,g_3 , h_3 ,f_4 , g_{4} , g_{4}' \right\rbrace_J$ (even though we will see that the time delay does not depend on $g_{4}'$ in either sector), and where $\Delta T_{b,I,\ell}^{(J)}$ are numerical factors depending on $\epsilon_1, \epsilon_2, \Omega, B, a_0, a_2, a_4$, and $a_6$ but \textit{not} on the Wilson coefficients $\mathcal{W}_J$. 

\begin{table}[h!]
	\centering
	\begin{tabular}{ c | c | c | c }
	Wilson coefficient & $(\omega \Delta T_{b,2,\ell}(\omega))$ & $(\omega \Delta T_{b,4,\ell}(\omega))$ & Sign-definiteness \\ \hline
	$f_2$ & $\epsilon_1^2 \mathcal{A}^+ > 0$ & $-\epsilon_1^2 \mathcal{A}^+ < 0$ & No \\
	$g_2$ & $\epsilon_1^2 \mathcal{A}^+ > 0$ & $\epsilon_1^2 \mathcal{A}^+ > 0$ & $(+)$ \\
	$f_3$ & $- \frac13 \epsilon_1^2 \epsilon_2^2 \mathcal{B}^+ < 0$ & $ \omega \Delta T_{b,4,\ell}^{(f_3)}(\omega)$ & No \\
	$g_3$ & $- \epsilon_1^2 \epsilon_2^2 \mathcal{B}^+ < 0$ & $-3 (\omega \Delta T_{b,4,\ell}^{(f_3)}(\omega))$ & No \\
	$h_3$ & non sign-definite & $4 (\omega \Delta T_{b,4,\ell}^{(f_3)}(\omega)) + 2 \epsilon_1^2 \epsilon_2^2 \frac{\mathcal{A}^+}{B^2}$ & No \\
	$f_4$ & $24 \epsilon_1^2 \epsilon_2^2 \Omega^2 \mathcal{C}^+ > 0$ & $-24 \epsilon_1^2 \epsilon_2^2 \Omega^2 \mathcal{C}^+ < 0$ & No \\
	$g_{4}$ & $12 \epsilon_1^2 \epsilon_2^2 \Omega^2 \mathcal{C}^+ > 0$ & $12 \epsilon_1^2 \epsilon_2^2 \Omega^2 \mathcal{C}^+ > 0$ & $(+)$ \\
	$g_{4}'$ & $0$ & $0$ & $0$ 
\end{tabular}
	\caption{Contributions to the time delay in the odd and even sectors from various Wilson coefficients.}
	\label{tab:signdef}
\end{table}

 We can see from Table \ref{tab:signdef} that the time delay is \textit{independent} of $g_4'$ and that the coefficients $g_2$ and $g_4$ produce sign-definite contributions to the time delay across both even and odd sectors. Any two-dimensional bound that does not involve $g_2$ and $g_4$ will be a two-sided bound that can lead to a compact causal region. In the following we show several representative two-dimensional bounds. First we analyze the $f_2$ and $g_2$ case which will allows us to fix the value of $g_2$. For the other bounds, we plot a two-dimensional slice of whole $6d$ parameter space ${f_2,f_3,g_3,h_3,f_4,g_4}$, that is, we fix the value of four of those coefficients and plot the bounds on the other two.

\subsubsection{Two-dimensional bounds}

\paragraph{Bounds on $f_2$ and $g_2$}
It is possible to obtain the strongest bounds on the leading order Wilson coefficients by considering a regime of validity of the EFT and WKB approximation in which all higher order corrections are highly suppressed. Instead of the hierarchy for the parameters in Eq.~\eqref{eq:HierarchyParam}, we can take
$\epsilon_2 \ll \epsilon_1  \ll 1$. Within this setting, one can show that
\begin{equation}
	(\omega \Delta T_{b,I,\ell}(\omega)) = \epsilon_1^2 X_I\mathcal{A}^+ + \mathcal{O}(\epsilon_1^2 \epsilon_2^2, \epsilon_1^2 \epsilon_2^2 \Omega^2)  \,,
\end{equation}
where 

\begin{equation}
	X_I = \begin{cases}
	g_2+f_2 \qquad &\text{for } I=2 \,, \\
		g_2-f_2 \qquad &\text{for } I=4 \,.
	\end{cases}
\end{equation}
Thus one can easily see that the time delay is sign-definite for both modes at leading order. The causality constraint  $(\omega \Delta T_{b,I,\ell}(\omega)) > -1$ translates into $X_I > - 1/(\epsilon_1^2 \mathcal{A}^+) \sim - 1/((\omega r_0) \epsilon_1^2)$ where $(\omega r_0) \epsilon_1^2 = \Omega \epsilon_2 (\epsilon_1/\epsilon_2)^2$ can be made arbitrarily large for $\epsilon_2 \ll \epsilon_1$ and hence we get

\begin{equation}
\boxed{	g_2+f_2 > 0\,, \qquad g_2-f_2 > 0 \,. }  \label{eq:boundg2f2}
\end{equation}
Together, this imply that $g_2$ is positive. Note that this constrain includes the predictions from the low energy limit of string theory in four dimensions \cite{TSEYTLIN1986391} in which $f_2=0$ since all the EFT corrections are proportional to powers of  $T_{\mu\nu}T^{\mu\nu}$, where $T^{\mu\nu}$ is the stress-energy tensor of the photon.

From now on, we will set $g_2=1$ which simply fixes the cutoff scale of the EFT to be determined solely by $\Lambda$. The $g_2=0$ case will be treated separately at the end of this section.

\paragraph{Bounds on $f_2$ and $g_3$}
As a first illustrative example, we focus on the bounds obtained in the $(f_2,g_3)$-plane.
In that case, there won't be any sign-definite contributions across both even and odd sectors in the, which allows for compact bounds, but we get one-sided bounds from the even sector, which can easily be explained.
\begin{figure}[!h]
	\begin{center}
		\includegraphics[width=0.5\textwidth]{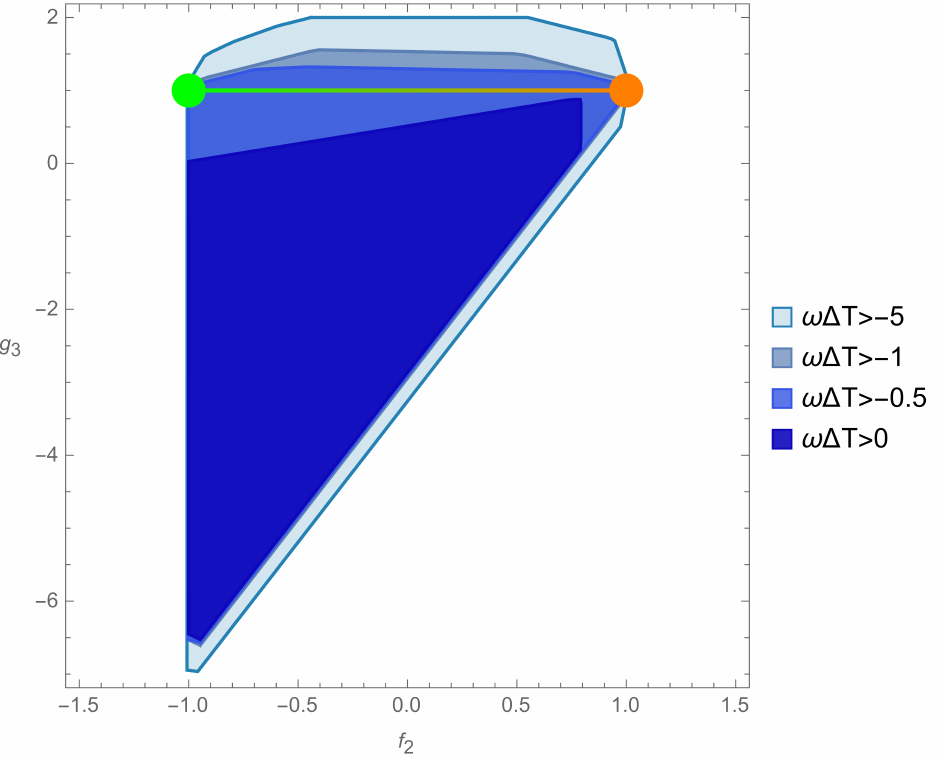}
	\end{center}
	\caption{Causality bounds in the $(f_2, g_3)$-plane for $f_3=3f_2$, $h_3=0$, $f_4=f_2/2$, and $g_4=1$ which are consistent with the values of the scalar and axion partial UV completions. One can see that the points denoting the values of the partial UV completions, scalar in green and axion in orange, lie in the boundary of the causal region $\omega \Delta T >-1$. The line connecting them involves values of a partial UV completion involving both the scalar and axion. Additionally, we show alternative bounds for a different constraint on the time delay. Depending on how negative we allow the time delay to be, we obtain weaker or stronger bounds. More importantly, if we were to require strict positivity of the time delay, this would rule out the axion and scalar partial UV completions, which indeed are causal.}
	\label{fig:plotf2g3SA}
\end{figure}

Starting with the scalar and axion UV completion as depicted in Fig.~\ref{fig:plotf2g3SA}, we have both $f_3$ and $f_4$ that are proportional to $f_2$ to allow for a smooth transition between the scalar and axion UV completions. The line corresponds to a partial UV completion involving an axion and a scalar whose coefficients are simply the sum of the axion and scalar cases with a parameter (namely $f_2$ in this case) tuning the contribution from each. More precisely, the coefficients in this line are given by 
\begin{equation}
	\mathcal{W}_J= \cos{\theta} \  \mathcal{W}_J^{\text{scalar}} + \sin{\theta} \  \mathcal{W}_J^{\text{axion}} \ , \quad  \theta\in [0,\pi/2] \ ,
    \label{eq:UVLineWJ}
\end{equation}
so that at $\theta=0$ we have the purely scalar case and at $\theta=\pi/2$ we have the purely axionic case. The line precisely lies within our causality bounds whereas the scalar and axion UV completions exactly sit on the boundary.

Along this line joining the scalar and axion UV completions, we set $f_3$ and $f_4$ to be proportional to $f_2$ and hence we have a time delay in the form of $(\omega \Delta T^{(f_2)} + 3\omega \Delta T^{(f_3)} + \omega \Delta T^{(f_4)}/2) f_2 + \omega \Delta T^{(fg_3)} g_3 + ({\rm constant})$. Interestingly, the term multiplying $f_2$ is strictly positive in the even sector, whereas the one in front of $g_3$ is negative. When imposing the causality constraint we end up with an equation of the form $g_3 < {\rm (positive)} f_2 + {\rm (constant)}$ and hence the even sector is responsible for the top and/or left bounds (the same configuration will arise in the $(f_2,f_3)$-plane in Fig.~\ref{fig:plotf2g3SA}), whereas we obtain the rest of the compact region thanks to the less rigid structure in the odd sector. Both UV partial completions and the whole line joining them are exactly contained within our causality bounds.

In Fig.~\ref{fig:plotf2g3SA}, we also plot alternative bounds on the time delay in order to appreciate how the boundary of the causal region changes. Our causality bound requires $\omega \Delta T>-1$, if instead, we allow for a more negative time delay, the bounds would slightly weaken as expected. Similarly, the bound get stronger if we force the time delay to be less negative. It is clear from Fig.~\ref{fig:plotf2g3SA} that if we require strict positivity of the time delay, that is, $\omega \Delta T>0$, then we would rule-out well behaved partial UV completions. This shows that strict positivity is too strong of a requirement and in fact, would rule out known causal theories. Although we show this only for a two-dimensional slice of the parameter space, this can be shown to be a generic feature for all slices. 

\begin{figure}[!h]
	\begin{center}
		\includegraphics[width=0.5\textwidth]{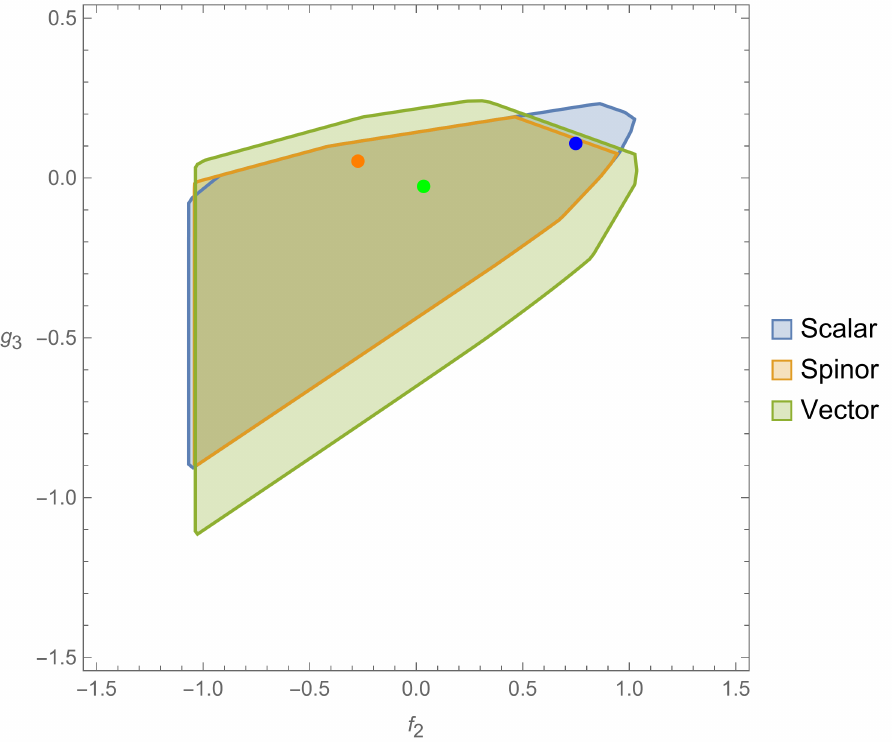}
	\end{center}
	\caption{Causality bounds in the $(f_2, g_3)$-plane for various values of $(f_3,h_3,f_4,g_4)$ that are consistent with the scalar, spinor and vector QED partial UV completions, respectively in blue, orange, and green, and are given in Table \ref{tab:UVcomp}. It is important to note that the three bounds are superimposed but each is derived for different values of $(f_3,h_3,f_4,g_4)$. Just as an indication, we have approximately $(f_3,h_3,f_4,g_4)\sim (0.36,0.04,0.01,0.10)$ for the scalar, $(-0.13,-0.01,0.00,0.02)$ for the spinor and $(0.02,0.00,0.00,0.01)$ for the vector.}
	\label{fig:plotf2g3QED}
\end{figure}
Let us now turn to 1-loop QED where three different set of coefficients are known to correspond to the scalar, spinor, and vector UV completions, as given in Table \ref{tab:UVcomp}. For each case independently, we leave $f_2$ and $g_3$ arbitrary and set the other values so that they match the ones of the relevant UV completion. It is clear that the structure is the same, as in the left and/or top bounds seen in Fig.~\ref{fig:plotf2g3QED} come from the even sector whereas the right and bottom ones arise from the odd sector.

It is interesting to note that the causality bounds for the three sets of different $(f_3,h_3,f_4,g_4)$ coefficients corresponding to the values they take in the scalar, spinor and vector one-loop QED partial UV completions are very similar. In fact, there is no noticeable difference in the even sector bounds. One main difference however is the lower bound, coming from the odd sector. It has the same slope for all three but is noticeably less constraining in the vector case. To understand this, let us turn to the generic equation governing such constraints. The slope is given by the coefficients of $f_2$ and $g_3$, but the vertical displacement of the bound is linear in $\sum_{J} \mathcal{W}_J (\omega \Delta T_{b,I,\ell}^{(J)}(\omega))$ where $J$ runs over the fixed values of $(f_3,h_3,f_4,g_4)$. From our analysis, we see that it is the $h_3$ contribution that is responsible for the discrepancy between the vector case and the other two. Let us focus on this and keep aside the other Wilson coefficients for now. Then, the equation for the lower bound is given by
\begin{equation}
	g_3 > {\rm (positive)} f_2 + \mathcal{H}_3 h_3 + {\rm (constant)} \,,
\end{equation}
where the numerical factor $\mathcal{H}_3$ is optimised such that the combination $\mathcal{H}_3 h_3$ is maximised for a tighter bound. This in turn means that the vertical position of the lower bound only depends on the absolute value $|\mathcal{H}_3 h_3|$. Turning now to Table \ref{tab:UVcomp}, it is easy to see that the value for $|h_3|$ is an order of magnitude smaller in the vector case than both the scalar and spinor and hence, this explains the fact that causality is less constraining in the vector case. There is also a noticeable difference in the bounds on the top right corner which again comes from the odd sector. In this case, the upper bound on $g_3$ is of the form 
\begin{equation}
	g_3 < {\rm (negative)} f_2 +{\rm (positive)} (f_3+4h_3)+{\rm (positive)} h_3+{\rm (positive)}(g_4-2f_4)+ {\rm (constant)} \,,
\end{equation}
where all the contributions from the parameters different from $f_2$ are positive and make the bound weaker. For the scalar QED, the values of the parameters in front of the positive definite contributions are positive and larger than in the spinor and vector QED cases so the bound is weaker in that case. For $f_2\leq 0.5$ the bound is dominated by the even sector constraints so we have no large difference in the bounds. 

\begin{figure}[!ht]
	\begin{center}
		\includegraphics[width=0.6\textwidth]{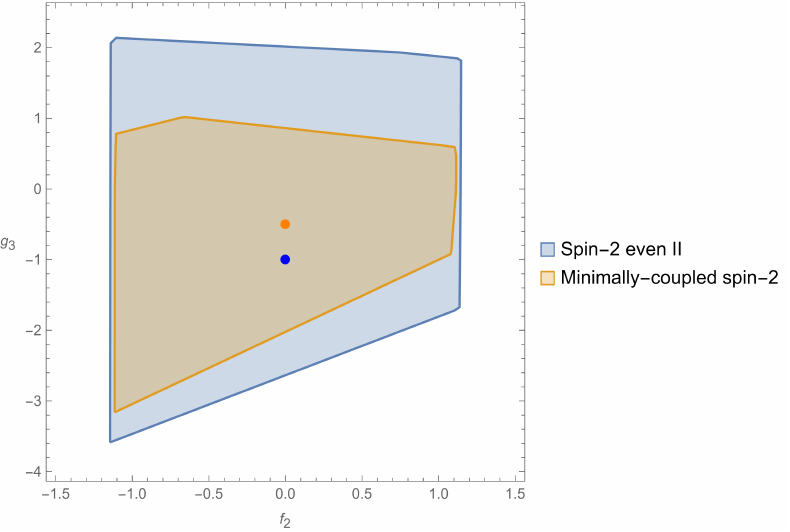}

	\end{center}
	\caption{Causality bounds in the $(f_2, g_3)$-plane with all other coefficients set to the values of the corresponding partial UV completions as in Table \ref{tab:UVcomp}. Precisely, we have $f_3=h_3=f_4=0$ for both and $g_4=1$ and $1/2$ for the blue and orange regions respectively. We see that the partial UV completions lie within the causal region. Note that the spin-2 even II partial UV completion is related to the minimally coupled graviton one by rescaling $g_2$ by a factor of $2$ and keeping all other parameters unchanged.}
	\label{fig:plotf2g3spin2}
\end{figure}

Last, we analyze tree-level UV completions involving spin-2 fields. More precisely, we will look into the partial UV completions constructed in \cite{Henriksson:2021ymi,Henriksson:2022oeu,Haring:2022sdp}. The spin-2 partial UV completion in \cite{Henriksson:2021ymi} is constructed by integrating out a massive spin-2 with a minimal coupling to the photon, that is, a coupling $h_{\mu\nu}T^{\mu\nu}$ where $T^{\mu\nu}$ is the stress-energy tensor of the photon. Meanwhile, the partial UV completions in \cite{Haring:2022sdp} are constructed by on-shell amplitude methods and requirements on the Regge behaviour. We can see that these partial UV completions lie within our causal region. There are additional partial UV theories constructed in \cite{Haring:2022sdp} that involve non-minimal couplings and are not consistent with causality which is not surprising in itself as these partial UV theories are not expected to admit a consistent full UV completion in the first place. We defer the analysis of those theories to Appendix~\ref{app:spin2}.

\begin{figure}[!h]
	\begin{center}
		\includegraphics[width=0.5\textwidth]{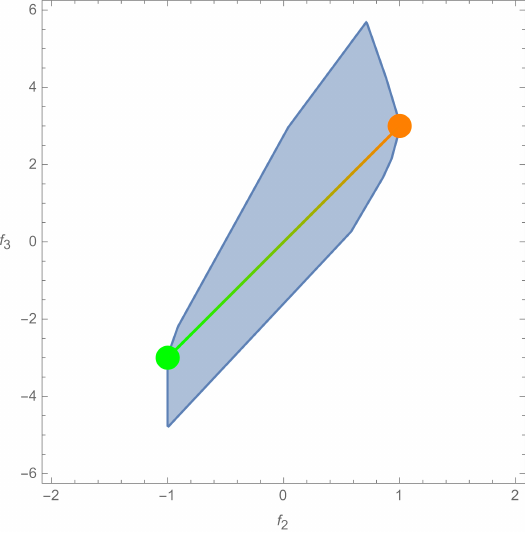}
	\end{center}
	\caption{Causality bounds in the $(f_2, f_3)$-plane for $g_3=1$, $h_3=0$, $f_4=f_2/2$, and $g_4=1$ which are consistent with the values of the scalar and axion partial UV completions. One can see that the points denoting the values of the partial UV completions, scalar in green and axion in orange, lie in the boundary of the causal region. The line connecting them corresponds to a partial UV completion involving both a scalar and an axion.}
	\label{fig:plotf2f3}
\end{figure}

\paragraph{Bounds on $f_2$ and $f_3$}
Neither $f_2$ nor $f_3$ contribute in a sign-definite way to the total time delay (see Table \ref{tab:signdef}) over both even and odd sectors and hence there are no obvious one-sided bounds. This freedom is welcome as it allows for compact causality regions in the $(f_2, f_3)$-plane as can be seen in Fig.~\ref{fig:plotf2f3}.

To better understand these bounds, we can analyse the even and odd sectors separately. Note that we have set $f_4=f_2/2$ so that one can use $f_2$ as a dialing parameter to extrapolate between the scalar and axion UV completions, as can be seen in Fig.~\ref{fig:plotf2f3}. The time delay now reads $(\omega \Delta T^{(f_2)} + \omega \Delta T^{(f_4)}/2)f_2 + \omega \Delta T^{(f_3)} f_3 + {\rm (constant)}$ with the terms multiplying $f_2$ and $f_3$ in this expression being respectively strictly positive and negative in the even sector, similarly to the previous $(f_2,g_3)$ analysis. This means that the even sector will provide upper and/or left bounds depending on the magnitude of the positive slope, $f_3 < {\rm (positive)} f_2 + {\rm (constant)}$. When carefully processing the bounds, we confirm that the left and top bounds come from the even sector, whereas the odd sector has a richer structure and produces the remaining constraints, hence explaining the hard changes of slopes on the right side.

Once again, we plot the whole line joining the scalar and axion UV completion points, parametrized by $f_2$ going from $1$ to $-1$, in Fig. \ref{fig:plotf2f3}. We have the same qualitative behavior, i.e. the end points corresponding to the scalar and axion UV completions are on the boundary of the causality bounds whereas the rest of the line is within.

\paragraph{Bounds on $f_2$ and $h_3$}
\begin{figure}[!h]
	\begin{center}
		\includegraphics[width=0.5\textwidth]{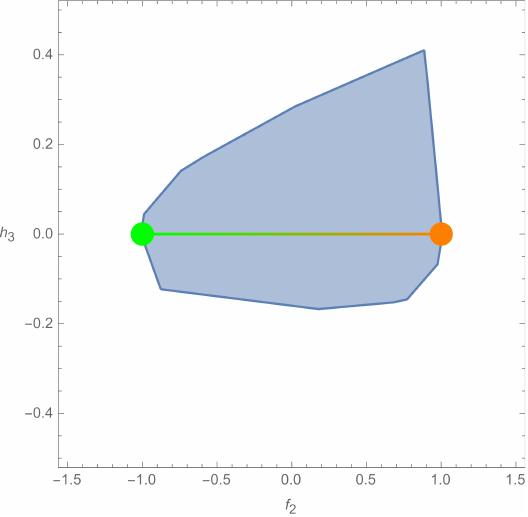}
	\end{center}
	\caption{Causality bounds in the $(f_2, h_3)$-plane for $f_3=3f_2$, $g_3=1$, $f_4=f_2/2$, and $g_4=1$ which are consistent with the values of the scalar and axion partial UV completions. The points denoting the values of the partial UV completions, scalar in green and axion in orange, lie in the boundary of the causal region as in the previous cases. Similarly, the line denoting the mixed partial UV completion of a scalar and axion also lies within the causal region.}
	\label{fig:plotf2h3SA}
\end{figure}
When going to the $(f_2,h_3)$-plane, we choose to set all remaining coefficients in such a way that they can satisfy either the scalar or the axion partial UV completions. This is done by allowing the remaining Wilson coefficients to depend on $f_2$, which will then once again act as a dialing parameter. One can extrapolate between both endpoints and obtain a full segment of partial UV completions as shown in Fig.~\ref{fig:plotf2h3SA}. The segment is fully contained within our causality bounds, with the scalar and axion points exactly lying on the boundary, as was previously the case in the $(f_2,f_3)$ and $(f_2,g_3)$-planes, respectively plotted in Figs.~\ref{fig:plotf2f3} and \ref{fig:plotf2g3SA}.

In this example, neither $f_2$ nor $h_3$ enjoy a sign-definite contribution to the time delay in either of the even and odd sectors. This way, the latter contributed respectively to left and right-sided bounds.

\begin{figure}[!h]
	\center
	\includegraphics[height=5cm]{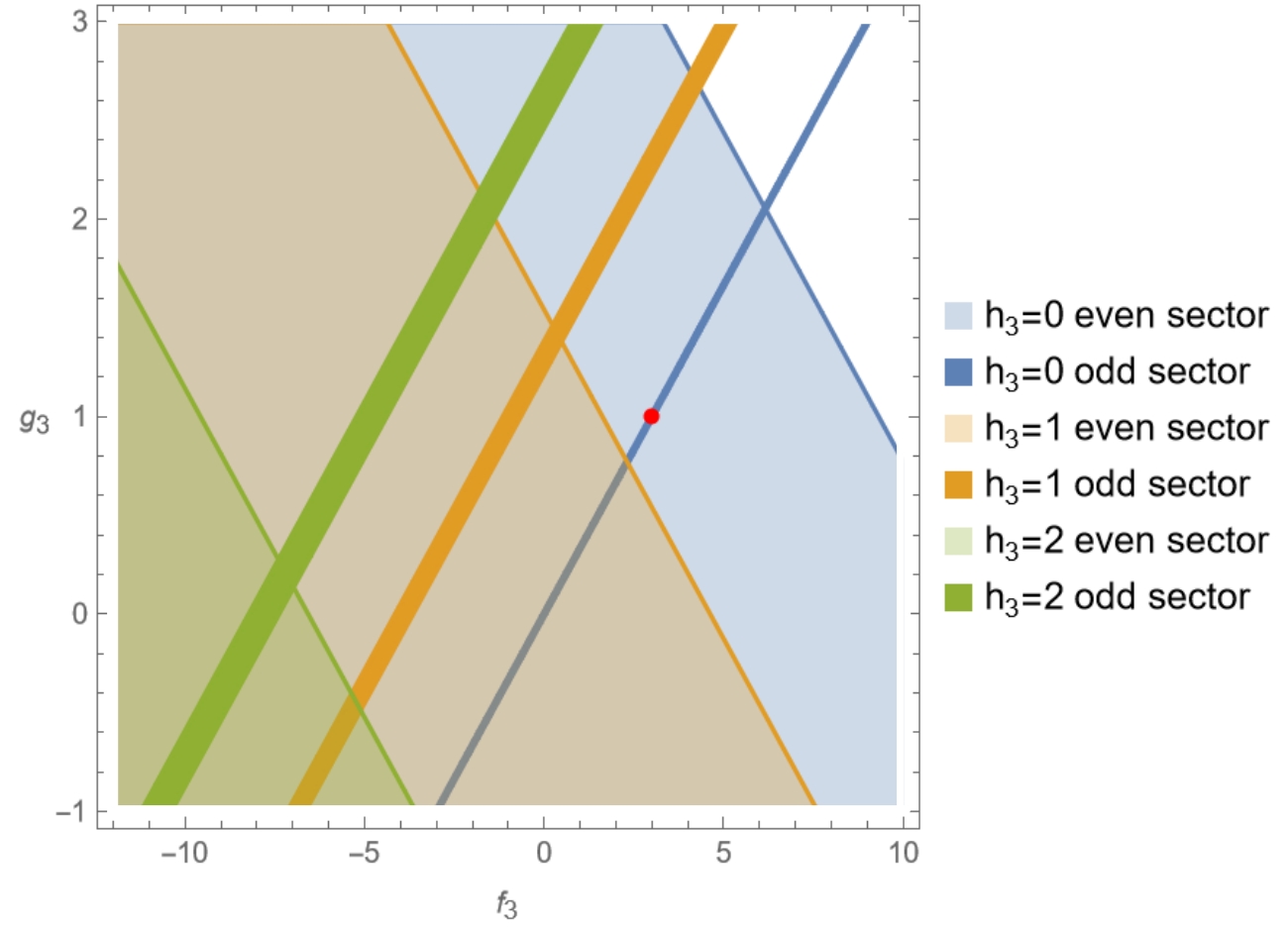}
	\hspace{0.3cm} \includegraphics[height=5cm]{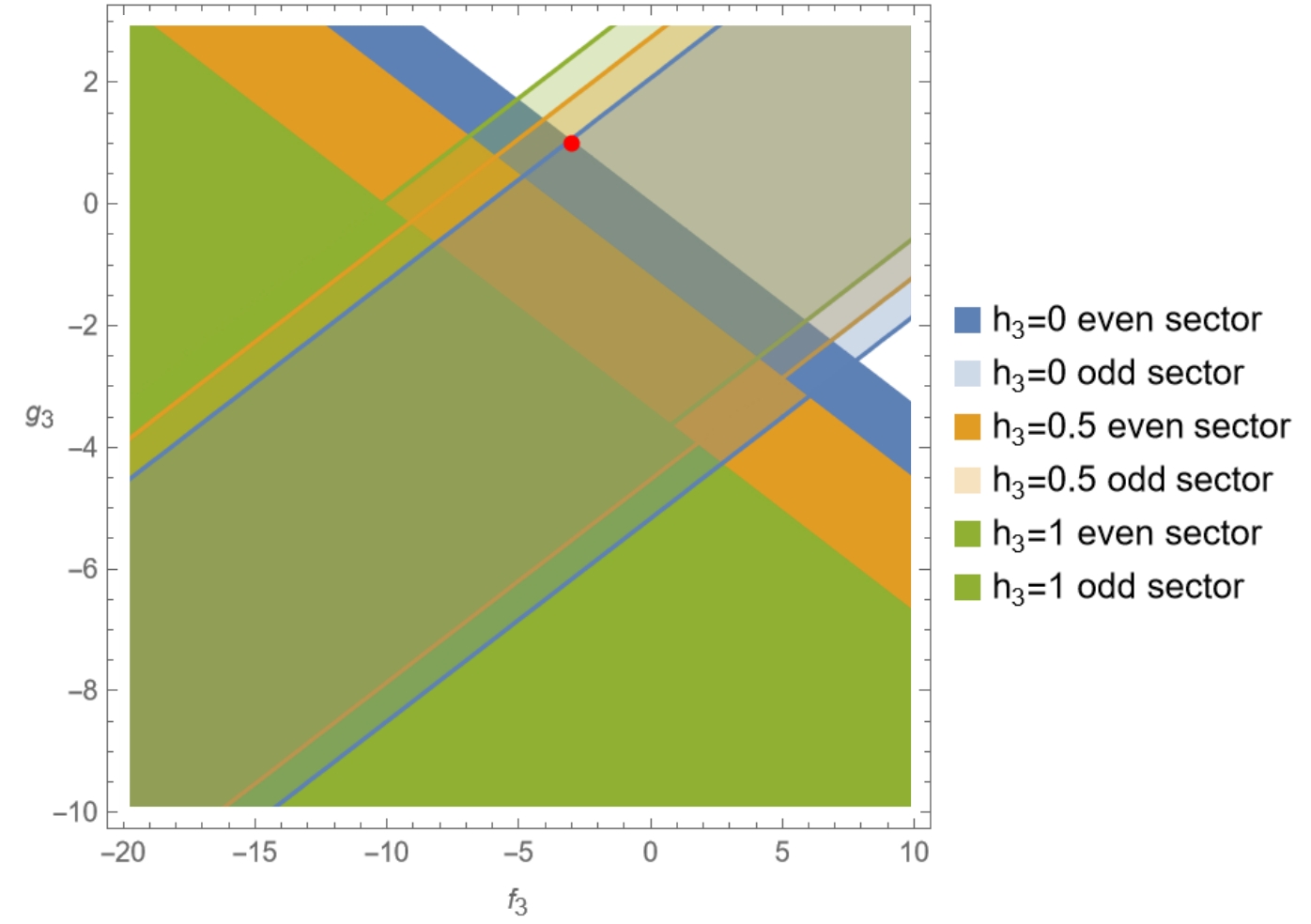}
	\caption{Causal bounds on $f_3-g_3$ plane with varying $h_3$ for the odd and even sectors with the values of the other coefficients fixed as in the scalar partial UV completion on the left (with $(f_2,f_4,g_4)=(1,1/2,1)$) and the axion one on the right (with $(f_2,f_4,g_4)=(-1,-1/2,1)$). The red dot indicates the value of the coefficients for the corresponding partial UV completion. The final causal region is obtained as the intersection of the causal regions for both the odd and even sectors. This region is not compact in the negative $f_3-g_3$ direction. 
 For clarity of the plots, we have chosen to depict 
the even sector as `transparent' and the odd as `solid' on the left plot corresponding the scalar UV completion while this is reverted on the right plot for the  axion completion.}
	\label{fig:plotF3G3}
\end{figure}

\begin{figure}[!h]
	\center
	\includegraphics[height=5cm]{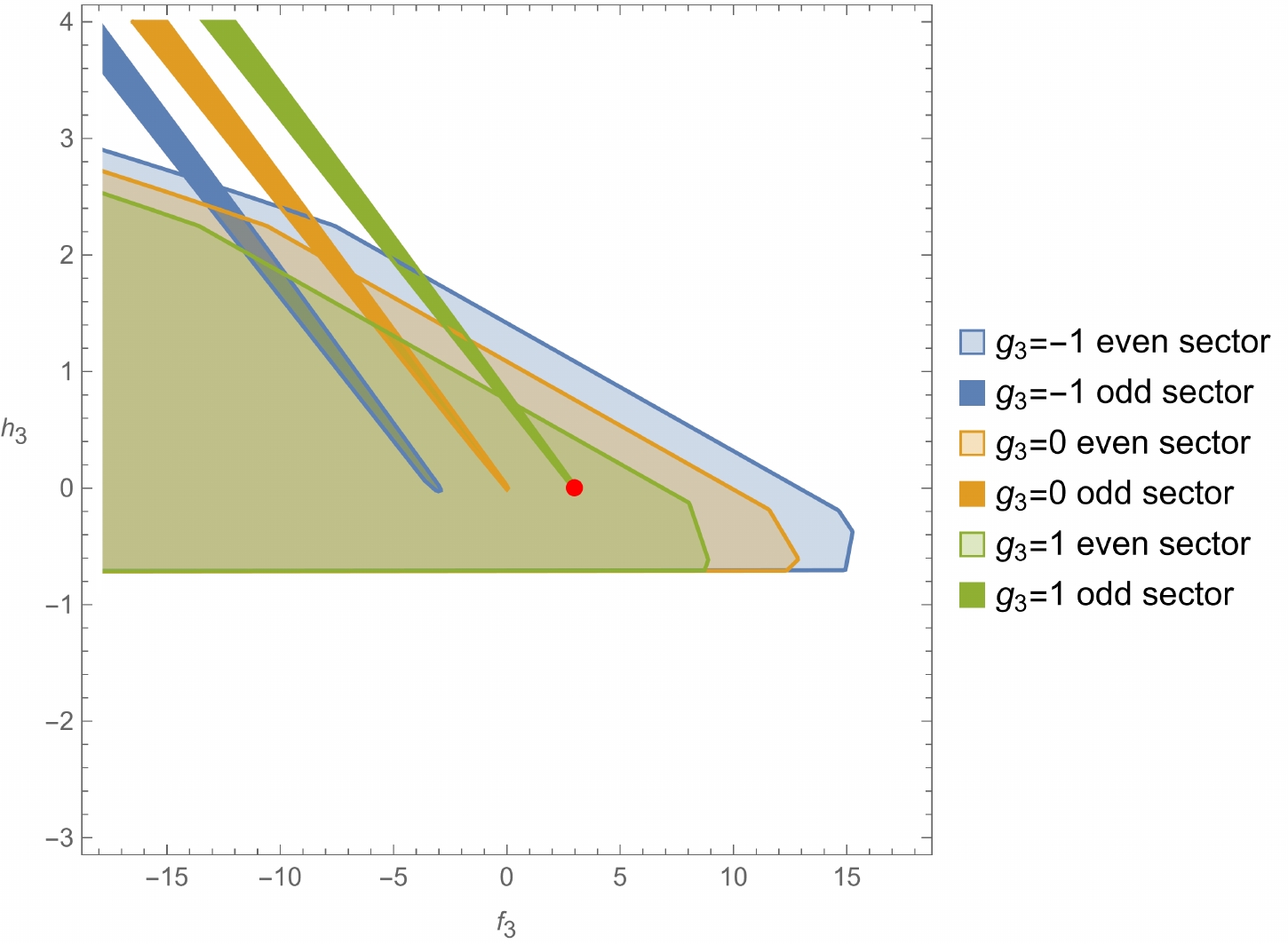}
	\hspace{0.3cm} \includegraphics[height=5cm]{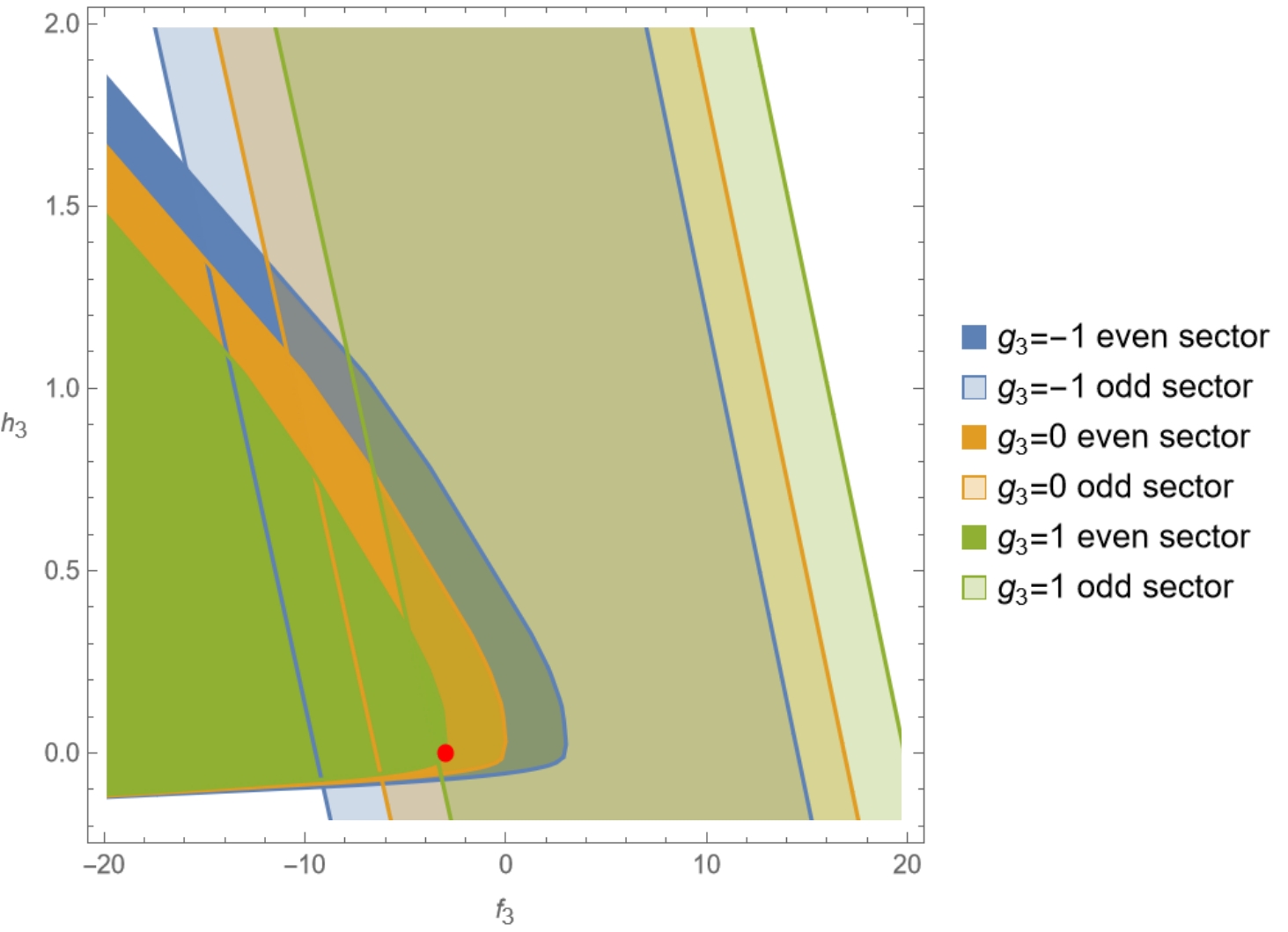}
	\caption{Causal bounds on $f_3-h_3$ plane for the odd and even sectors with varying $g_3$ and all other parameters set by the values of the scalar (left) or axion (right) partial UV completions. More precisely, we have $(f_2,f_4,g_4)=(1,1/2,1)$ and $(-1,-1/2,1)$ for the scalar and axion respectively. The final causal region is obtained as the intersection of the causal regions for both the odd and even sectors. The red dot represents the corresponding values of the partial UV completions and lies in the boundary of the causal region.}
	\label{fig:plotF3H3}
\end{figure}

\begin{figure}[!h]
	\center
	\includegraphics[height=5cm]{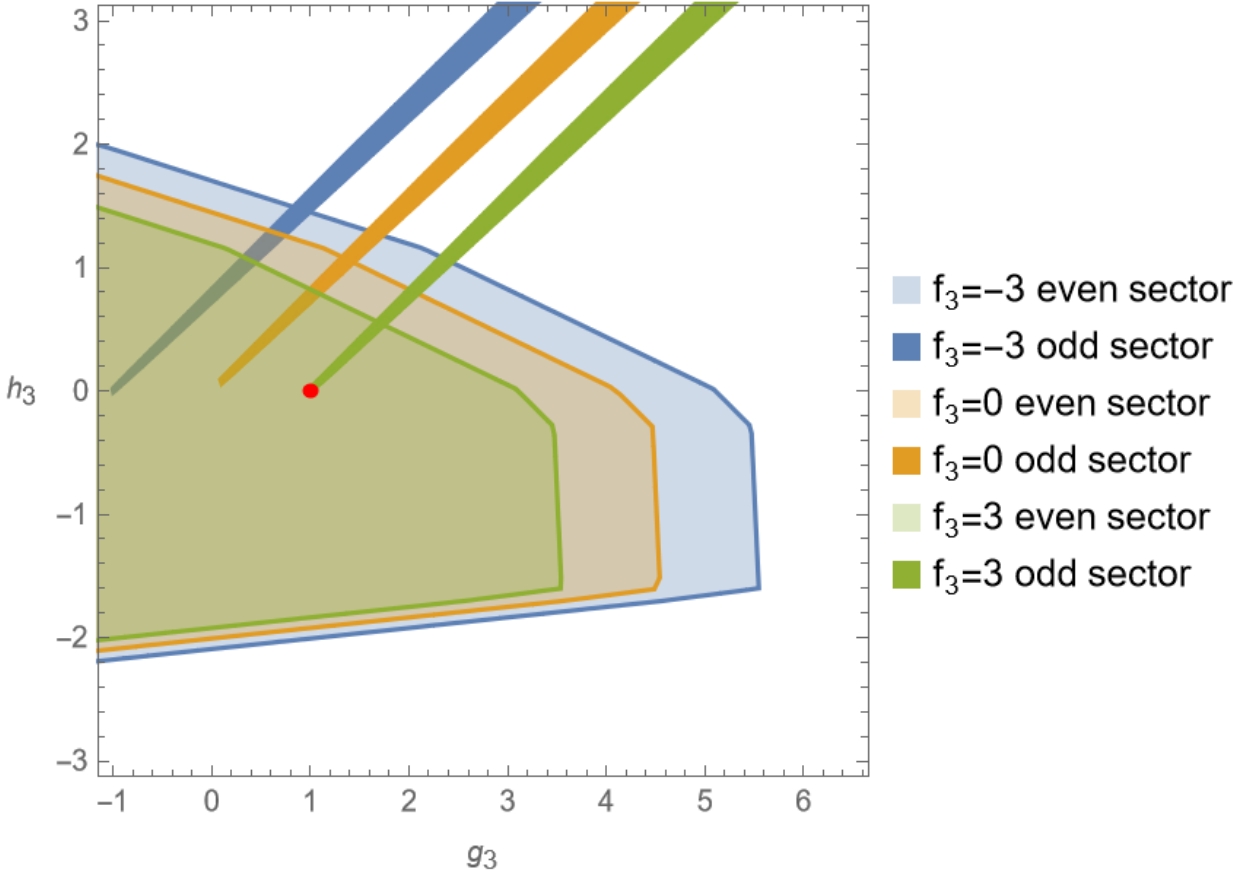}
	\hspace{0.3cm} \includegraphics[height=5cm]{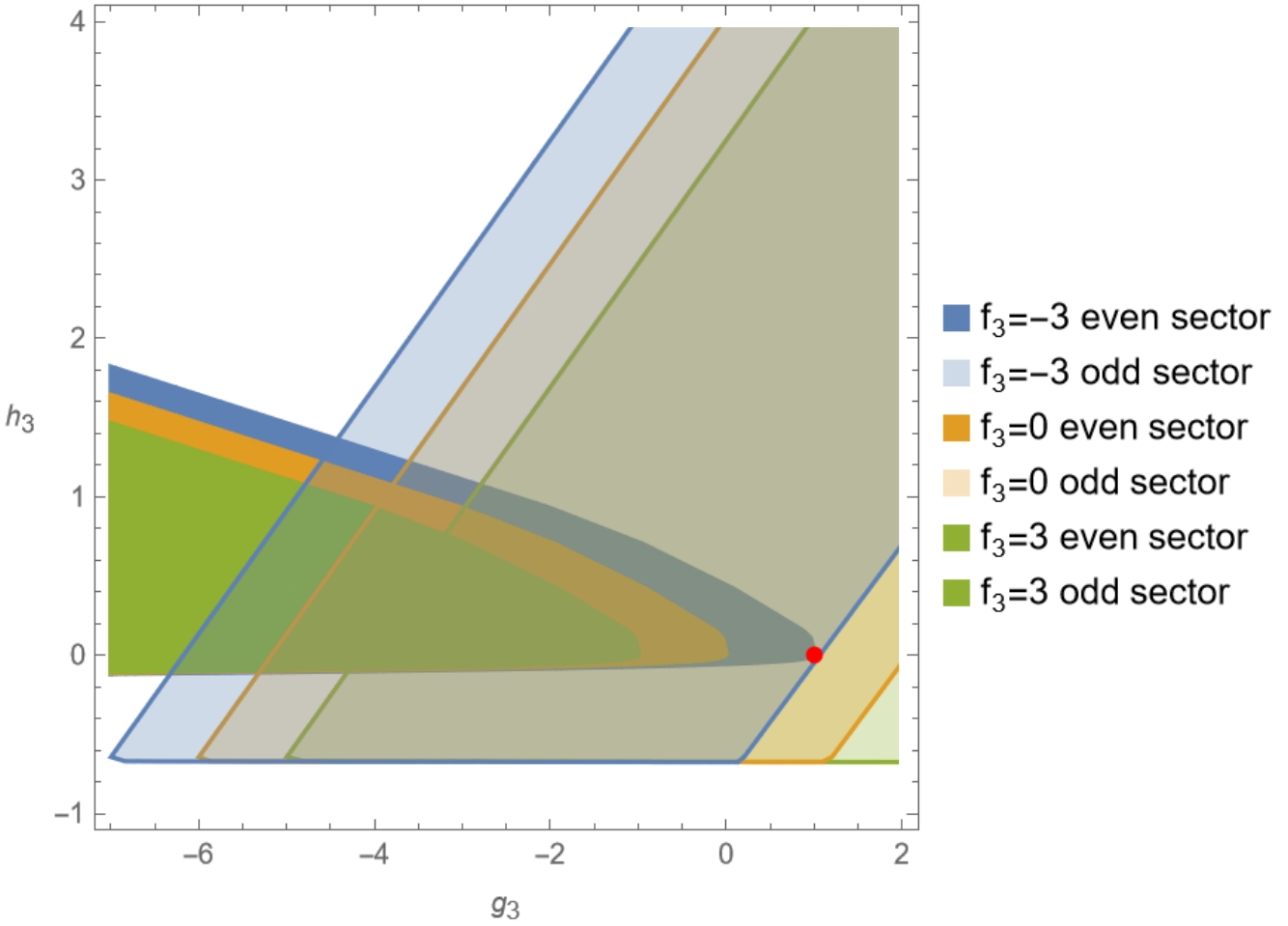}
	\caption{Causal bounds on $g_3-h_3$ plane for the odd and even sectors with varying $f_3$ and all other parameters set by the values of the scalar (left) or axion (right) partial UV completions. Once again, we have $(f_2,f_4,g_4)=(1,1/2,1)$ and $(-1,-1/2,1)$ for the scalar and axion respectively. As in the previous cases, the final causal region is obtained as the intersection of the causal regions for both the odd and even sectors and the red dot represents the corresponding partial UV completion.}
	\label{fig:plotG3H3}
\end{figure}

\paragraph{Bounds on $f_3$, $g_3$, and $h_3$}
We now turn to analyze the bounds for the dimension-$10$ coefficients. For these parameters, we obtain bounds of the following form
\begin{align}
	f_3+3g_{3}&<X_{u^\ell_{2}}+Y_{u^\ell_{2}}h_3 \ ,  \label{eq:f3g3h3even}\\
 -X_{u^\ell_{4}}-Y_{u^\ell_{4}}h_3<f_3-3g_{3}+4h_3&<X_{u^\ell_{4}}+Y_{u^\ell_{4}}h_3\ ,  \label{eq:f3g3h3odd}
\end{align}
where $X_{u^\ell_{2,4}}$ depends on the other amplitude parameters as observed in Figs.~\ref{fig:plotF3G3}, \ref{fig:plotF3H3}, and \ref{fig:plotG3H3} as well as on the specific background. Thus, the value of $X_{u^\ell_{4}}$ and $Y_{u^\ell_{4}}$ can be different for the upper and lower bounds since it will be optimized to have the tightest bounds. This analysis simplifies in the even sector; for the choice of the amplitude parameters as in the axion partial UV completion, we get $X_{u^\ell_{2}}=1/(\epsilon_1^2 \epsilon_2^2 \mathcal{B}^+)$ and in the odd sector for the scalar partial UV completion we have $X_{u^\ell_{4}}=1/|\omega \Delta T_{b,4,\ell}^{(f_3)}(\omega)|$. 

On the $f_3-g_3$ plane, we see that the even sector only gives rise to an upper bound due to the sign definite contribution from both $g_3$ and $f_3$. This upper bound becomes stronger as we increase $h_3$ since this can contribute with a negative time delay that needs to be balanced by a more positive contribution from $f_3$ and $g_3$. Meanwhile, the odd sector contributions are not sign definite and give both upper and lower bounds. The strength of these two-sided bounds does not change drastically when varying $h_3$ since the background is optimized so that the width of the bound on the $f_3$ direction is the smallest possible bound. While the strength of the bounds in the odd sector does not vary drastically with $h_3$, the location does change and moves towards the negative $f_3$ direction as $h_3$ is increased.  Note that the values of $h_3$ considered here are chosen so that we have a non-vanishing causal region, which in these cases requires $h_3\gtrsim0$ as can be seen below.

In the $f_3-h_3$ plane, we should expect upper and lower bounds. This is observed in Fig.~\ref{fig:plotF3H3} where the bounds arising from the even and odd sectors are plotted separately. We can appreciate that each sector separately does not give a compact region; it is the combination of both that gives rise to the compact causal region. On the even sector, the contribution of $f_3$ is sign definite, so we can only get an upper found, while in the odd sector, we have both an upper and lower bound. As in the $f_3-g_3$ bounds, we have tight bounds on the odd sector for the scalar partial UV completion case since the contributions from all the other amplitude parameters vanish and the actual values of the UV completion lie on the boundary. Similarly, we observe that the axion partial UV completion is in the boundary of the even sector causal region. The bounds are not as strong as in the scalar completion case due to the sign-definite contribution of $f_3$. If we were to change the contributions from the dimension-$12$ coefficients, we would be able to tune the background to obtain a negative contribution to the odd sector time delay which leads to a stronger bound. The effect of this is to move the odd sector causal region further up so that it no longer intersects with the even sector causal region. This can be read as further constraints on the Wilsonian coefficients of the dimension-12 operators. 

Last, we analyze the bounds on the $g_3-h_3$ plane. The analysis above ($f_3-h_3$ case) is identical for the $g_3-h_3$ bounds since the contribution of $g_3$ is degenerate with the one in $f_3$. In the even sector, the $f_3$ and $g_3$ contributions have the same sign, but in the odd sector, they have opposite signs, so the only difference will be the change in the sign of the slope of the odd sector bounds as seen in Fig.~\ref{fig:plotG3H3}.

\paragraph{Bounds on $f_4$, $g_4$, and $g'_4$}
The leading order contribution to the time delay from the dimension-$12$ operators is
\begin{equation}
	\omega \Delta T_{b,\{2,4\},\ell} \supset 12 \left(g_{4}\pm 2f_4\right)\epsilon_1^2 \epsilon_2^2 \Omega^2 \mathcal{C}^+  \, , \label{eq:Dim12Contrib}
\end{equation}
where $\mathcal{C}^+>0$ is defined in Eq.~\eqref{eq:positiveInt} and the $+$($-$) sign comes from the even (odd) sector. Thus, we expect to get bounds of the form:
\begin{equation}
	g_{4}+2f_4>Z_{u^\ell_{2}} \ , \quad g_{4}-2f_4>Z_{u^\ell_{4}} \ ,
\end{equation}
where $Z_{u^\ell_{2,4}}$ are numbers determined by the optimization of the background which also depend on the other scattering amplitude parameters. They depend very weakly on $f_2$ and only become slightly tighter as we approach $f_2=\pm 1$ boundaries. Together this give rise to a lower bound $g_4>(Z_{u^\ell_{2}}+Z_{u^\ell_{4}})/2$ as predicted in Table~\ref{tab:signdef}. By optimizing our bounds, we are able to find 
\begin{equation}
	g_4>0 \ ,
\end{equation}
this lower bound can be stronger for specific $f_3, g_3, h_3$ values.

\begin{figure}[!h]
	\center
	\includegraphics[height=5cm]{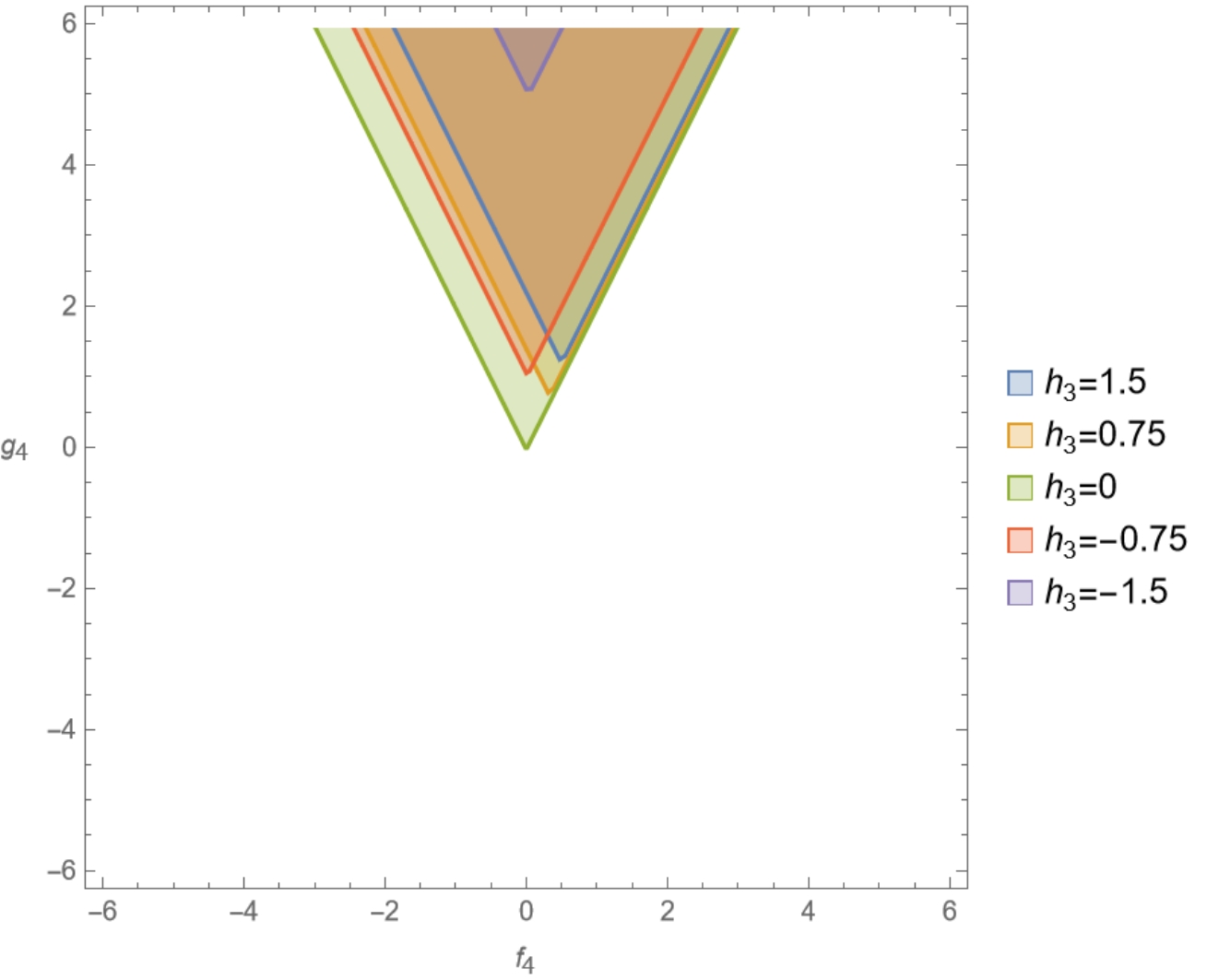}
	\includegraphics[height=5cm]{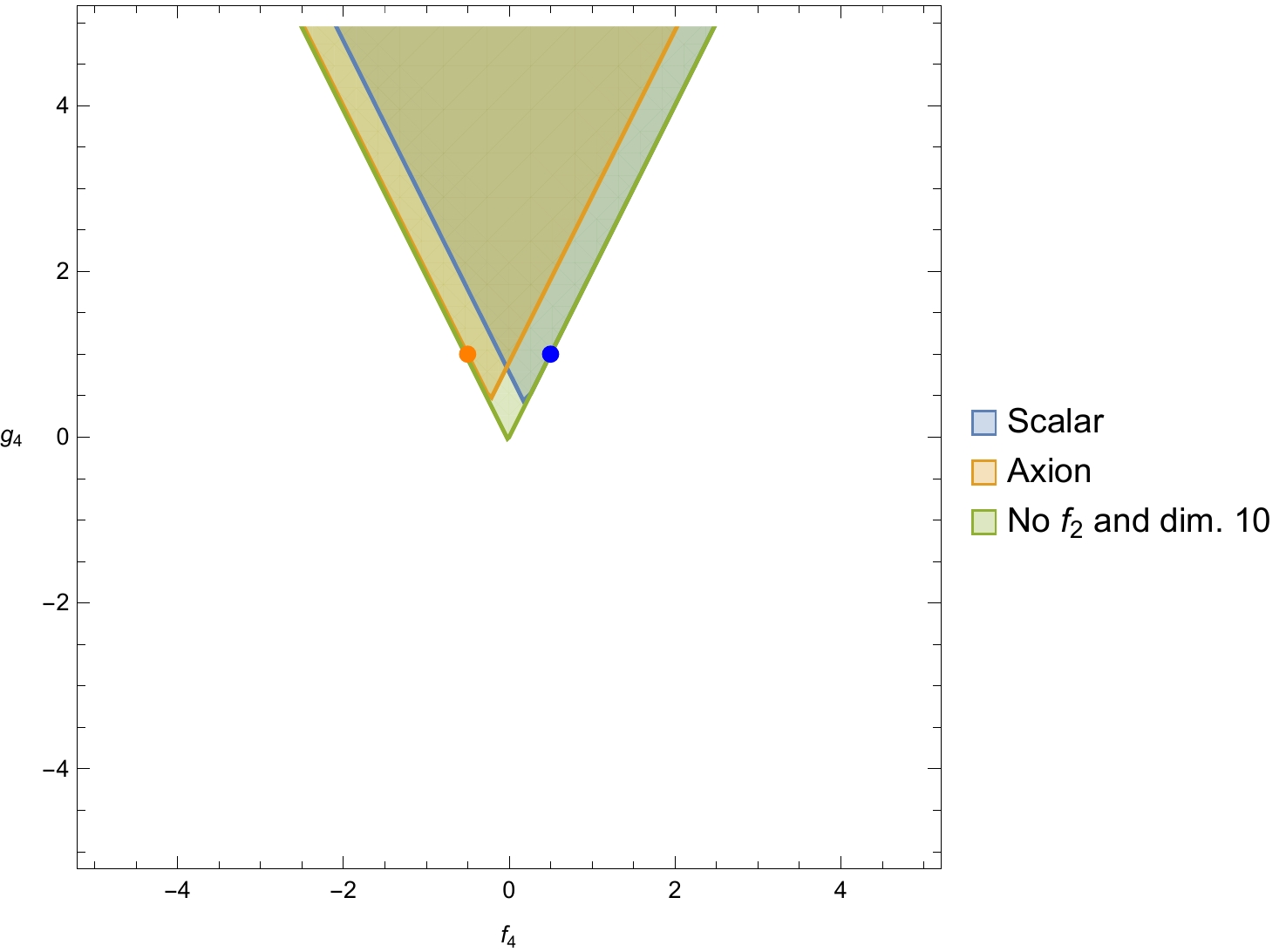}
	\caption{Bound on the $f_4-g_4$ plane. The left-hand side has $f_2=f_3=g_3=0$ and varying $h_3$. On the right-hand side, we choose the parameters not being plotted to have the corresponding values of the partial UV completion as in Table \ref{tab:UVcomp}. The scalar (blue) has $(f_2,f_3,g_3,h_3)=(1,3,1,0)$, the axion (orange) $(f_2,f_3,g_3,h_3)=(-1,-3,1,0)$, and the green region $f_2=f_3=g_3=h_3=0$. Note that in principle, the plots extend infinitely to the top since we cannot bound $g_4$ from above; however, the EFT makes sense at the cutoff $\Lambda$ only if all Wilsonian coefficients are taken to at most roughly of order 1.}
	\label{fig:Boundsf4g4}
\end{figure}

In Fig.~\ref{fig:Boundsf4g4}, we observe the bounds on the $f_4-g_4$ plane. The lower bound on $f_4$ becomes stronger when $f_3+3g_3>0$ since these terms can give a negative contribution to the time delay of $u_2^\ell$ which in turn requires a larger positive value of $f_4$ to not obtain an observable time advance. Meanwhile, a non-zero $f_3-3g_3$ contribution can be tuned to give a negative time delay for $u_2^\ell$ so that the upper bound on $f_4$ becomes stronger. Both the lower and upper bounds on $f_4$ are improved for a non-zero $h_3$. This is a large effect compared to that of $f_3$ and $g_3$ since these parameters have a suppression of $B^2/R^2$ in the integrand of the time delay with respect to a part of the $h_3$ contribution, (see Table~\ref{tab:signdef}). In the odd sector, the non-suppressed $h_3$ contribution is positive definite so a positive $h_3$ does not improve largely the upper bounds on $f_4$. This is not the case in the even sector so a positive $h_3$ does improve significantly the lower bounds. For a negative $h_3$ the non-suppressed contributions from both the even and odd sectors can be made negative and largely improve both upper and lower bounds.\\

Note that up to the EFT order we are working on, we have no contribution from the Wilson coefficient $c_7$, or equivalently from the scattering amplitude parameter $g'_4$. The contribution of $g'_4$ ($c_7$) starts at order $\epsilon_1^2 \epsilon_2^4$. If we choose to include this contribution, but still neglect all the WKB corrections (since they are not calculable in our setting), it would require that we neglected $\epsilon_1^2\epsilon_2^2/\Omega^2$ corrections, but include $\epsilon_1^2 \epsilon_2^4$, $\epsilon_1^2 \epsilon_2^4\Omega^2$, $\epsilon_1^2 \epsilon_2^6\Omega^2$ terms. The latter contributions arise from operators with dimension-$14$ and $16$. In this new setting, the validity of the EFT will require a large $\Omega$ bounded as $\epsilon_2^{-1}\ll\Omega \ll \epsilon_2^2$ which will naturally enhance the new contributions from the operators of dimension-$14$ and $16$. While we can tune the background solution to decrease the effect of these operators and enhance that of $g'_4$, these explorations are beyond the scope of this work. Instead, we can parameterise the amplitude as in \cite{Henriksson:2021ymi,Henriksson:2022oeu}, in which case we can obtain bounds for both $g_{4,1}$ and $g_{4,2}$ parameters defined in Table \ref{tab:conversion}. This simply follows from their leading order contribution to the time delay which will give bounds of the form
\begin{equation}
	2f_4+g_{4,1}+2g_{4,2}>A_{u^\ell_{2}} \ , \quad -2f_4+g_{4,1}+2g_{4,2}>A_{u^\ell_{4}} \ .
\end{equation}
\subsubsection{Case $g_2=0$}
\begin{figure}[!tbh]
	\center
	\includegraphics[width=0.4 \textwidth]{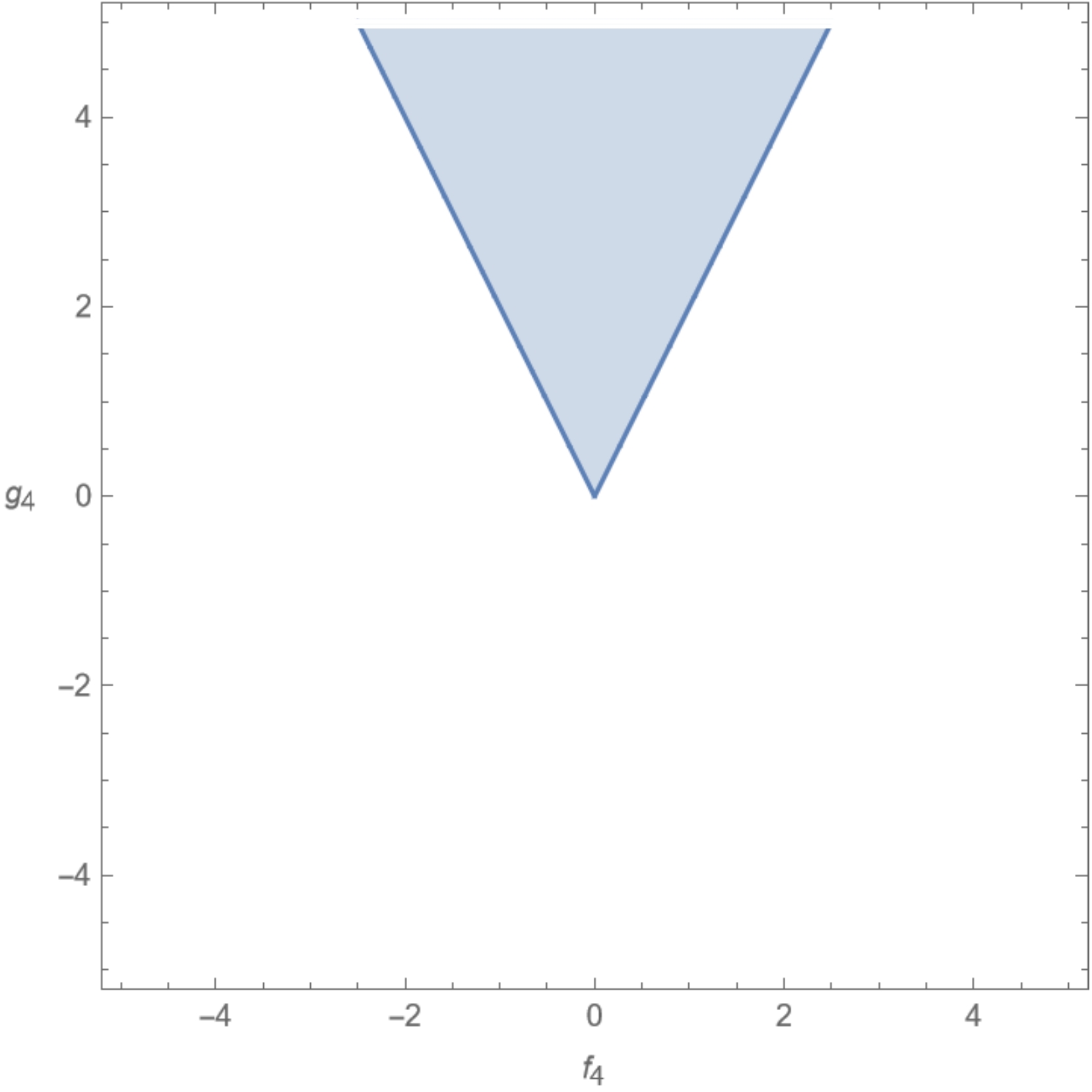}
	\caption{Bound on the $f_4-g_4$ plane for the case $g_2=0$. These bounds are independent on the value of all the other scattering amplitude parameters.}
	\label{fig:Boundsf4g4G2isZero}
\end{figure}

One can show that when $g_2=0$ causal propagation implies\footnote{Strictly speaking we obtain bounds of the form $-\epsilon<\mathcal{W}_J<\epsilon$, where $\epsilon$ is smaller than the WKB and EFT contributions that we are neglecting so we can effectively take $\epsilon=0$.} 
\begin{equation}
	f_2=f_3=g_3=h_3=0 \,.
\end{equation}
Once both $f_2$ and $g_2$ vanish, the requirements for the validity of the EFT in Eq.~\eqref{eq:HierarchyParam} change. Instead, we have a less restrictive situation where:
\begin{equation}
	\epsilon_1,\epsilon_2   \ll 1\ , \quad  \epsilon_2^{1/2}\ll\Omega \ll \frac{1}{\epsilon_2}  \ . \label{eq:HierarchyParamG2Zero}
\end{equation}
Using this new setup, we can find the causal bounds on the $f_4-g_4$ plane as shown in Fig.~\ref{fig:Boundsf4g4G2isZero}. We can see that the bounds are equal to the $g_2\neq0$ case with all other amplitude parameters set to zero, that is, they are given by
\begin{equation}
	2f_4+g_{4}>0 \ , \quad -2f_4+g_{4}>0 \ ,
\end{equation}
which implies that $g_4$ is positive. Note that since we can only obtain a one-sided bound for the $g_4$ coefficient within our EFT setup, we are only able to constrain its sign. One should note that $f_4$ will be required to vanish if $g_4$ were to vanish.

\section{Positivity and Causality}

\begin{table}[h!]
\begin{adjustbox}{max width=1\textwidth,center}
	\begin{tabular}{| c | c | c|}
 \hline {\bf Property} &
	{\bf Causality Bounds} & {\bf Positivity Bounds}  \\ 
 \hline
 Lorentz &  \tabitem Lorentz invariant EFT&   \tabitem  Invariant EFT and UV completion  \\ 
invariance
& & \tabitem Crossing symmetry \\
 \hline
 Unitarity &  \tabitem Hermitian Hamiltonian: &  \tabitem Positive discontinuity  \\
& real Wilson coefficients & of the EFT and UV amplitude
\\
 \hline
 Causality & \tabitem No resolvable time advance &  \tabitem Analyticity of amplitude  \\
& & in the complex $s$ plane for fixed $t$ 
\\
 \hline
Locality&     \tabitem IR theory is local  &    \tabitem IR and UV theories are local\\
& &  \tabitem Froissart-like bound in the UV \\
 \hline
 Other  & \tabitem  EFT and WKB expansions under control  &  \\ 
 assumptions& \tabitem  Background  generated by    & \tabitem IR EFT is under perturbative control \\ 
& localized external source & \\
    \hline
\end{tabular}
\end{adjustbox}
	\caption{Summary of the assumptions underlying the positivity and causality bounds. 
 }
\label{tab:assumptions}
\end{table}

After deriving the causality and positivity bounds for the low-energy parameter space of the massless photons EFT in Eq.~\eqref{eq:Lagrangian}, we proceed to compare these results and discuss their complementary. In all of the plots in this section, the regions constrained by IR causality and positivity will be bounded by a thick line and a dashed line respectively.  Recall that in this (and previous sections) we have set $g_2=1$. We proceed to compare the strength of causality and positivity bounds for different slicings of the parameter space showing the cases where positivity bounds can give stronger results than causality in Section~\ref{sec:strongpos} and the cases where positivity and causality complement each other in Section~\ref{sec:complemen}. Finally, we comment on the case $g_2=0$ in Section~\ref{sec:g20}.

It is worth highlighting that given the set of assumptions that we make in deriving either the IR causality bounds or the positivity bounds (summarized in Table~\ref{tab:assumptions}), we do not claim to possess the absolute most restrictive bounds that could possibly arise from said assumptions. In particular, this means that where regions allowed by causality and positivity do not overlap, it does not imply that somehow the two approaches are in contradiction. For example, if a region of parameter space is allowed by current positivity bounds but excluded by the causality bounds we have derived, this would not imply that the assumptions on the positivity side allow for causality-violating parameter values nor that there is any contradiction. Rather it would simply indicate that at the current level, the analytic positivity bounds we have derived are not the most optimal nor the most restrictive constraints and signal that pushing further positivity bounds beyond the current state-of-the-art one would likely be able to prove that this region of parameter space is no longer allowed by positivity. When this occurs, this should be read as a way to read from causality how one would expect future positivity bounds to evolve as they become more restrictive.

\subsection{Positivity bounds are stronger then causality.} \label{sec:strongpos}

\begin{figure}%
    \centering
    \subfloat[\centering ]{\includegraphics[width=0.55\textwidth]{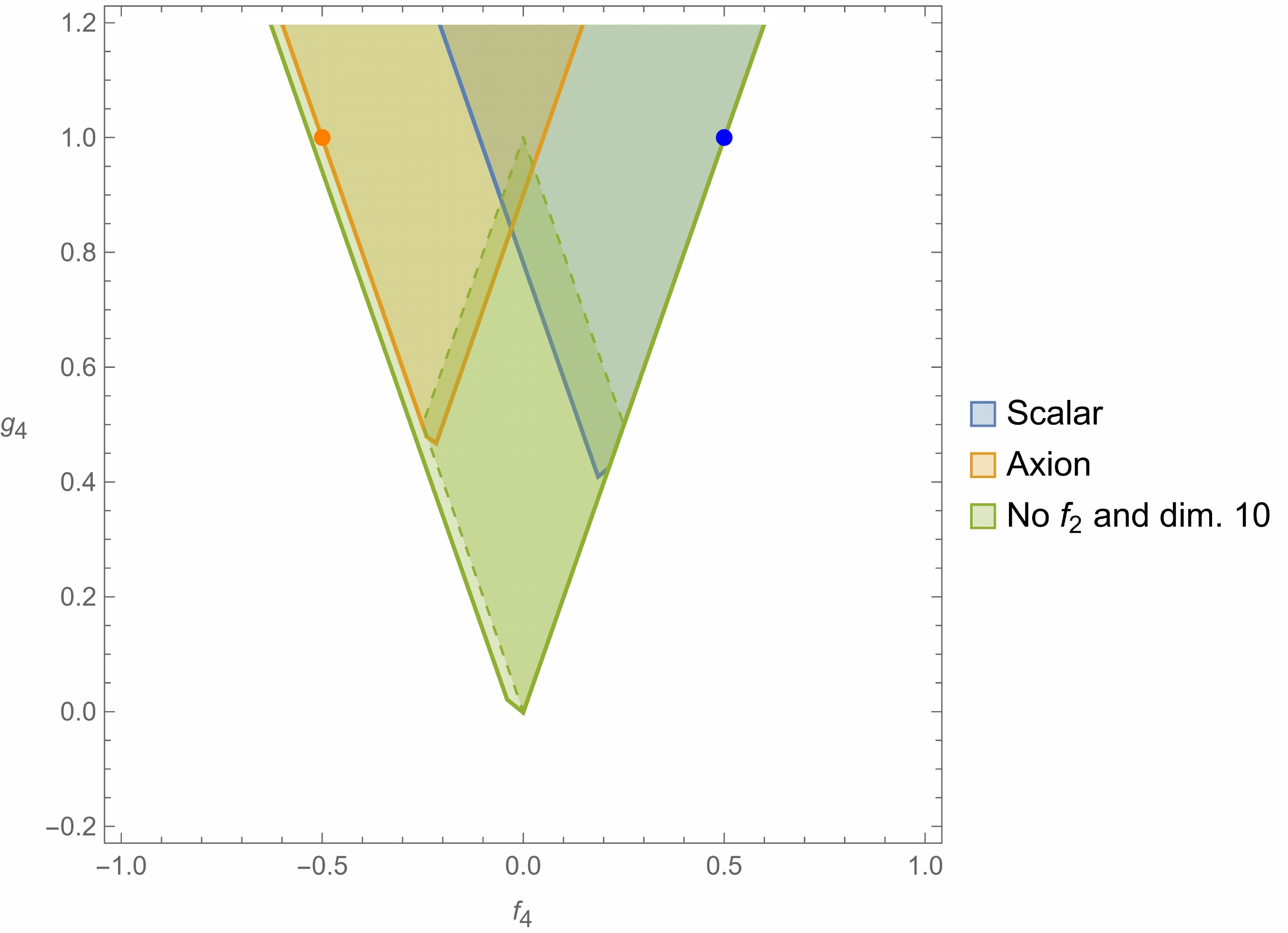}}%
    \qquad
    \subfloat[\centering ]{\includegraphics[width=0.39\textwidth]{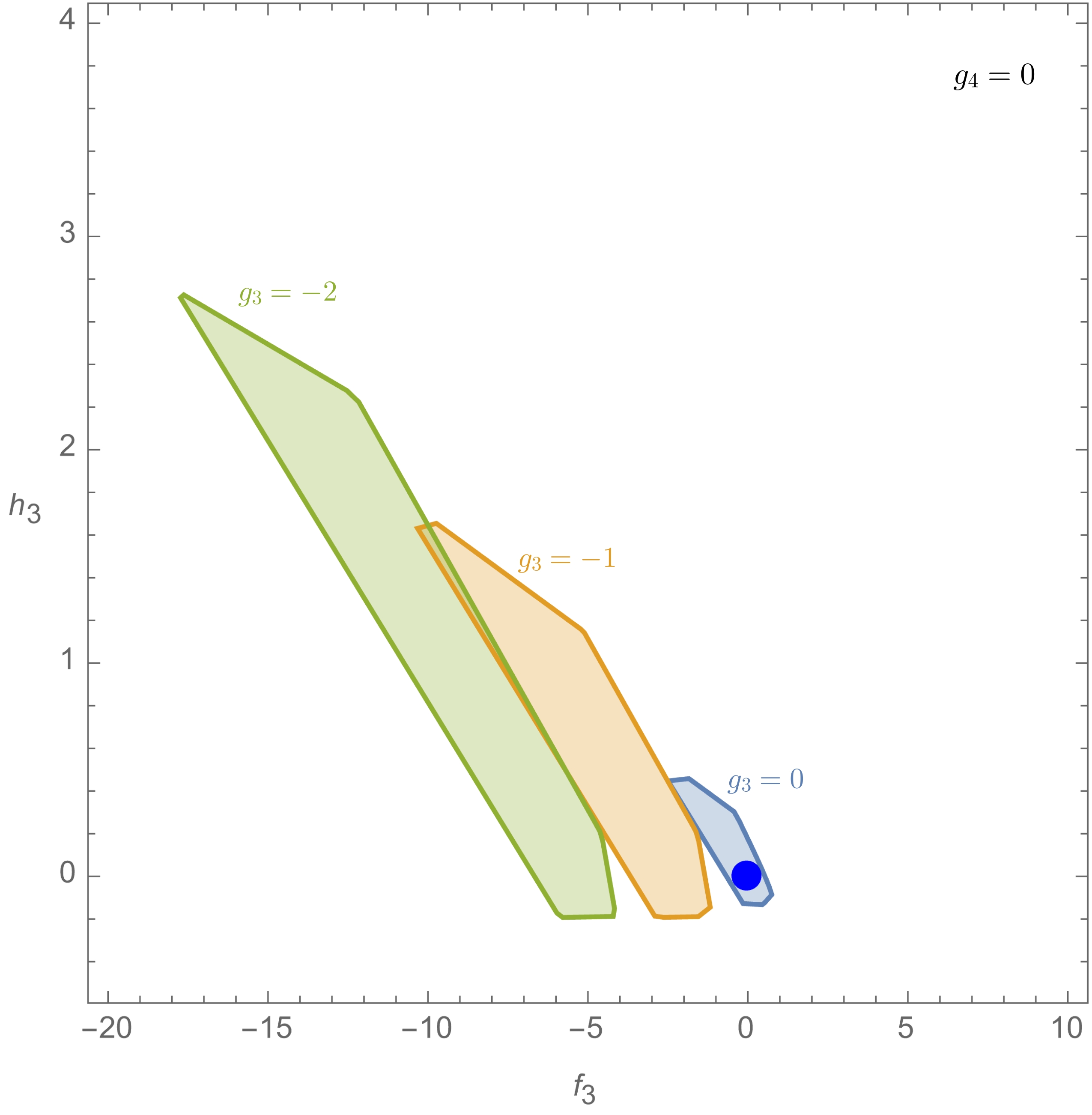}}%
    \caption{(a) Causality (solid) and positivity (dashed) bounds for $g_4$--$f_4$. The green region corresponds to $(f_2,f_3,g_3,h_3)=(0,0,0,0)$, the blue to $(1,3,1,0)$ (values taken by the scalar UV completion) and the orange to $(-1,-3,1,0)$ (values taken by the axion UV completion). The blue and orange points are correspond to values of $(f_4,g_4)$ produced by the scalar and axion UV completions respectively. (b) $f_3$--$h_3$ plane for varying $g_3=0,-1,-2$ and all other coefficients zero. Positivity bounds leave $g_3=h_3=f_3=0$ as the only possibility in this case. The causality bounds are compact in $f_3$ and $h_3$ for any given $g_3$ but generically, the direction $f_3+3g_3$ is not bounded from below, see Eq.~\eqref{eq:f3g3h3even}.}%
    \label{fig:compareg4f4}%
\end{figure}

\paragraph{Bounds on $f_4$ and $g_4$}
Unlike positivity bounds, causality constraints cannot provide compact bounds on all the parameter space in EFT for fixed $g_2$. As mentioned previously, due to technicalities of the WKB approximation, there is no upper bound on $g_4$ from causality, see Fig.~\ref{fig:compareg4f4}~(a). Positivity bounds provide the upper limit on $g_4<g_2$. The form of the allowed region depends on the value of $f_2$, shrinking to a single line for the extreme values $f_2=\pm 1$. The parameter range allowed by causality is much larger in this case. Remarkably, its form resembles the same cone as obtained from positivity.

\paragraph{Bounds for $g_4=0$.}
The non-linear bounds in \eqref{NL1},\eqref{NL2} only allow for null values of $g_3,~f_3,~h_3$. The causality constraints are weaker in this case, still allowing for certain ranges expanding with $g_3$, as it is shown in Fig.~\ref{fig:compareg4f4} (b). The coefficient $g_3$ is bounded only from above, and $f_3$, $h_3$ are in the compact ranges which grow when $g_3$ is decreasing. 

\begin{figure}
 \centering
    \subfloat[\centering ]{\includegraphics[width=0.3\textwidth]{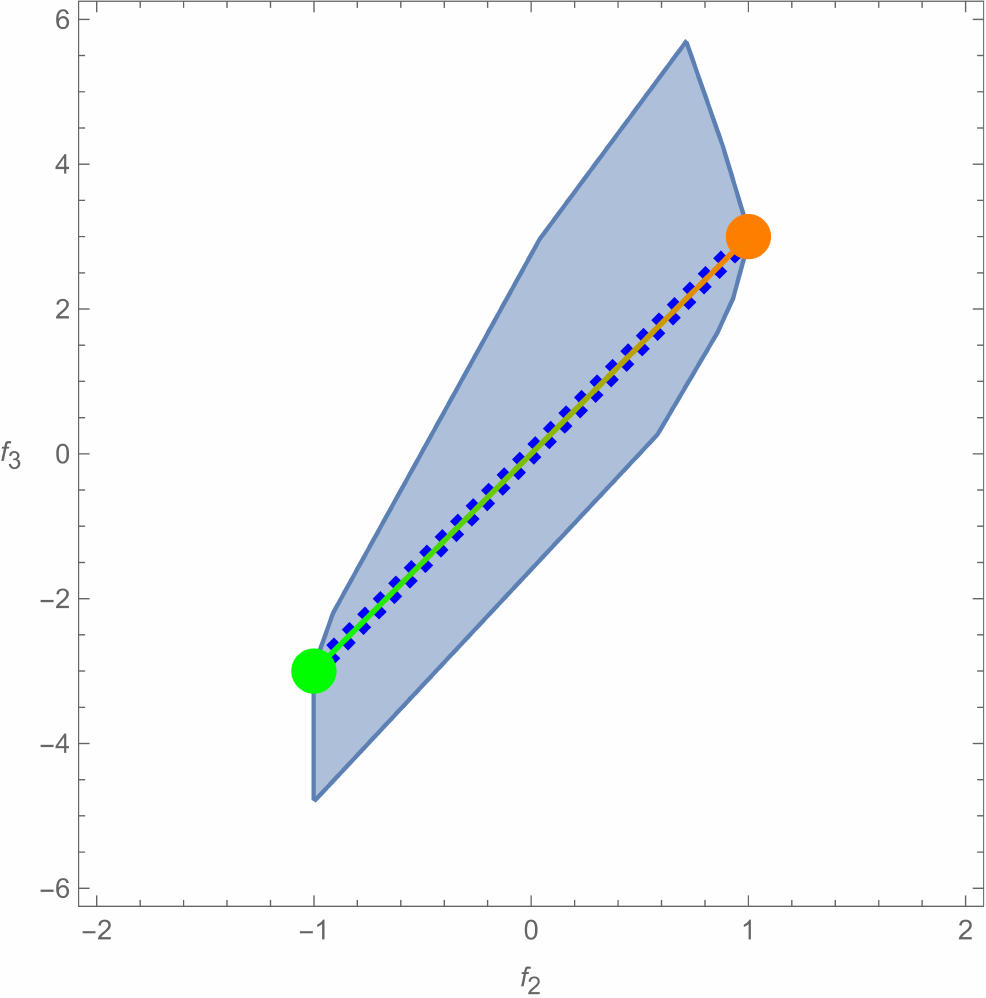}}%
    \qquad
    \subfloat[\centering ]{\includegraphics[width=0.3\textwidth]{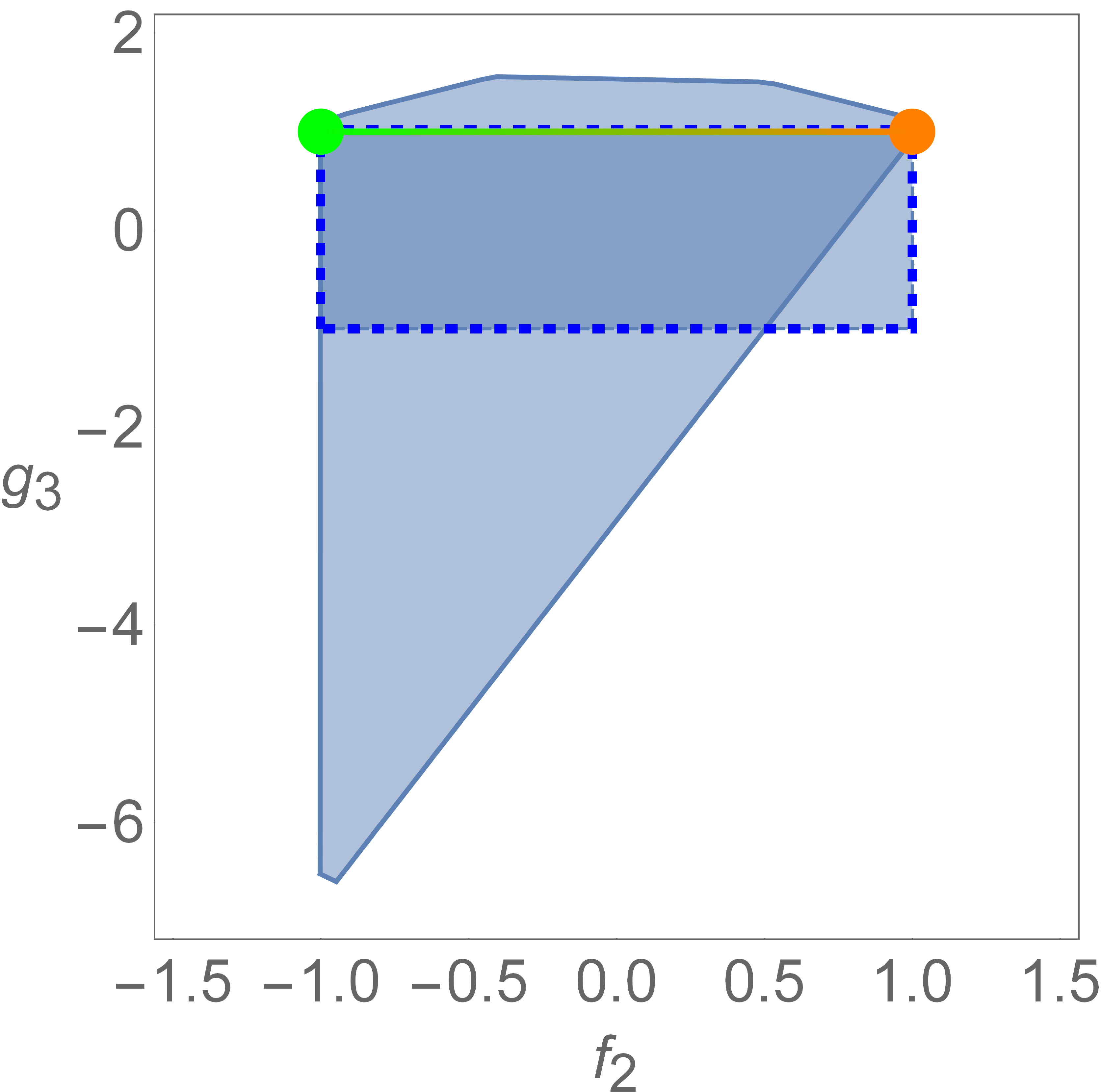}}%
\qquad
    \subfloat[\centering ]
    {\includegraphics[width=0.29\textwidth]{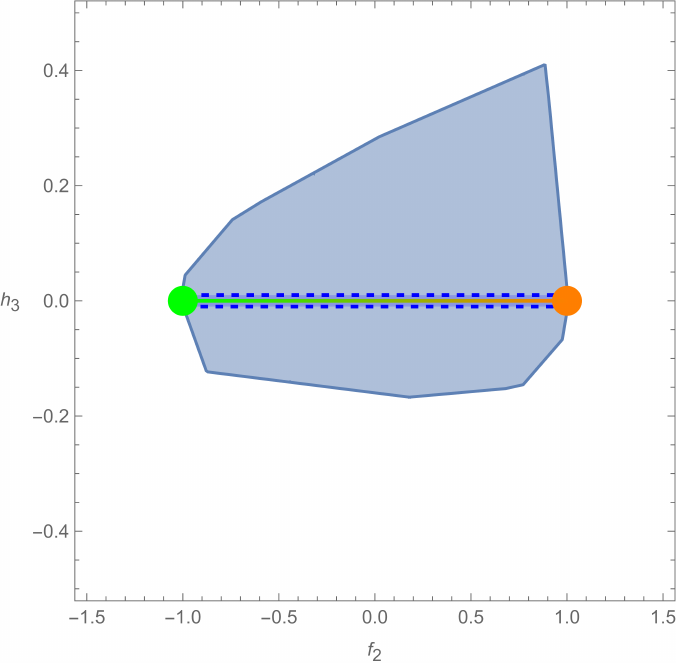}}
    \caption{We plot the causality (thick line) and positivity (dashed line) bounds for different slicings of our parameter space and set the parameters not being plotted to values consistent with the scalar and axion partial UV completions. The slices we plot are: (a)$(f_2, f_3)$-plane for $g_3=1$, $h_3=0$, $f_4=f_2/2$, and $g_4=1$;  (b) $(f_2, g_3)$ plane for $f_3=3f_2$, $h_3=0$, $f_4=f_2/2$, and $g_4=1$; (c) $(f_2, h_3)$ plane for $f_3=3f_2$, $g_3=1$, $f_4=f_2/2$, and $g_4=1$. One can see that the points denoting the values of the partial UV completions, scalar in green and axion in orange, lie in the boundary of the causal region. The line connecting them corresponds to a partial UV completion involving both a scalar and an axion.}
	\label{fig:UVLineComparison}
\end{figure}

\paragraph{Bounds on $(f_2,f_3)$ and $(f_2,h_3)$ planes} 
In both the $(f_2,f_3)$ and $(f_2,h_3)$ planes the positivity bounds exactly reduce to the line interpolating between the scalar and axion UV completions as seen in Fig.~\ref{fig:UVLineComparison}. In these cases, the positivity bounds coincide with a one-dimensional region of the parameter space that admits a partial UV completion and are thus stronger than the causality bounds, which give an allowed area surrounding the UV completion segment. 

\subsection{Complementarity of the positivity and causality bounds.} \label{sec:complemen}

We present here several slices showing that causality bounds, being more sensitive to the different combinations of the EFT parameters, can give additional constraints compared to the known positivity bounds. Typically, the positivity constraints on $f_3$ and $h_3$ are relatively weak, see Fig.~\ref{fig:g3h3f3}. 

\paragraph{Bounds on $f_2$ and $g_3$}
In the $(f_2,g_3)$ plane we see that both partial UV completions and the line joining them lie within the region allowed by causality and positivity bounds. The positivity bounds, denoted by the dashed boundary, are more constraining overall but allow for parameter values that are forbidden by causality bounds.  Here we see for the first time that causality bounds provide additional information on the parameter space than that obtained by positivity alone. Once again we clarify that this does not mean that theories satisfying positivity bounds are not causal, it just shows that these bounds can probe different regions of the parameter space.

\paragraph{Around QED UV completions.}
\begin{figure}%
    \centering
    \subfloat[\centering Scalar ]{\includegraphics[width=0.3\textwidth]{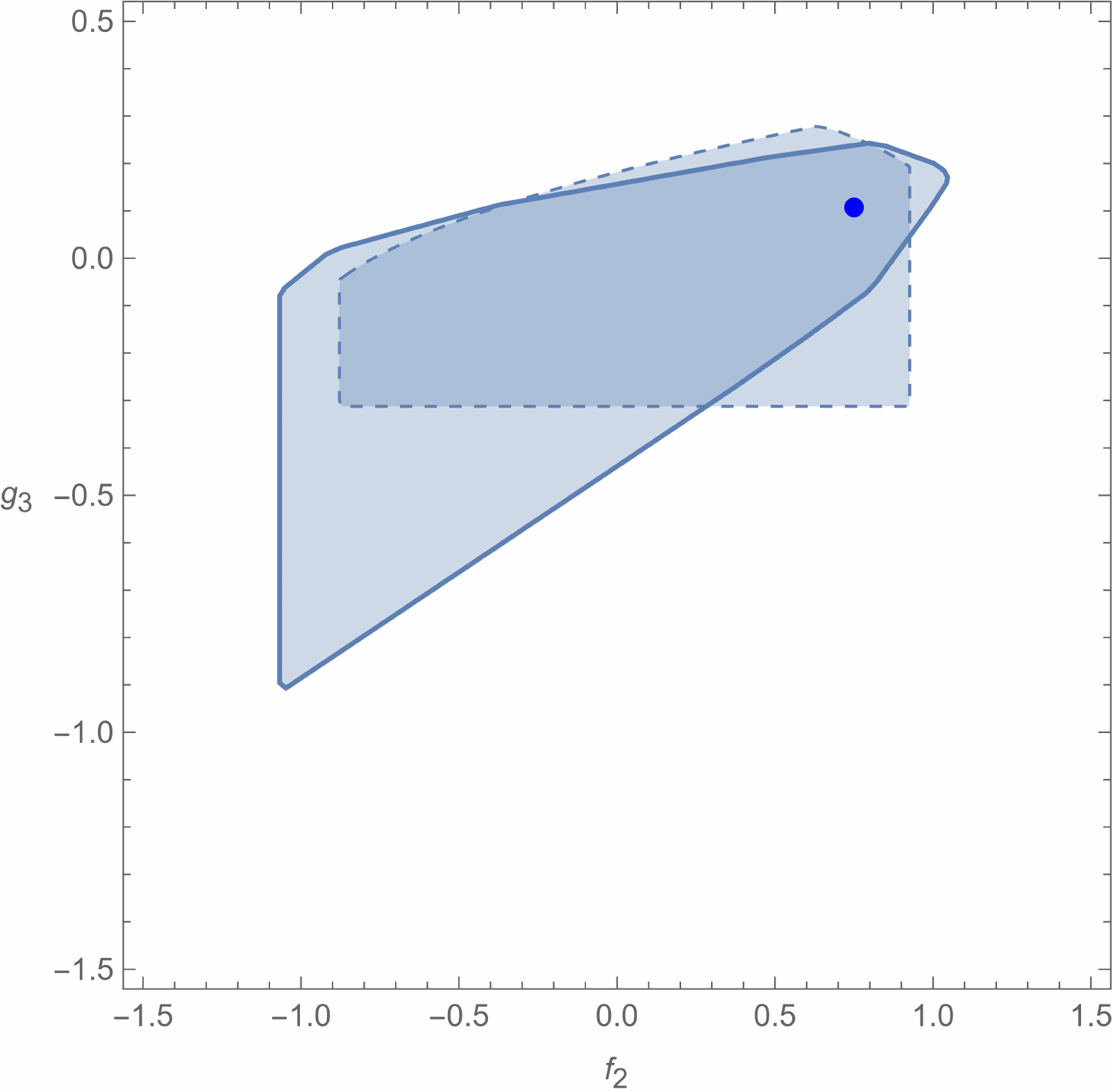}}%
    \qquad
    \subfloat[\centering Spinor]{\includegraphics[width=0.3\textwidth]{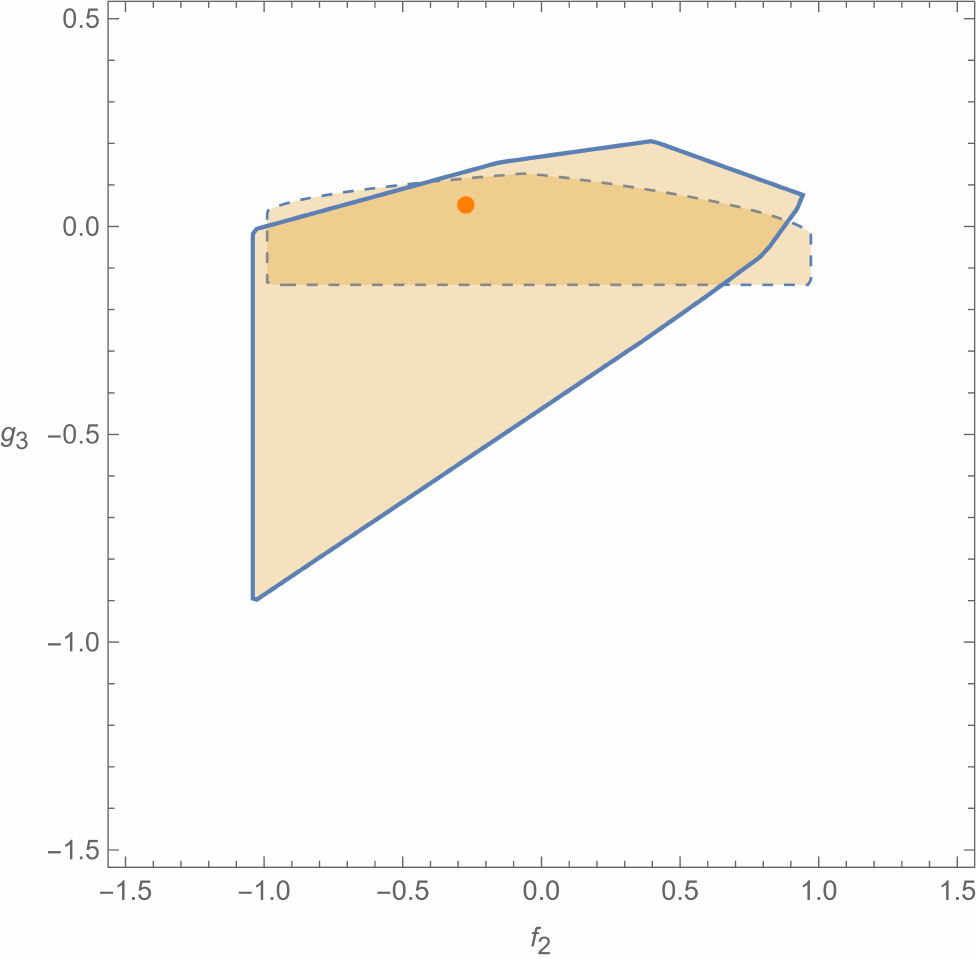}}%
     \qquad
    \subfloat[\centering Vector]{\includegraphics[width=0.3\textwidth]{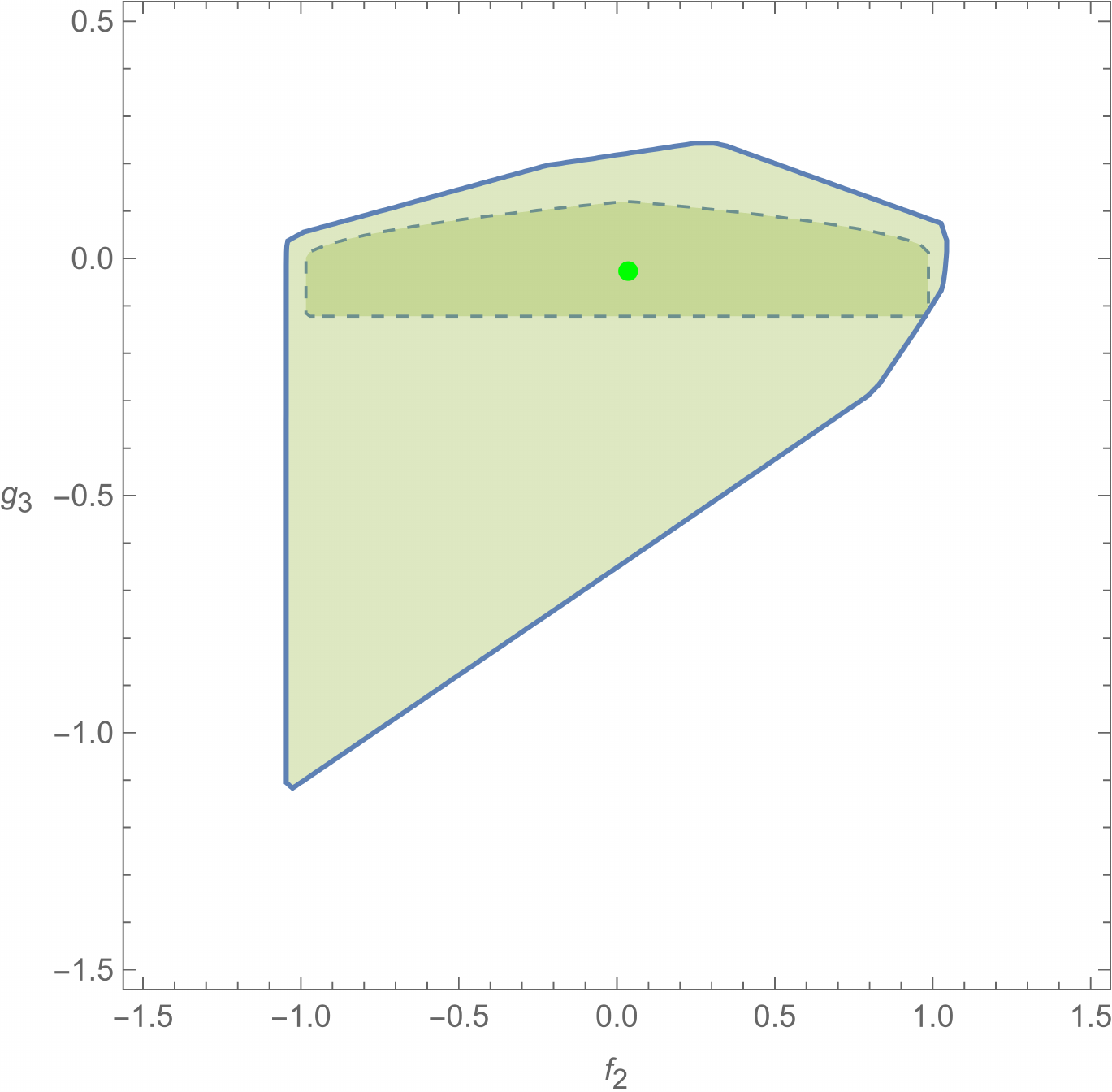}}%
    \caption{Causality (thick line) and positivity (dashed line) bounds in the $(f_2, g_3)$ plane for various values of $(f_3,h_3,f_4,g_4)$ that are consistent with the scalar, spinor and vector QED partial UV completions, namely $(f_3,h_3,f_4,g_4)\sim (0.36,0.04,0.01,0.10)$ for the scalar, $(-0.13,-0.01,0.00,0.02)$ for the spinor and $(0.02,0.00,0.00,0.01)$ for the vector. The exact values are given in Table \ref{tab:UVcomp}}%
    \label{fig:f2g3QEDComparison}%
\end{figure}

Turning now to the loop level QED partial UV completions, the causality bounds are compared to the region compatible with positivity in Fig.~\ref{fig:f2g3QEDComparison}. While positivity bounds give a smaller allowed region than causality bounds, we see that they actually probe different regions for the scalar and spinor QED cases. Thus, combining both bounds will ultimately reduce the allowed region. Note finally that all QED partial UV completions lie within both causality and positivity bounds, as expected for consistency.

\begin{figure}[!h]
	\center
	\includegraphics[height=5cm]{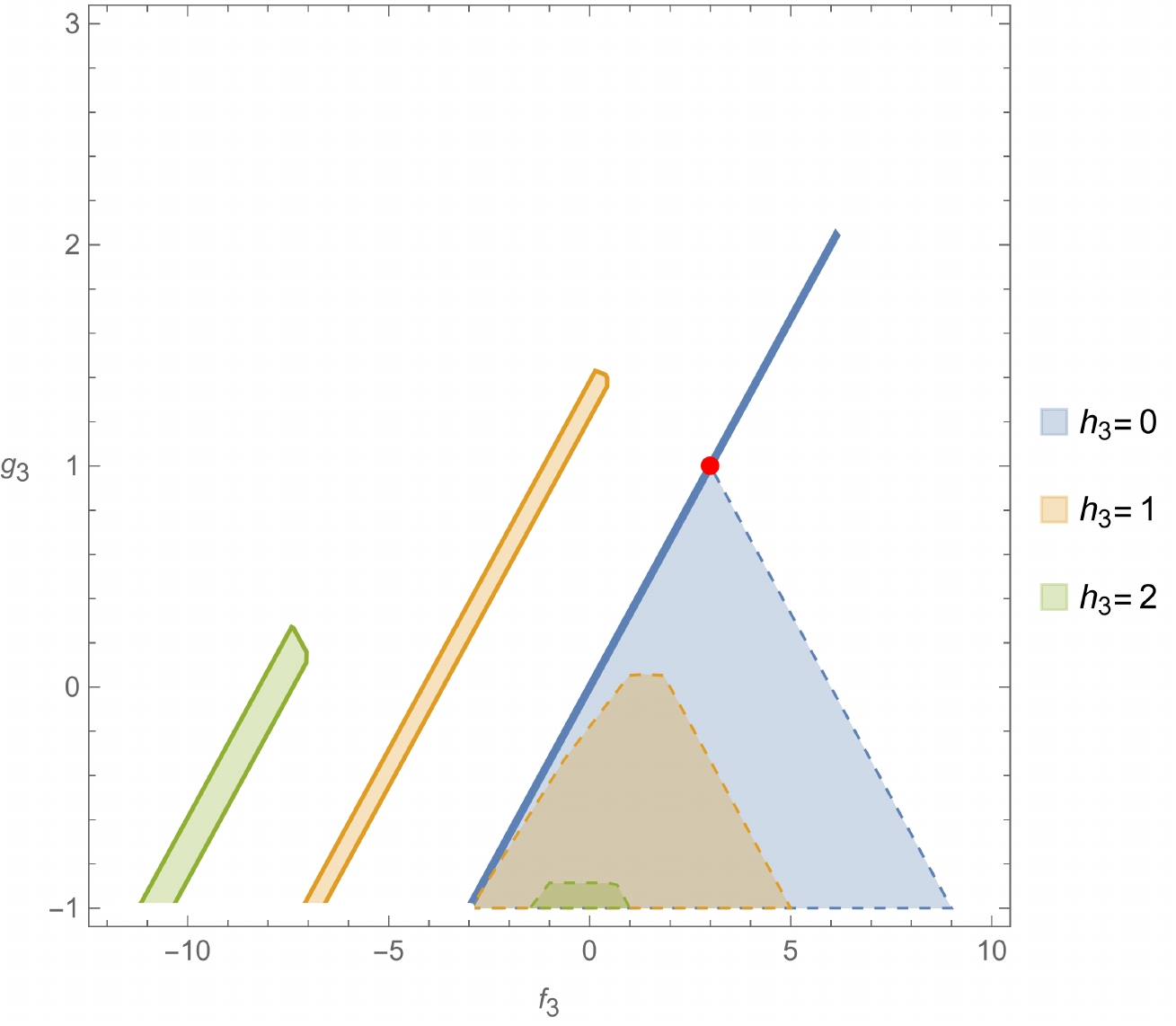}
	\hspace{0.3cm} \includegraphics[height=5cm]{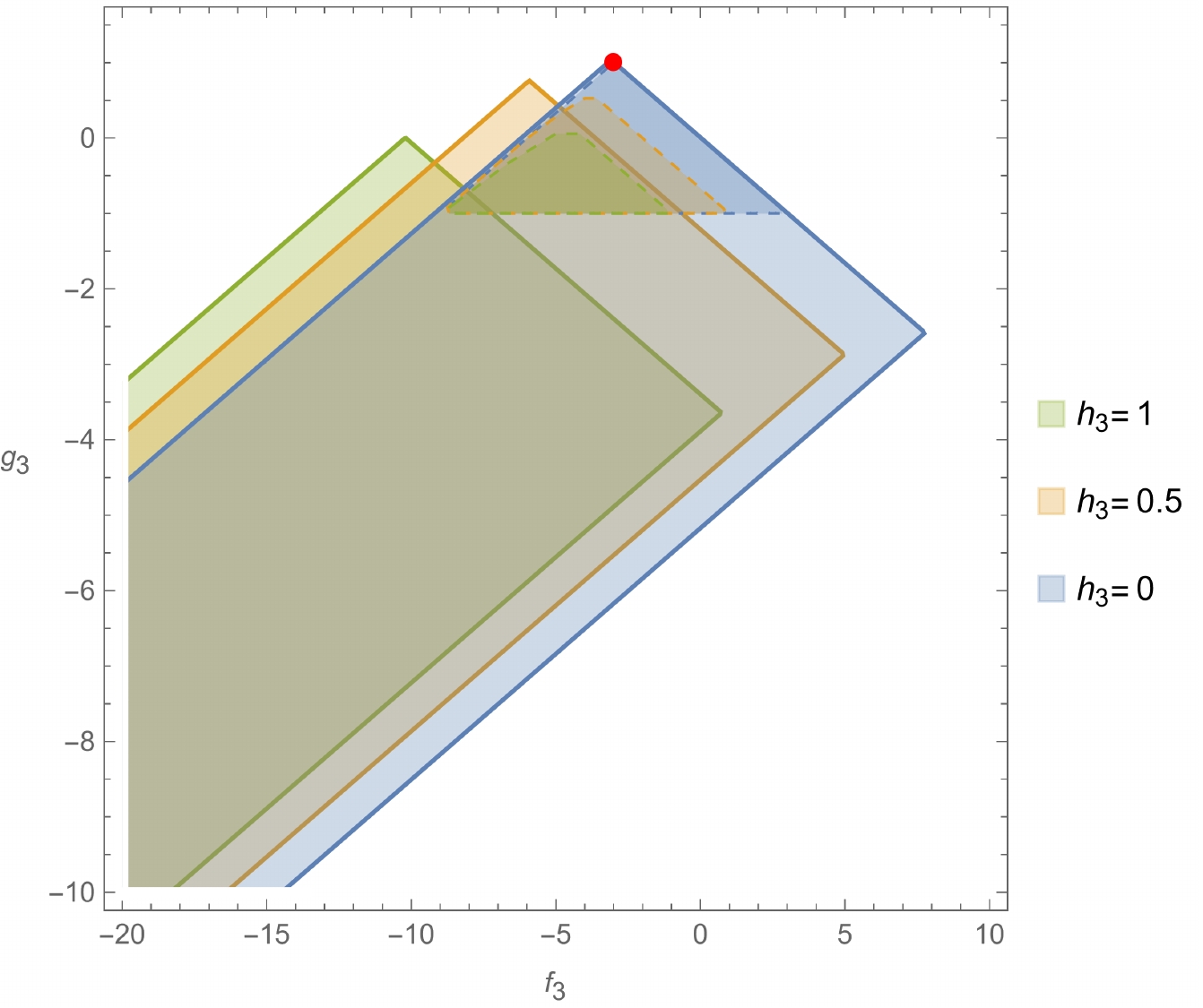}
	\caption[$f_3 - g_3$ causality and positivity bounds for several values of $h_3$.]{Causality (thick line) and positivity (dashed line) bounds on $(f_3,g_3)$ plane with varying $h_3$ and the values of the other coefficients fixed as in the scalar partial UV completion on the left and the axion one on the right. In particular, the left panel has $(f_2,f_4,g_4)=(1,1/2,1)$ and the right one $(f_2,f_4,g_4)=(-1,-1/2,1)$.}
	\label{fig:f3g3ScalAxionComparison}
\end{figure}

\paragraph{Bounds on $f_3$, $g_3$, and $h_3$.}
Let us start by discussing the $(f_3,g_3)$ plane where all other coefficients are set to the values of either the scalar or axion UV completions found in Table \ref{tab:UVcomp}. The bounds are reported in Fig.~\ref{fig:f3g3ScalAxionComparison}. The first thing to note is that even though allowed regions exist both from causality and positivity bounds for $h_3=1$ and $2$, their union is empty and hence these values are discarded in the sense that they cannot correspond to theories endowed with causal propagation in both the IR and UV. The more interesting case, $h_3=0$, corresponding to the scalar or axion partial UV completions is allowed by both methods. For the left-hand side (scalar UV completion parameters), the causality bounds are much more constraining than their positivity counterparts in the $f_3-3g_3$ direction. As explained previously, this is due to having the additional parameter fixed as in the scalar UV completion which implies that they do not contribute to the time delay in the odd sector. On the other hand, positivity bounds provide a slightly lower upper value than causality in the $f_3+3g_3$ direction and also a lower bound that we cannot obtain from causality.  

For the right side of Fig.~\ref{fig:f3g3ScalAxionComparison}, we see that the axion partial UV completion is consistent with both methods as expected. This time, the positivity bounds are stronger than the bounds obtained from IR causality but it is still interesting to compare them since, for instance, their union becomes smaller as $h_3$ is raised. Thus, utilizing both bounds allows us to rule out theories with values of $h_3$ further away from the axion UV completion value, $h_3=0$.

\begin{figure}[!h]
	\center
	\includegraphics[height=5cm]{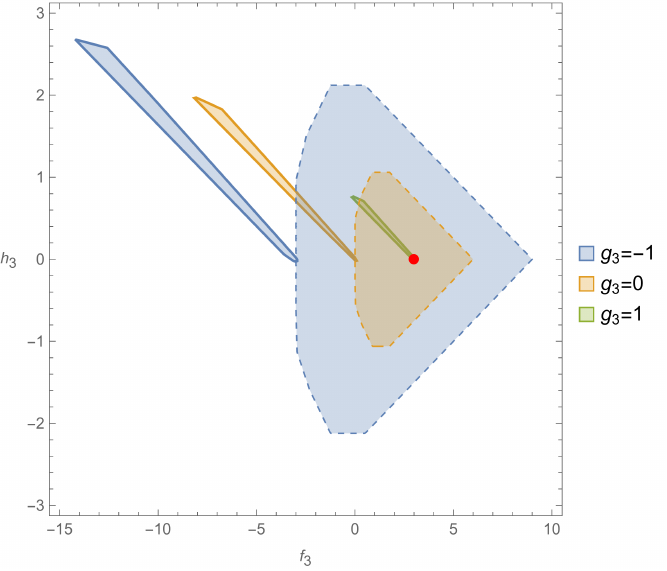}
	\hspace{0.3cm} \includegraphics[height=5cm]{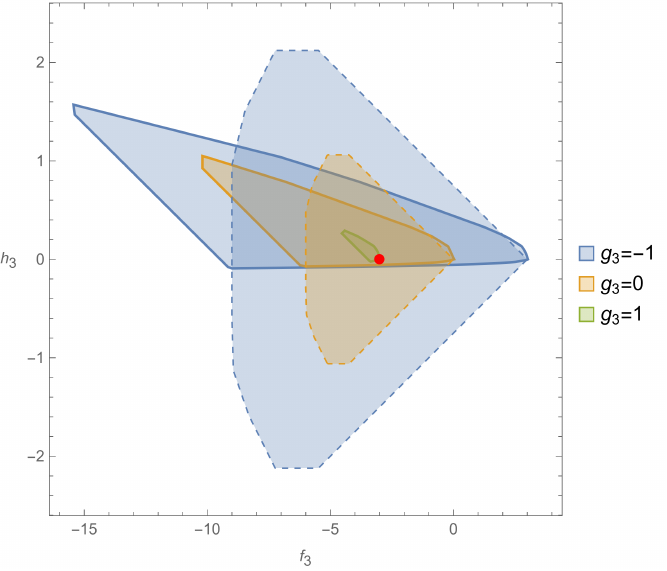}
	\caption[$f_3 - h_3$ causality and positivity bounds for varying $g_3$.]{Causality (thick line) and positivity (dashed line) bounds on $(f_3,h_3)$ plane with varying $g_3$ and all other parameters set by the values of the scalar (left) or axion (right) partial UV completions. In particular, the left panel has $(f_2,f_4,g_4)=(1,1/2,1)$ and the right one $(f_2,f_4,g_4)=(-1,-1/2,1)$.}
	\label{fig:f3h3ScalAxionComparison}
\end{figure}

Let us now analyze the $(f_3,h_3)$ plane, where the values of all other parameters are consistent with the scalar partial UV completion (left panel of Fig.~\ref{fig:f3h3ScalAxionComparison}). One can see that the causality bounds are much more constraining than their positivity counterparts, except in the case $g_3=1$ where the latter reduce to the single scalar UV completion point. For the other cases, even though both sets of bounds have a finite size, their union also reduces to a single point, once again proving the power of this combined approach. Turning now to the right panel (axion UV completion parameters), both causality and positivity bounds have similar strength, with the causality ones being slightly more constraining this time. The scalar and axion UV partial UV completions are consistent with both methods and lie in the boundary of the bounds.

\begin{figure}[!h]
	\center
	\includegraphics[height=5cm]{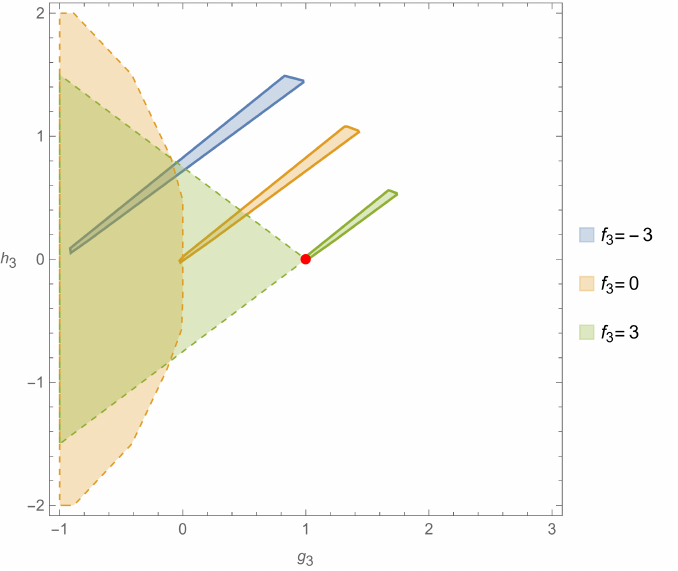}
	\hspace{0.3cm} \includegraphics[height=5cm]{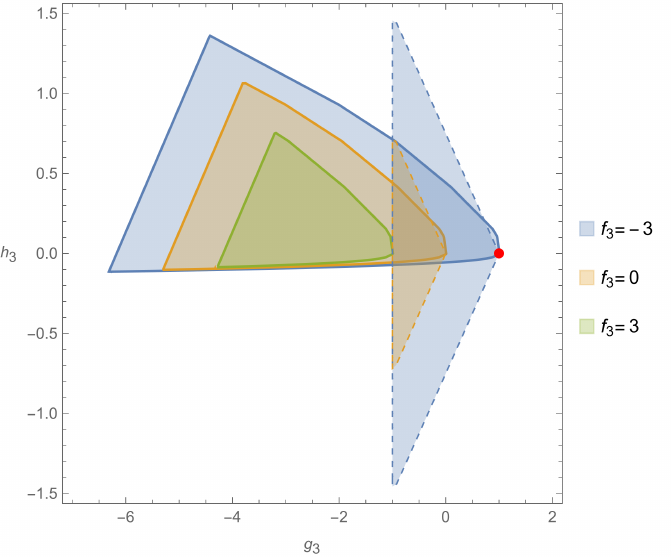}
	\caption[$g_3 - h_3$ causality and positivity bounds for several values of $f_3$.]{Causality (thick line) and positivity (dashed line) bounds on $(g_3,h_3)$ plane with varying $f_3$ and all other parameters set by the values of the scalar (left) or axion (right) partial UV completions. In particular, the left panel has $(f_2,f_4,g_4)=(1,1/2,1)$ and the right one $(f_2,f_4,g_4)=(-1,-1/2,1)$.}
	\label{fig:g3h3ScalAxionComparison}
\end{figure}

The $(g_3,h_3)$ case is very interesting. When focusing on the value $f_3=3$ on the left panel of Fig.~\ref{fig:g3h3ScalAxionComparison}, which is precisely the value that is compatible with the scalar partial UV completion, the intersection of the causality and positivity bounds reduce to this single point. This example perfectly illustrates the power of combining the two methods, which can dramatically reduce the space of causal theories, to the point of converging to a point-like region.

The right panel shows the axion case. The causality bounds are slightly less constraining than the positivity ones but their union is smaller. It is worth noting that the axion partial UV completion is consistent with both bounds. If $f_3=3$ (with all other coefficients set to the axion completion values) then positivity bounds shrink the allowed region to a single point allowing only for $(g_3,h_3)=(0,-1)$, which is also consistent with causality.

\begin{figure}[!h]
	\center
	\includegraphics[height=4.6cm]{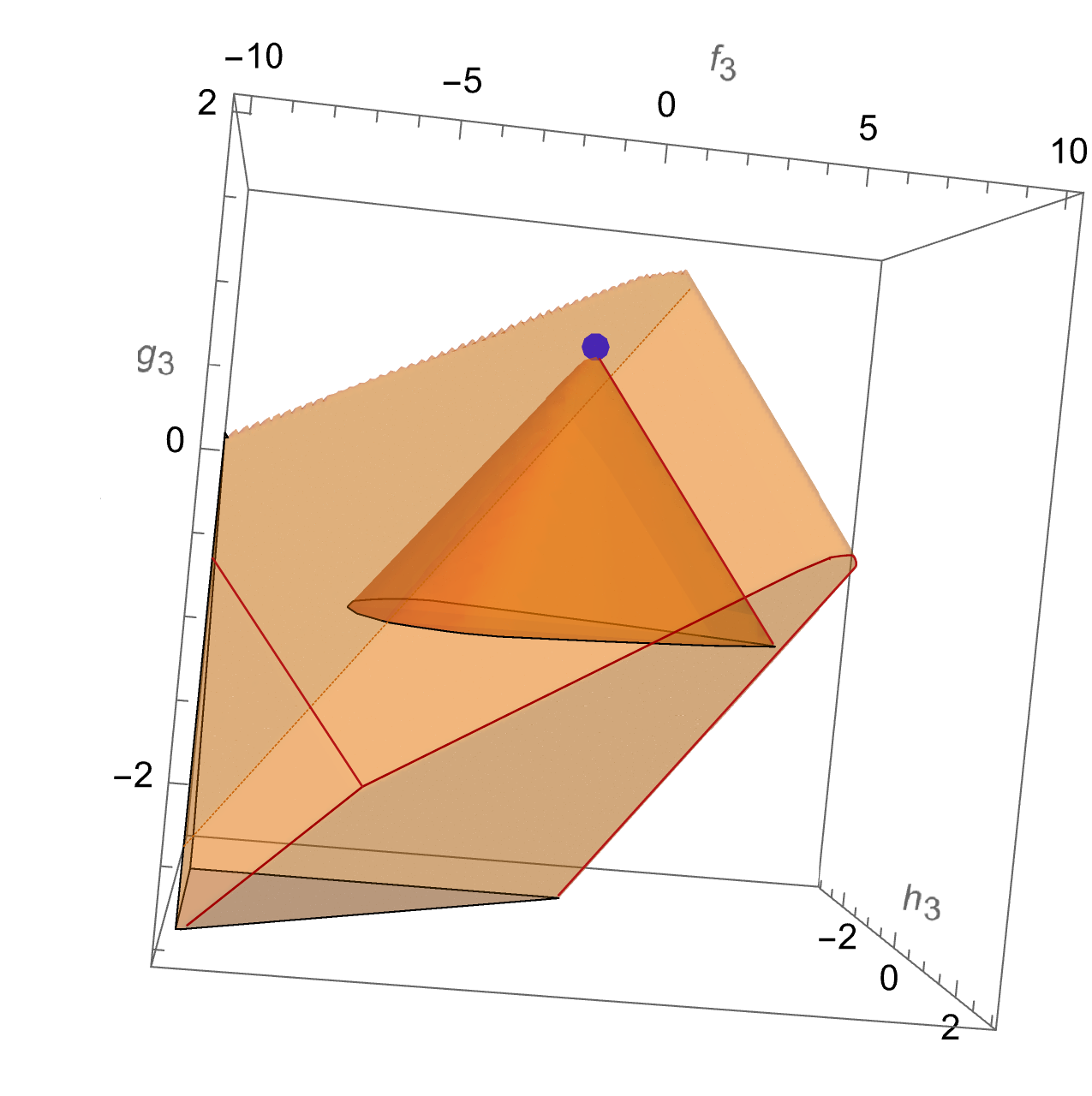}
	\hspace{0.2cm} \includegraphics[height=4.6cm]{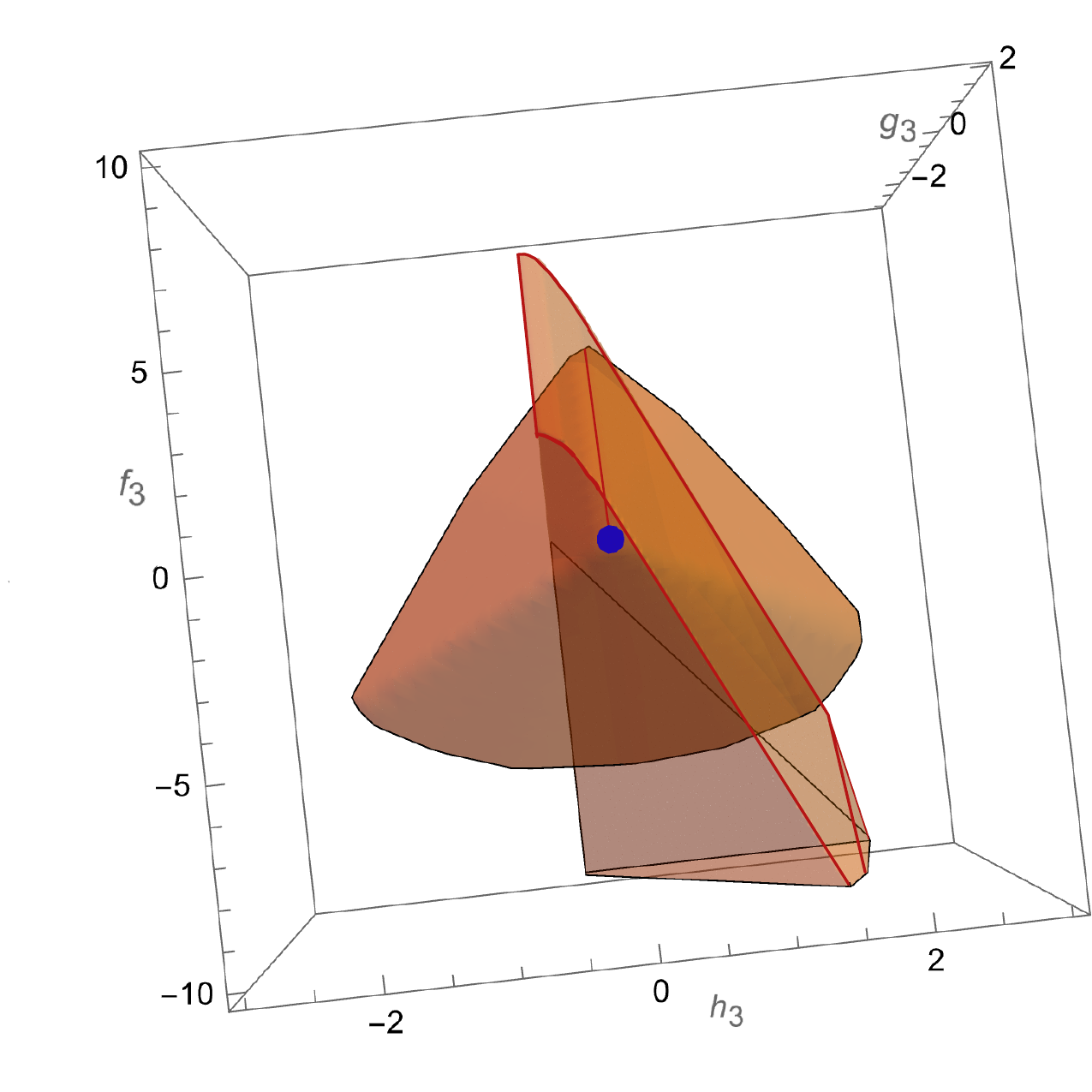}
 \hspace{0.2cm} \includegraphics[height=4.6cm]{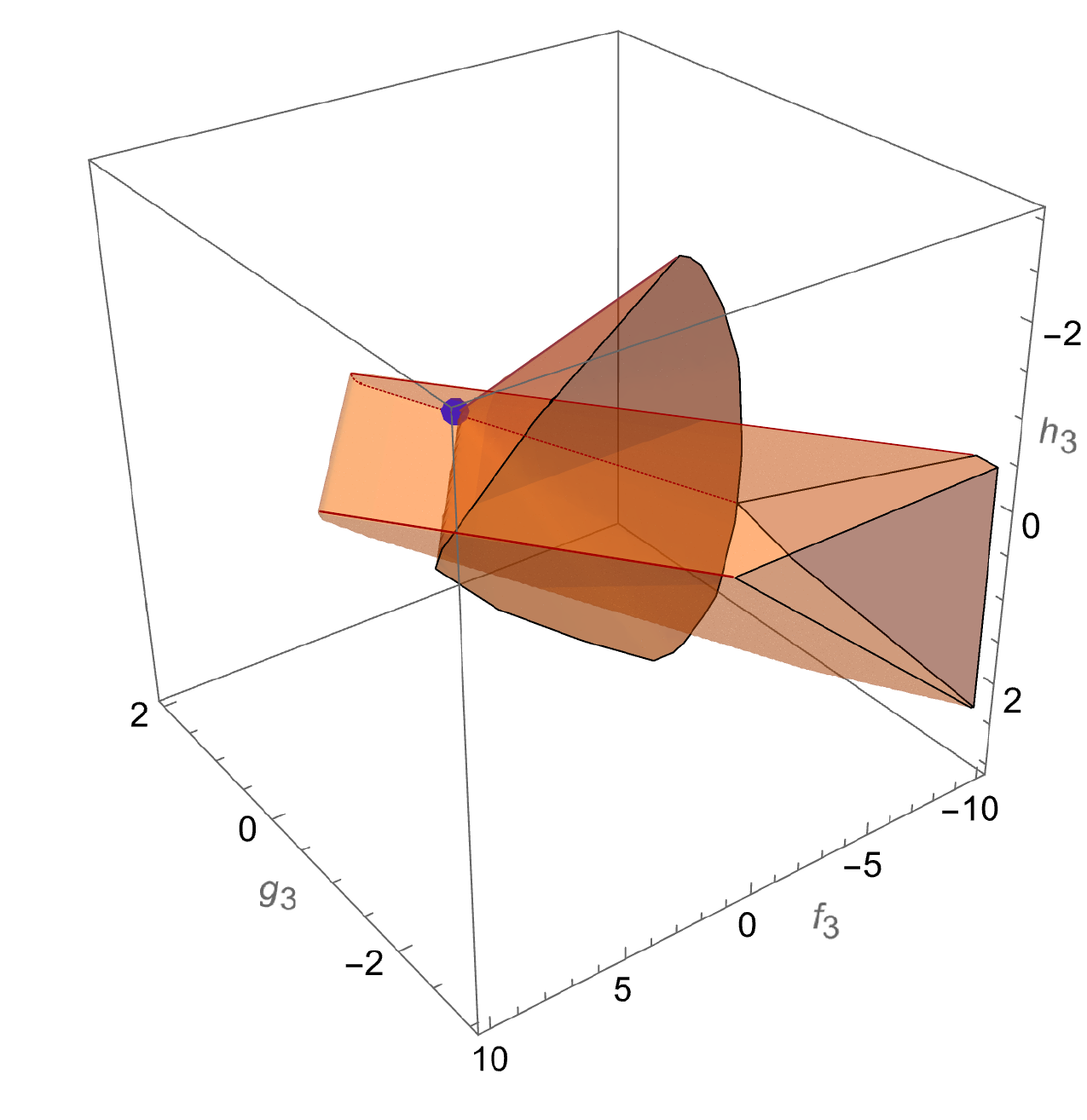}
	\caption{3-dimensional plot of positivity and causality bounds on $g_3,~h_3,~f_3$ for $f_2=f_4=0$, $g_4=1$ seen from different directions. Positivity bounds are shown in a solid colour while the causality constraint is plotted with a transparent lighter colour. The blue point corresponds to a partial UV completion with an axion and scalar with the same coupling strength.}
	\label{fig:3d}
\end{figure}

So far we were looking at extreme parameter choices close to the values of either scalar or axion UV completions. We know that these values are at the boundary of the 6-dimensional allowed region. To visualize what happens inside the constrained volume, instead of in a slice, we present the 3D plots for $h_3-g_3-f_3$ parameters when $f_2=f_4=0$ which illustrate the complementarity of the positivity and causality bounds, see Fig.~\ref{fig:3d}. It is interesting to observe from these plots that causality can constrain $h_3$ direction stronger than positivity, especially for the lower bound on $h_3$. 

\subsection{Case $g_2=0$.} \label{sec:g20}

As it was discussed in Section 3, positivity bounds do not allow to have $g_2=0$ in an interacting theory. In particular, the non-linear bounds in \eqref{NL12} force all the EFT coefficients that we considered to be zero when $g_2=0$. 
On the other hand, causality bounds are independent of any assumptions about the UV theory and allow us to probe low-energy properties. On the contrary, causality bounds still allow an interacting theory with $g_2=0$. More precisely, the causality requirements lead to $f_2=f_3=g_3=h_3=0$. However, causality conditions imply only $g_4>0$ and $-g_4/2<f_4<g_4/2$, see Fig.~\ref{fig:Boundsf4g4G2isZero}. We can see that this is consistent with having all scattering amplitude parameters vanishing in the free theory as in the results obtained from positivity bounds. In our current setting, we cannot obtain an upper bound on $g_4$ since the non-sign definite contributions cannot be included consistently up to the order in the WKB expansion that we can compute. If we were able to find a consistent setting to include these corrections, we would obtain an upper bound on $g_4$ that very likely will impose $g_4=0$ and hence $f_4=0$.


\section{Conclusions}

In this work, we have derived two sets of consistency conditions on the photon EFT parameters (defined in \eqref{eq:Lagrangian}) based on different assumptions, namely, positivity bounds and causality constraints. The latter are obtained by requiring the absence of a resolvable time advance for photon modes propagation on top of a non-trivial background generated by an external source. Meanwhile, the positivity bounds are based on analytic properties of the scattering amplitudes, which can be inferred by assuming microcausality. However, they have additional requirements on the UV completion, such as a Froissart-like bound. The assumptions of each method are summarized in Table~\ref{tab:assumptions}. Note that, while positivity bounds require the assumption of unitarity, they do not use its full power. Thus, the bounds on the Wilson coefficients could be improved by imposing full non-linear unitarity with numerical techniques. The upshot of our work is that we do not need to use numerical techniques to obtain strong constraints on the Wilson coefficients. Instead, we can combine different analytic methods (positivity and causality bounds) to obtain strong constraints. It is also worth noting that as a consistency check, we have verified that known, healthy, partial UV completions lie within the allowed region of both positivity and causality bounds.

Interestingly, we have shown that applying strict positivity of the time delay would rule out known (partial) UV completions. In other words, known consistent (partial) UV completion lead to a negative contribution to the time delay which is unresolvable and hence consistent with our notion of causality but would be misidentified if the appropriate resolvability condition had not been properly accounted for.

When comparing the positivity and causality bounds obtained in this paper, we found that in certain regions of the parameter space positivity leads to stronger bounds, but in other regions causality and positivity are complementary. As an example of the former, we found that the causality bounds derived here are predominantly double-sided and compact, with a few exceptions, especially the $g_4$ case that does not admit any upper bound in our current setup. On the other hand, positivity bounds always lead to compact regions. Nevertheless, there are other cases where the bounds are complementary and their combination can lead to a great reduction of the allowed parameter space as seen in \ref{sec:complemen}. An interesting case worth highlighting are the bounds on $h_3$. The positivity bounds are symmetric with respect to $h_3\rightarrow -h_3$, but the causality bounds are not. From the causality point of view, the time delay of the odd modes has a large positive $h_3$ contribution which generically implies a strong lower bound on $h_3$, but not so strong upper bound. Thus, causality can improve the lower bounds on $h_3$.

In some slices of the parameter space, we have found that the positivity and causality allowed regions are disjoint, thus those parameters are ruled out. We highlight that this does not imply that a  theory satisfying positivity bounds is acausal in the IR. Both our causality and positivity bounds do not exhaust the full power of the assumptions used to derive them. These extreme examples simply show that our bounds probe different regions of the parameter space and that combining them can be extremely powerful. It would be interesting to combine all the existing constraints on photon EFTs from previous works such as \cite{Henriksson:2021ymi,Henriksson:2022oeu,Haring:2022sdp} together with the present bounds to find even stronger bounds on the EFT Wilson coefficients. Similarly, it is worth noting that causality bounds can be applied to general theories involving operators with higher-point interactions. Thus, while positivity bounds mostly focus on $2-2$ scattering, causality can also be complementary to these explorations by constraining higher-point interactions. Generically, combining different techniques to bound EFT parameters can be very fruitful and the full power of combining causality and positivity bounds is yet to be proved. 

\section*{Acknowledgements}
We would like to thank Calvin Chen, Greg Kaplanek, Aoibheann Margalit and Andrew J. Tolley for useful discussions. The work of MCG, AT and CdR at Imperial is supported by STFC grant ST/T000791/1. VP is funded by the Imperial College President's Fellowship. SJ is supported by an
STFC studentship. CdR is also supported by a Simons Investigator award 690508.  

\appendix
\section{Equations of Motion} \label{ap:eom}
The equation of motion for the vector field $A^{\mu}$ with Lagrangian given by Eq.~\eqref{eq:Lagrangian} and sourced by an arbitrary current $J^\nu$ is
\begin{equation}
	\partial^\mu \mathcal{E}^{(F)}\mn =g J_\nu \ , 	\label{eq:eomA}
\end{equation}
where
\begin{align}
\mathcal{E}^{(F)}\mn \equiv&\ F\mn - 8 \frac{c_1}{\Lambda^4} F\mn F^{\alpha \beta} F_{\alpha \beta} - 8 \frac{c_2}{\Lambda^4} F_{\mu \alpha} F_{\nu \beta} F^{\alpha \beta} \label{eq:eomF} \\
	& + 2 \frac{c_3}{\Lambda^6} \left[ F\du{\mu}{\alpha} \left( \p_{\alpha} F\du{\nu}{\beta} \p_{\gamma} F\du{\beta}{\gamma} + F^{\beta \gamma} \p_{\gamma} \p_{\alpha} F_{\nu \beta} \right) - F^{\alpha\beta} \p_{\alpha} F\du{\mu}{\gamma} \left( \p_{\gamma} F_{\nu \beta} + \p_{\nu} F_{\beta\gamma} \right) \right] - \left( \mu \leftrightarrow \nu \right) \nonumber \\
	& + \frac{c_4}{\Lambda^6} \left[ \frac12 F\du{\mu}{\alpha} F\du{\nu}{\beta} \left( \p_{\gamma} \p_{\beta} F\du{\alpha}{\gamma} - \p_{\gamma} \p_{\alpha} F\du{\beta}{\gamma} \right) - F\du{\mu}{\alpha} \left( \left( \p_{\alpha} F_{\beta\gamma} - \p_{\beta} F_{\alpha\gamma} \right) \p^{\gamma} F\du{\nu}{\beta} + \p_{\gamma} F\du{\beta}{\gamma} \p_{\nu} F\du{\alpha}{\beta} \right. \right. \nonumber \\
	&\left. \left. \qquad \, \, \vphantom{+ \p_{\gamma} F\du{\beta}{\gamma} \p_{\nu} F\du{\alpha}{\beta}} + F^{\beta\gamma} \p_{\nu} \p_{\gamma} F_{\alpha\beta} \right) - F^{\alpha\beta} \left( \p^{\gamma} F_{\mu\alpha} \left( \p_{\beta} F_{\nu \gamma} - \p_{\nu} F_{\beta\gamma} \right) + \p_{\alpha} F\du{\mu}{\gamma} \p_{\nu} F_{\beta\gamma} \right) \vphantom{\frac12} \right]- \left( \mu \leftrightarrow \nu \right) \nonumber \\
	&+ \frac{c_5}{\Lambda^6} \left[ F\du{\mu}{\alpha} \left( \p_{\alpha} F_{\beta\gamma} \p_{\nu} F^{\beta\gamma} + F^{\beta\gamma}\p_{\nu}\p_{\alpha}F_{\beta\gamma} \right) + F^{\alpha\beta} \left( \p_{\gamma} F_{\alpha\beta} \p_{\mu} F\du{\nu}{\gamma} + \p_{\gamma} F\du{\mu}{\gamma} \p_{\nu} F_{\alpha\beta} \right) \right. \nonumber \\
	&\left. \qquad \, \, \, \, + F\mn \left( F^{\alpha\beta} \p_{\gamma} \p_{\beta} F\du{\alpha}{\gamma} + \p_{\beta} F_{\alpha\gamma} \p^{\gamma} F^{\alpha\beta} \right) \right] - \left( \mu \leftrightarrow \nu \right)  \nonumber \\
	&+ \frac{1}{2} \frac{c_6}{\Lambda^8} \left[ - F\du{\mu}{\alpha} \p_{\nu} \p_{\alpha} \left( \p_{\rho} F_{\beta \gamma} \p^{\rho} F^{\beta \gamma} \right) - \p_{\mu} F\du{\nu}{\alpha} \p_{\alpha} \left( \p_{\rho} F_{\beta \gamma} \p^{\rho} F^{\beta \gamma} \right) - \p_{\alpha} F\du{\mu}{\alpha} \p_{\nu} \left( \p_{\rho} F_{\beta \gamma} \p^{\rho} F^{\beta \gamma} \right)  \right. \nonumber \\
	&\qquad \qquad + 2 \p^{\alpha} F\mn \left( 2 \p^{\rho} F^{\beta\gamma} \p_{\gamma}\p_{\alpha} F_{\beta\rho} + \p_{\alpha} F^{\beta\gamma} \p_{\gamma}\p^{\rho} F_{\beta\rho} + F^{\beta\gamma} \p_{\alpha} \p_{\gamma} \p^{\rho} F_{\beta\rho} \right) \nonumber \\
	&\qquad \qquad \left. + 4 \Box F\mn \p_{\beta} \left( F_{\alpha\gamma} \p^{\gamma} F^{\alpha\beta} \right) \right] - \left( \mu \leftrightarrow \nu \right)  \nonumber \\
	&+ \frac{c_7}{\Lambda^8} \left[ \p_{\mu} F^{\alpha\beta} \left( \p_{\gamma} F^{\gamma\rho} \p_{\nu}\p_{\rho} F_{\alpha\beta} - \p_{\alpha} F^{\gamma\rho} \p_{\nu} \p_{\beta} F_{\gamma\rho} \right)  + \p^{\alpha} F\mn \p^{\rho} \left( \p_{\beta} F^{\beta\gamma} \p_{\gamma} F_{\alpha\rho} \right) \right. \nonumber \\
	&\qquad \qquad \left. +2 \p_{\alpha}\p_{\gamma} F\mn \p_{\rho} \left( F^{\alpha\beta} \p_{\beta} F^{\gamma\rho} \right) + F^{\alpha\beta} \p_{\mu} \p_{\alpha} F^{\gamma \rho} \p_{\nu} \p_{\beta} F_{\gamma \rho} \right] - \left( \mu \leftrightarrow \nu \right)  \nonumber \\
	&+  \frac{c_9}{\Lambda^8} \left[ -F_{\mu\alpha} \p_{\gamma} \p^{\rho} \left( \p^{\alpha} F_{\beta\rho} \p_{\nu} F^{\beta\gamma} \right) +2 \p_{\mu} F^{\alpha\beta} \p^{\rho} F_{\alpha\gamma} \p^{\gamma} \p_{\beta} F_{\nu \rho} + 2F^{\alpha\beta} \p_{\mu}\p^{\gamma} F_{\beta\rho} \p^{\rho} \p_{\alpha} F_{\nu\gamma} \right. \nonumber \\
	&\qquad \qquad \left. +  \p_{\beta} F_{\mu\alpha} \p^{\rho} \left(  \p^{\alpha} F_{\gamma\rho} \p_{\nu} F^{\beta\gamma} + \p^{\alpha} F^{\beta\gamma} \p_{\nu} F_{\gamma\rho} - F_{\gamma\rho} \p^{\alpha} \p_{\nu} F^{\beta\gamma} - F^{\beta\gamma} \p^{\alpha} \p_{\nu} F_{\gamma\rho} \right) \right]  - \left( \mu \leftrightarrow \nu \right)  \,.\nonumber
\end{align}
In Section~\ref{sec:spherical}, we focus on the perturbations around a spherically symmetric background by perturbing the field as in Eq.~\eqref{eq:Aspherical}. In this setting, one can find that one of Maxwell's equations lead to a constraint and fixes one of the unphysical modes. To find this constraint, we perform a field-redefinition of $u_1^{\ell}$ and write
\begin{align}
	v_1^{\ell} \equiv& \frac{i\omega  u_1^{\ell}}{r}-\left(u_2^{\ell}\right)' -\frac{16}{\Lambda ^4} \left(2 c_1+c_2\right) A_0' A_0'' u_2^{\ell} \label{eq:constraint} \\
	&- \frac{2}{\Lambda^6} \frac{1}{r} \left\lbrace \vphantom{\frac{A_0'}{r}} c_3 A_0' (A_0' +4r A_0'') \left(u_2^{\ell}\right)'' + 2c_3 \left( 2r(A_0'')^2 + A_0' \left( 3 A_0'' + 2r A_0^{(3)} \right) \right) \left(u_2^{\ell}\right)' \right. \nonumber \\
	& + \left[ c_3 \left( \omega^2 - \frac{2}{r^2} \right) (A_0')^2 + A_0'' \left\lbrace 2(3c_3-c_4-2c_5)A_0'' +3(c_3-c_4-2c_5)r A_0^{(3)} \right\rbrace \right. \nonumber \\
	& \left. \left. + \frac{A_0'}{r} \left\lbrace 2 \left( 2c_3(r^2 \omega^2-1) + c_4+2c_5 \right) A_0'' + 2 (3c_3 -c_4 -2c_5) r A_0^{(3)} + (c_3 -c_4 -2c_5) r^2 A_0^{(4)} \right\rbrace \right] \right\rbrace u_2^{\ell} \nonumber \\
	&+ \frac{1}{\Lambda^8} \sum_{n=0}^4 d_n (u_2^{\ell})^{(n)} \,, \nonumber
\end{align}
where
\begin{eqnarray}
	d_0 &=& 8\left(-c_8\left(L^2-3\right)+2c_6+c_7\right) \frac{(A_0')^2}{r^5} + 4\left(\left(3c_6+c_8\right)L^2-11c_6-4c_7-9c_8\right) \frac{A_0'A_0''}{r^4} \\
	& + &\left(\left(2c_6+c_8\right)L^2+2c_7+10c_8\right) \frac{\omega^2 (A_0')^2}{r^3} + 4\left(7c_6+3c_7+2c_8\right) \frac{A_0' A_0^{(3)}}{r^3} +4\left(7c_6+2c_7+3c_8\right) \frac{(A_0'')^2}{r^3} \nonumber \\
	& +& \left(-c_8\left(L^2+12\right)-16c_6-20c_7\right) \frac{\omega^2 A_0' A_0''}{r^2} - 2 \left(c_6\left(L^2+3\right)+2c_7-c_8\right) \frac{A_0'A_0^{(4)}}{r^2} \nonumber \\
	&-& 2 \left(c_6\left(3L^2+7\right)+6c_7-c_8\right)\frac{A_0'' A_0^{(3)}}{r^2} - \left(2c_6+c_8\right) \frac{\omega^4 (A_0')^2}{r} \nonumber \\
	& +&2\left(-c_6+7c_7+5c_8\right) \frac{\omega^2 A_0' A_0^{(3)}}{r} +2 \left(9c_7+5c_8\right) \frac{\omega^2 (A_0'')^2}{r} \nonumber \\
	& -&2\left( \left(c_6+c_8\right) \frac{A_0' A_0^{(5)}}{r} + 2\left(3c_6-c_7+2 c_8\right) \frac{ A_0'' A_0^{(4)}}{r} + \left(5c_6-2c_7+3c_8\right) \frac{(A_0^{(3)})^2}{r} \right) \nonumber \\
	& +&c_8\omega^4A_0'A_0'' + \omega^2\left(\left(2c_6+8c_7+5c_8\right)A_0^{(4)}A_0'+2\left(3
	c_6+8c_7+4c_8\right)A_0^{(3)}A_0''\right) \nonumber \\
	& - &2 \left(c_6+c_8\right)\left(A_0'' A_0^{(5)} + 3A_0^{(3)}A_0^{(4)}\right)+128(2c_1+ c_2)^2(A_0')^3 A_0'' \,, \nonumber 
 \end{eqnarray}
 \begin{eqnarray}
	d_1 &=& 4\left(c_7-c_8\left(L^2+2\right)\right) \frac{(A_0')^2}{r^4} + 4\left(c_8\left(L^2+4\right)+2c_6-c_7\right) \frac{A_0'A_0''}{r^3} \\
	& - &\left(c_8 \left(2L^2+5 \right)+8c_6-2c_7\right) \frac{A_0' A_0^{(3)}}{r^2} - 2\left(c_8\left(L^2+4\right)+4c_6\right) \frac{(A_0'')^2}{r^2}+\left(-4c_6+2c_7+c_8\right) \frac{\omega^2 (A_0')^2}{r^2} \nonumber \\
	& -&\frac{A_0'}{r}\left(\left(4c_6+2c_7+c_8\right)\omega^2
	A_0''+\left(-4c_6+4c_7+3c_8\right)A_0^{(4)}\right) +\frac{3\left(4c_6-3c_8\right)A_0^{(3)}A_0''}{r} \nonumber \\
	&+&2\left(\left(c_6+c_8\right) A_0' A_0^{(5)}+\left(4c_6-c_8\right) A_0''A_0^{(4)} +\left(3c_6-2c_8\right) (A_0^{(3)})^2\right) \nonumber \\
	& +&\omega^2\left(\left(2c_7+3c_8\right) A_0'A_0^{(3)} +c_8 (A_0'')^2\right) \vphantom{\frac{(A_0')^2}{r^4}} \,, \nonumber 
 \end{eqnarray}
 \begin{eqnarray}
	d_2 &=& \left(\left(2 c_6 + c_8  \right) L^2 - 6 c_7 + 6 c_8 \right) \frac{(A_0')^2}{r^3} - \left(c_8\left(L^2 + 2\right) + 16 c_6 -8 c_7\right) \frac{A_0' A_0''}{r^2}  \\
	& - &\frac{2}{r}\left(\left(2 c_6 + c_8 \right)\omega^2 (A_0')^2 + \left(c_6 + 3 c_7 -2 c_8 \right) A_0' A_0^{(3)} + \left(c_7 - 2 c_8 \right) (A_0'')^2 \right) \nonumber \\
	& + & A_0'\left(2 c_8\omega^2 A_0'' + \left(2 c_6 + 4 c_7 +7 c_8 \right) A_0^{(4)} \right) + 2\left(3 c_6 + 2 c_7 + 3 c_8 \right) A_0'' A_0^{(3)} \vphantom{\frac{(A_0')^2}{r^3}} \,, \nonumber
 \end{eqnarray}
 \begin{eqnarray}
	d_3 &=& \left(-4 c_6 + 2 c_7 +  c_8 \right) \frac {(A_0')^2} {r^2} - \left(4 c_6 + 2 c_7 + c_8 \right) \frac {A_0' A_0''} {r} + 2 \left(2 c_7 + 3 c_8 \right) A_0' A_0^{(3)} + 2 c_8 (A_0'')^2 \,, \\
	d_4 &=& c_8 A_0' A_0'' - \left(2 c_6+c_8\right) \frac{\left(A_0'\right){}^2}{r} \,.
\end{eqnarray}
  Note that we only need to go up to order $1/\Lambda^8$ for the terms $c_1^2, c_1 c_2$ and $c_2^2$ in order to get the $\epsilon_1^4$ contributions at leading order. In terms of this new variable we find that the constraint is given by
 \begin{equation}
 	v_1^{\ell} = 0 \ .
 \end{equation}
\section{Explicit expressions for propagation around spherically symmetric backgrounds}\label{ap:SphericalExpr}
The equations of motion for the propagating modes are given by Eq.~\eqref{eq:ModesEOM} with $W_{I,\ell}$ given by
\begin{align}
	W_{2,\ell} =& 1 -\frac{B^2}{R^2} + 16 \epsilon_1^2 \frac{B^2}{R^2} \left(2c_1+c_2\right) f'(R)^2  \label{eq:W2}\\
	& + 2 \epsilon _1^2 \epsilon _2^2 \left\lbrace (-5c_3+c_4+2c_5) \frac{B^2}{R^2}\frac{f'(R)^2}{R^2} + \left( \frac{B^2}{R^2}(7c_3+c_4+2c_5)-4c_3 \right) f''(R)^2 \right. \nonumber \\
	&\qquad -2 \left. \left( \left( \frac{B^2}{R^2}(c_3+c_4+2c_5) + 3c_3 \right) f''(R) +2c_3 \left(1-2\frac{B^2}{R^2}\right)R f^{(3)}(R) \right) \frac{f'(R)}{R} \right\rbrace \nonumber \\
	&+2\epsilon_1^2 \epsilon_2^2 \Omega^2 (2c_6-c_8) \frac{B^2}{R^2} \left\lbrace \left(2\frac{B^2}{R^2}-1\right)\frac{f'(R)^2}{R^2} +\left(1-\frac{B^2}{R^2}\right)f''(R)^2 \right. \nonumber \\
	&\qquad - \left. \left( \frac{B^2}{R^2} f''(R) + \left(1-\frac{B^2}{R^2}\right)R f^{(3)} \right) \frac{f'(R)}{R} \right\rbrace \,, \nonumber
\end{align}
and
\begin{align}
	W_{4,\ell} =& 1 -\frac{B^2}{R^2} + 8 \epsilon_1^2 \frac{B^2}{R^2} c_2 f'(R)^2  \label{eq:W4} \\
	& + 2 \epsilon _1^2 \epsilon _2^2 \left\lbrace (-3c_3+c_4) \frac{B^2}{R^2}\frac{f'(R)^2}{R^2} + \left( \frac{B^2}{R^2}(c_3-c_4)-4c_3 \right) f''(R)^2 \right. \nonumber \\
	&\qquad + \left. \left( 2 c_3 \left( \frac{B^2}{R^2}-3 \right) f''(R) + \left(\frac{B^2}{R^2}(c_3-c_4)-4c_3\right)R f^{(3)}(R) \right) \frac{f'(R)}{R} \right\rbrace \nonumber \\
	&-2\epsilon_1^2 \epsilon_2^2 \Omega^2 c_8 \frac{B^2}{R^2} \left\lbrace \left(2\frac{B^2}{R^2}-1\right)\frac{f'(R)^2}{R^2} +\left(1-\frac{B^2}{R^2}\right)f''(R)^2 \right. \nonumber \\
	&\qquad - \left. \left( \frac{B^2}{R^2} f''(R) + \left(1-\frac{B^2}{R^2}\right)R f^{(3)} \right) \frac{f'(R)}{R} \right\rbrace \,. \nonumber
\end{align}

Let us now write down the turning point for both physical modes
\begin{equation}
	R^{t}_{I,\ell} = B \left[ 1 - \epsilon_1^2 \Psi_{I,\ell}^{(1)}(B) - \epsilon_1^2 \epsilon_2^2 \Psi_{I,\ell}^{(2)}(B) - \epsilon_1^2 \epsilon_2^2 \Omega^2 \Psi_{I,\ell}^{(3)}(B) \right] \,,
\end{equation}
where
\begin{eqnarray}
	\Psi_{2,\ell}^{(1)}(B)&=& 8 (2c_1+c_2) f'(B)^2 
	= (f_2+g_2) f'(B)^2 \,, \nonumber \\
	\Psi_{2,\ell}^{(2)}(B) &=& -(5c_3-c_4-2c_5) \frac{f'(B)^2}{B^2} -2(4c_3+c_4+2c_5) \frac{f'(B)f''(B)}{B} +(3c_3+c_4+2c_5) f''(B)^2 \nonumber \\
	& +&4c_3 f'(B) f^{(3)}(B) \nonumber \\
	&=& - \frac13 \left\lbrace (f_3+3(g_3-4h_3)) \frac{f'(B)^2}{B^2} -2(f_3+3(g_3+2h_3)) \frac{f'(B)f''(B)}{B} \right. \nonumber \\
	&+&\left. (f_3+3g_3+4h_3) f''(B)^2 +8h_3 f'(B) f^{(3)}(B) \vphantom{\frac{f'(B)^2}{B^2}} \right\rbrace \,, \nonumber \\
	\Psi_{2,\ell}^{(3)}(B) &=& (2c_6 -c_8) \left( \frac{f'(B)^2}{B^2} - \frac{f'(B)f''(B)}{B} \right) 
	= 2(2f_4+ g_{4}) \left( \frac{f'(B)^2}{B^2} - \frac{f'(B)f''(B)}{B} \right) \,,\nonumber
\end{eqnarray}
and
\begin{align}
	\Psi_{4,\ell}^{(1)}(B) =& 4c_2 f'(B)^2 = -(f_2-g_2) f'(B)^2 \,, \nonumber \\
	\Psi_{4,\ell}^{(2)}(B) =& -(3c_3-c_4) \frac{f'(B)^2}{B^2} -4c_3 \frac{f'(B)f''(B)}{B} -(3c_3+c_4) \left( f''(B)^2 + f'(B) f^{(3)}(B) \right) \nonumber \\
	=& \frac13 \left\lbrace -(f_3-3g_3-8h_3) \frac{f'(B)^2}{B^2} +8h_3 \frac{f'(B)f''(B)}{B} \right. \nonumber \\
    &\left. \vphantom{\frac{f'(B)^2}{B^2}} \qquad + (f_3-3g_3+4h_3) \left( f''(B)^2 + f'(B) f^{(3)}(B) \right) \right\rbrace \,, \nonumber \\
	\Psi_{4,\ell}^{(3)}(B) =& - c_8 \left( \frac{f'(B)^2}{B^2} - \frac{f'(B)f''(B)}{B} \right) = - 2(2f_4 - g_{4}) \left( \frac{f'(B)^2}{B^2} - \frac{f'(B)f''(B)}{B} \right) \,,
\end{align}
where, in the last equalities, we have converted the Wilson coefficients from the Lagrangian to the scattering amplitude parameters by using Eq.~\eqref{eq:conversion}. The functions $U_{I,\ell}$ appearing in the phase shift and time delay expressions are given by
\begin{align}
	U_{I,\ell} =& \frac{B^2}{R^2} \left[ \epsilon_1^2 \left( \Psi_{I,\ell}^{(1)}(R) - \Psi_{I,\ell}^{(1)}(B) \right) + \epsilon_1^2 \epsilon_2^2 \left( \Psi_{I,\ell}^{(2)}(R) - \Psi_{I,\ell}^{(2)}(B) \right) + \epsilon_1^2 \epsilon_2^2 \Omega^2 \left( \Psi_{I,\ell}^{(3)}(R) - \Psi_{I,\ell}^{(3)}(B) \right) \right] \nonumber \\
	&+ \left( 1 - \frac{B^2}{R^2} \right) \left( \epsilon_1^2 \epsilon_2^2 \Upsilon_{I,\ell}^{(1)}(R) + \epsilon_1^2 \epsilon_2^2 \Omega^2 \Upsilon_{I,\ell}^{(2)}(R) \right) \,,
	\label{eq:defUIell}
\end{align}
where it is clear that the first term in square bracket vanishes when $R \rightarrow B$ which ensures the convergence of the time delay. The analytical expressions for the functions entering Eq.~\eqref{eq:defUIell} above are given by
\begin{align}
	\Upsilon_{2,\ell}^{(1)}(R) =& -2c_3 \left( 3 \frac{f'(R)f''(R)}{R} +2 \left( f''(R)^2 + f'(R)f^{(3)}(R) \right) \right) \nonumber \\
	=& \frac43 h_3 \left( 3 \frac{f'(R)f''(R)}{R} +2 \left( f''(R)^2 + f'(R)f^{(3)}(R) \right) \right) \,, \nonumber \\
	\Upsilon_{2,\ell}^{(2)}(R) =& -(2c_6-c_8) \left\lbrace 2 \frac{f'(R)^2}{R^2} - \frac{f'(R)f''(R)}{R} - f''(R)^2 + f'(R)f^{(3)}(R) \right\rbrace \nonumber \\
	=& -2(2f_4+g_{4}) \left\lbrace 2 \frac{f'(R)^2}{R^2} - \frac{f'(R)f''(R)}{R} - f''(R)^2 + f'(R)f^{(3)}(R) \right\rbrace \,,
\end{align}
and
\begin{align}
	\Upsilon_{4,\ell}^{(1)}(R) =& \Upsilon_{2,\ell}^{(1)}(R) \,, \nonumber \\
	\Upsilon_{4,\ell}^{(2)}(R) =& c_8 \left\lbrace 2 \frac{f'(R)^2}{R^2} - \frac{f'(R)f''(R)}{R} - f''(R)^2 + f'(R)f^{(3)}(R) \right\rbrace \nonumber \\
	=& 2(2f_4-g_{4})) \left\lbrace 2 \frac{f'(R)^2}{R^2} - \frac{f'(R)f''(R)}{R} - f''(R)^2 + f'(R)f^{(3)}(R) \right\rbrace \,.
\end{align}
Finally, the expression for the time delay is
\begin{align}
	&(\omega \Delta T_{b,I,\ell}(\omega)) \nonumber \\
	&= 2(\omega r_0) \left[ \int_{B}^{\infty} \frac{\p_{\omega} \left( \omega U_{I,\ell}(R) \right)}{\sqrt{1- \frac{B^2}{R^2}}} \mathrm{d}R + \frac{\pi}{2} \left( B - \p_{\omega} \left( \omega R^{t}_{I,\ell} \right) \right) \right] \nonumber \\
	&= 2(\omega r_0) \left[ \int_{B}^{\infty} \frac{B^2}{R^2} \frac{\left[ \eps_1^2 \Psi_{I,\ell}^{(1)}(R) + \eps_1^2 \epsilon_2^2 \Psi_{I,\ell}^{(2)}(R) + 3 \eps_1^2 \epsilon_2^2 \Omega^2 \Psi_{I,\ell}^{(3)}(R) \right] - \left[ R \leftrightarrow B \vphantom{\Psi_{I,\ell}^{(1)}(R)} \right]}{\sqrt{1- \frac{B^2}{R^2}}} \mathrm{d}R \right. \nonumber \\
	& \qquad \qquad \quad + \int_{B}^{\infty} \sqrt{1- \frac{B^2}{R^2}} \left( \epsilon_1^2 \epsilon_2^2 \Upsilon_{I,\ell}^{(1)}(R) + 3 \epsilon_1^2 \epsilon_2^2 \Omega^2 \frac{B^2}{R^2} \Upsilon_{I,\ell}^{(2)}(R) \right) \mathrm{d}R \nonumber \\
	& \qquad \qquad \quad \left. + \frac{\pi}{2} B \left( \epsilon_1^2 \Psi_{I,\ell}^{(1)} + \epsilon_1^2 \epsilon_2^2 \Psi_{I,\ell}^{(2)} + 3 \epsilon_1^2 \epsilon_2^2 \Omega^2 \Psi_{I,\ell}^{(3)} \right) \vphantom{\frac{\left( \epsilon_1^2 \Phi_{I,\ell}^{(1)} + \epsilon_1^2 \epsilon_2^2 \Phi_{I,\ell}^{(2)} + 3 \epsilon_1^2 \epsilon_2^2 \Omega^2 \Phi_{I,\ell}^{(3)} \right)}{\sqrt{1- \frac{B^2}{R^2}}}} \right] \nonumber \\
	&= 2(\omega r_0) \left[ \int_{B}^{\infty} \frac{B^2}{R^2} \frac{\left[ \epsilon_1^2 \Psi_{I,\ell}^{(1)}(R) + \epsilon_1^2 \epsilon_2^2 \Psi_{I,\ell}^{(2)}(R) + 3 \epsilon_1^2 \epsilon_2^2 \Omega^2 \Psi_{I,\ell}^{(3)}(R) \right]}{\sqrt{1- \frac{B^2}{R^2}}} \mathrm{d}R \right. \nonumber \\
	& \qquad \qquad \quad \left. + \int_{B}^{\infty} \sqrt{1- \frac{B^2}{R^2}} \left( \epsilon_1^2 \epsilon_2^2 \Upsilon_{I,\ell}^{(1)}(R) + 3 \epsilon_1^2 \epsilon_2^2 \Omega^2 \frac{B^2}{R^2} \Upsilon_{I,\ell}^{(2)}(R) \right) \mathrm{d}R \right] \,. 
\end{align}
Note that the integral over the constant $\Psi_{I,\ell}^{(j)}(B)$ can be performed and exactly cancels out the extra term that is free of any integration.
\section{Optimization method}
\label{ap:method}
We describe here the algorithm used to optimize the causality bounds. For any two-dimensional plot in the $(\mathcal{W}_J, \mathcal{W}_K)$-plane, we start by fixing all the remaining coefficients collectively denoted by $\{ \mathcal{W}_L \}$ to constants corresponding to a particular UV completion or another interesting case. This is necessary in order to reduce the complexity of the optimization by only allowing $2$ of the $8$ coefficients to vary\footnote{Note however that the causality bounds derived in this paper are insensitive to $g_4'$ and that $g_2$ is set to either $1$ or $0$ without loss of generality.}.

The vector causality constraints reduce to $2$ constraints in the even and odd sector and the extremization procedure can be done in each independently. When presenting the results, we can choose to show the causality bounds of each sector individually or to show the final result which is achieved by taking their union. In the following, we forget about the two sectors and assume we are specialising to a given one, only to take the union of both at the end.

The boundary of the causality constraint reads $(\omega \Delta T) = -1$ and can be solved for $\mathcal{W}_J$ as a function of $\mathcal{W}_K$ and all the other parameters of the problem (other Wilson coefficients, background parameters, and parameters defining the validity of the EFT and WKB expansions). Next, we discretise the direction $\mathcal{W}_K$ by letting this coefficient take values in the interval $\left[ \mathcal{W}_K^{\rm (min)} ,\mathcal{W}_K^{\rm (max)} \right]$ divided in $n_K$ equal steps of length $\Delta \mathcal{W}_K = (\mathcal{W}_K^{\rm (max)} - \mathcal{W}_K^{\rm (min)})/n_K$,

\begin{equation}
    \mathcal{W}_K = \left\{ \mathcal{W}_K^{\rm (min)}, \mathcal{W}_K^{\rm (min)} + \Delta \mathcal{W}_K , \cdots , \mathcal{W}_K^{\rm (max)} \right\} \,.
\end{equation}
For each such value of $\mathcal{W}_K$, we find the maximal (if it exists) value that $\mathcal{W}_J$ can take by extremizing over the parameters of the background profile and the ones controlling the EFT expansion, under the constraint that one remains in the regime of validity of the EFT and the WKB approximations simultaneously. This way, we get a set of extremal parameters per discretised value of $\mathcal{W}_K$ that we plug back into the equation $(\omega \Delta T) = - 1$, which in turn gives an optimised straight line in the $(\mathcal{W}_J, \mathcal{W}_K)$-plane separating the causality-violating region from the allowed one. One gets such a line for each discretised value of $\mathcal{W}_K$ and hence can form an envelope by imposing all the constraints derived in this way. This procedure gives an upper (lower) bound on $\mathcal{W}_J$ when the coefficient in front of $\mathcal{W}_J$ is negative (positive). Since in some cases, the coefficient is not sign-definite (as seen in Table~\ref{tab:signdef}), we obtain but upper and lower bounds. Finally, we refine our bounds by performing the same procedure once again after swapping the role of $\mathcal{W}_J$ and $\mathcal{W}_K$, i.e. discretising the other axis and getting left/right bounds rather than upper/lower.

As mentioned earlier, this is done independently in the even and odd sectors of the vector. The theory is only causal if neither mode propagate in a causality-violating way, hence the final causality bounds are obtained by imposing causality on both sectors simultaneously.

\section{Spin-2 partial UV (in)completions} \label{app:spin2}

Here, we focus on the partial UV (in)completions suggested in \cite{Haring:2022sdp}. These are constructed by on-shell amplitude methods and requirements on the Regge behaviour. First, they construct the residue at the spin-2 pole and split this in two cases: parity even and odd. They allow for the freedom of adding arbitrary contact terms which are then fixed so that the amplitude has the desired Regge limit (growing as $\mathcal{O}\left(s^2, u^2, t^2\right)$). Without taking into account these additional contributions from contact terms, the even and odd partial UV completions only propagate even or odd modes, but these additional contact terms do not respect the parity and the partial UV completions in \cite{Haring:2022sdp} propagate both even and odd modes. From a Lagrangian perspective, the construction of \cite{Haring:2022sdp} includes higher derivative couplings between the photon and the massive spin-2 field. 

\begin{figure}[!h]
	\begin{center}
		\includegraphics[width=0.48\textwidth]{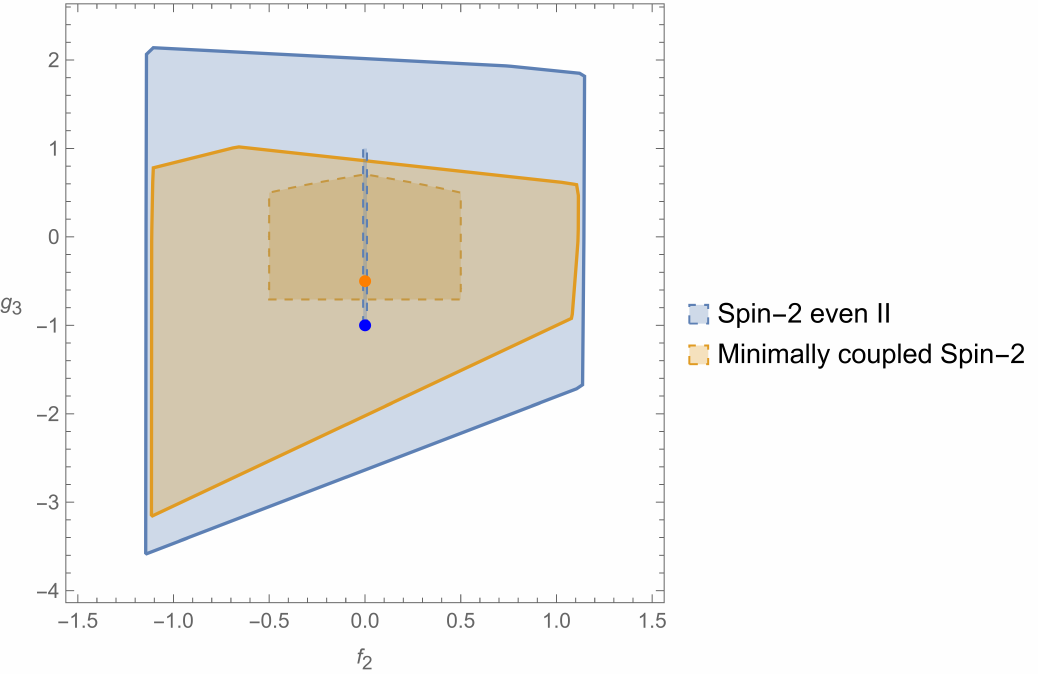}
  \includegraphics[width=0.4\textwidth]{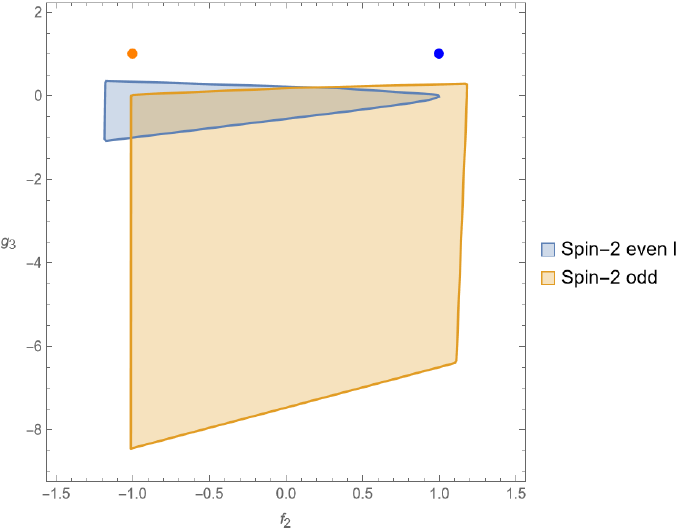}
	\end{center}
        	\caption[$f_2 - g_3$ causality and positivity bounds, spin-2 partial UV completions.]{ Causality (thick line) and positivity (dashed line) bounds in the $(f_2, g_3)$ plane with all other coefficients set to the values of the corresponding partial UV completions as in Table \ref{tab:UVcomp}. On the left side, we have $f_3=h_3=f_4=0$ for both regions and $g_4=1$ and $1/2$ for the blue and orange regions respectively. Meanwhile, the right panel has $f_3=h_3=0$, $g_4=1$ for both regions and $f_4=1/2$ and $-1/2$ for the blue and orange regions respectively. The partial UV completions with causal propagation appear on the left whereas the ones on the right do not agree with either of the causality or positivity bounds. In the left-hand plot the parameter values for the spin--2 even UV completion give an allowed region that is described by $f_2=0\,\cup|g_3|<1$ giving a one-dimensional region, depicted as a slim two-dimensional region for visibility. On the right-hand plot, there is no region allowed by positivity bounds.}
	\label{fig:f2g3Spin2Comparison}
\end{figure}
The causality and positivity bounds for four different spin-2 partial UV (in)completions are presented in Fig.~\ref{fig:f2g3Spin2Comparison}. According to Table \ref{tab:UVcomp}, all dimension-$8$ and $10$ free coefficients (i.e. with the exception of $f_2$ and $g_3$ in this specific case) are identical in each of the four proposed partial UV completions. For the two UV completions that are causal, they also lie within the positivity bounds and so are consistent UV completions. The remaining two possible `UV completions' (on the right side) which do not satisfy causality, also do not satisfy positivity as there is in fact no region allowed by positivity in the slice in which they live. We highlight again that the causality bounds do not require any UV assumptions. In other words, the implications of our result are that the non-minimal coupling between the photon and the spin-2 in these partial UV completions leads to acausal propagation.

The fact that these spin-2 partial UV (in)completions lie outside positivity bounds was already observed in 
\cite{Haring:2022sdp}. Note that the fact that spin-2 fields lead to an $s^2$ growth of the amplitude is not the reason why positivity bounds fail. The Froissart-like bound requirement that the amplitude should grow strictly softer than $s^2$ is a requirement in the $|s|\to \infty$ limit, that is, in the full UV theory not on the EFT nor the partial UV completion. Just as is the case of gravity, when considering massive spin-2 fields coupled to photons, we can also assume that at very high energies there is a UV completion with the desired behavior.

\section{Reducing the angle dependence for the positivity bounds}\label{app:angle}

Linear positivity bounds for indefinite helicity amplitudes typically have the form 
\begin{equation}
\label{angle cond}
\begin{split}
 & A_1+ A_2 \cos{(2\theta)}\cos{(2\chi)}+A_3 (\cos{\phi}\sin{(2\theta)}+\cos{\phi}\sin{(2\chi)})+
  \\
&+A_4 \cos{(\phi-\psi)}\sin{(2\theta)}\sin{(2\chi)}+A_5\cos{(\phi+\psi)}\sin{(2\theta)}\sin{(2\chi)}>0,
  \end{split}
\end{equation}
which must be valid for all values of angles $\phi,\psi,\chi,\theta$. This requirement implies several non-linear bounds on the coefficients $A_1,A_2,A_3,A_4,A_4$ depending on the coupling constants.

\begin{figure}%
    \centering
    \subfloat[\centering ]{\includegraphics[width=0.45\textwidth]{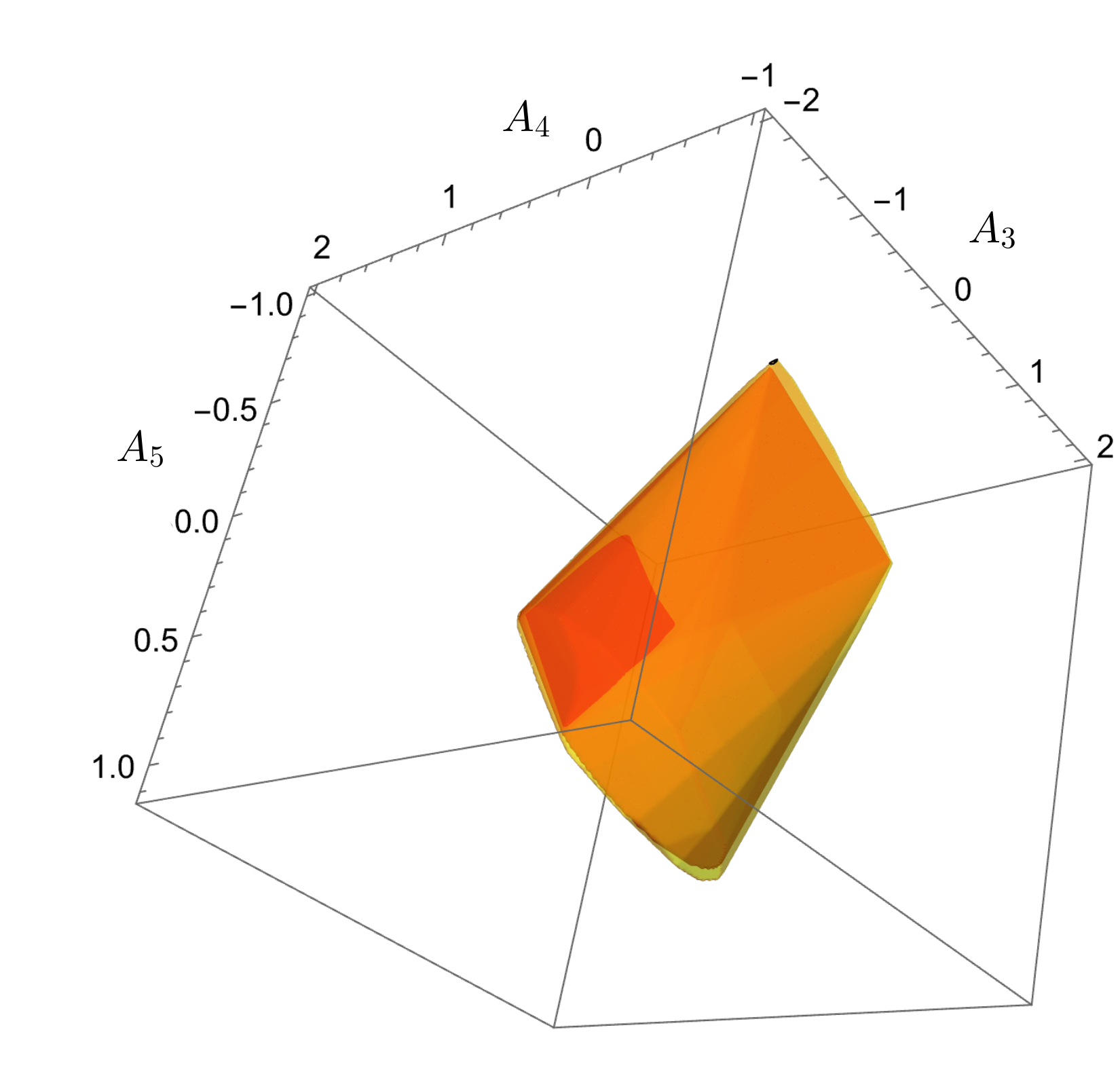}}%
    \qquad
    \subfloat[\centering ]{\includegraphics[width=0.45\textwidth]{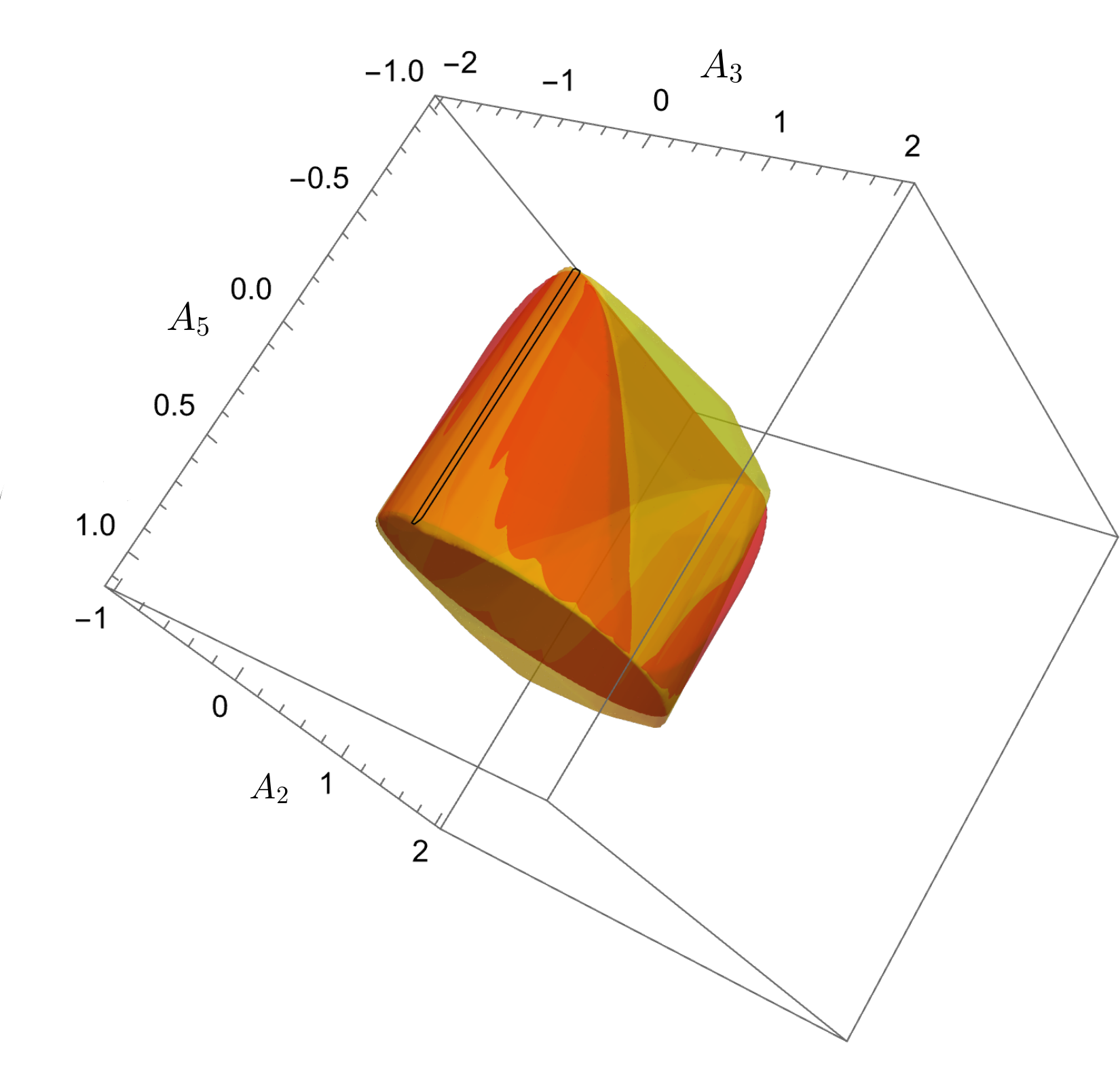}}%
    \caption{(a) Allowed region for $A_3,~A_4,~A_5$ when $A_1=1$ and $A_2=1/2$ are taken. The red volume presents the analytic bound while the yellow area corresponds to the result of numerical optimization of angles. (b) Allowed region for $A_2,~A_3,~A_5$ for $A_1=1$, $A_4=1/2$. The meaning of the colours is the same as in the plot (a).} 
    \label{fig:Aplot}%
\end{figure}

Taking the particular case $\theta=0$ one obtains
\begin{equation}
    A_1+A_2 \cos{(2\chi)}+A_3 \cos{\psi}\sin{(2\chi)}>0.
\end{equation}
This can be transformed to 
\begin{equation}
    A_1+\sqrt{A_2^2+A_3^2\cos^2{\psi}}\cos{(2\chi+\chi_0)}>0.
\end{equation}
Here $\chi_0$ depends on $A_2, A_3$ but the concrete expression is irrelevant for the derivation. The strongest bound corresponds to $\cos{(2\chi+\chi_0)}=-1$. Thus,
\begin{equation}
    A_1-\sqrt{A_2^2+A_3^2\cos^2{\psi}}>0,
\end{equation}
or
\begin{equation}
    A_1^2>A_2^2+A_3^2\cos^2{\psi}.
\end{equation}
This condition is satisfied for any $\psi$ if
\begin{equation}
\label{cond1}
     A_1^2>A_2^2+A_3^2.
\end{equation}
Thus, the angle dependence is completely reduced and the obtained condition is a non-linear bound on the couplings. However, numerical analysis shows that Eq.~\eqref{angle cond} in fact leads to stronger constraints than Eq.~\eqref{cond1}. To derive them, it seems to be enough to take $\theta=\chi=\pi/4$,
\begin{equation}
 A_1+ A_3 (\cos{\phi}+\cos{\psi})+A_4 \cos{(\phi-\psi)}+
A_5\cos{(\phi+\psi)}>0,
\end{equation}
or
\begin{equation}
    A_1+A_3\cos{\psi}+\cos{\phi}(A_3+(A4+A5)\cos{\psi})+\sin{\phi}\sin{\psi}(A_4-A_5)>0.
\end{equation}
Repeating the previous procedure we obtain
\begin{equation}
    (A_1+A_3\cos{\psi})^2-(A_3+(A4+A5)\cos{\psi})^2-\sin^2{\psi}(A_4-A_5)^2>0.
\end{equation}
After some simplifications, one can get
\begin{equation}
    (A_3^2-A_4 A_5)y^2+2(A_1 A_2-2 A_3(A_4+A_5))y+A_1^2-A_3^2-A_4^2+2A_4A_5-A_5^2>0.
\end{equation}
Here $y=\cos{\psi}$ is between $-1$ and $1$. This bound can be brought to the form
\begin{equation}
 A y^2+B y+C>0 \,.
\end{equation}
 This is satisfied for all angles if one of the following four conditions is satisfied,
\begin{itemize}
    \item $A>0$, $D=B^2-4 A C<0$,
    \item $A<0$, $D>0$ and $x_1<-1$ and $x_2>1$ where $x_1<x_2$ are two real roots of the quadratic polynomial,
    \item $D>0$, $A>0$, $x_1<-1$ or $x_2>1$,
    \item $A=0$, $\mid B\mid<C$.
\end{itemize}

In Fig.~\ref{fig:Aplot} we show the coincidence of the allowed ranges obtained by numerical scanning over angles and analytic result described here. The numerical region (yellow-shaded) is a bit wider reflecting the fact that scanning does not always provide the optimal angles for the strongest bound.

\section{Comparison of conventions with other works}

In this Appendix, we provide a conversion chart that enables the reader to go from our conventions to the ones of \cite{Henriksson:2021ymi,Henriksson:2022oeu} and \cite{Haring:2022sdp}. 
\begin{table}[h!]
	\centering
	\begin{tabular}{ c | c | c | c }
		EFT dim & This Paper & \cite{Henriksson:2021ymi,Henriksson:2022oeu} & \cite{Haring:2022sdp} \\
		\hline
		\multirow{2}{*}{$8$} & $c_1$ & $f_2 = 8c_1 +2c_2$ & $f_2 = 8c_1 +2c_2$ \\ 
		& $c_2$ & $g_2 = 8c_1 +6c_2$ & $g_2 = 8c_1 +6c_2$ \\
		\hline
		\multirow{3}{*}{$10$} & $c_3$ & $f_3 = -3(c_3+c_4+c_5)$ & $f_3 = -3(c_3+c_4+c_5)$ \\
		& $c_4$ & $g_3 = -c_5$ & $g_3 = -c_5$ \\
		& $c_5$ & $h_3 = -\frac32 c_3$ & $h_3 = - \frac32 c_3$ \\
		\hline
		\multirow{3}{*}{$12$} & $c_6$ & $f_4 = \frac14 c_6$ & $f_4 = \frac14 c_6$ \\
		& $c_7$ & $g_{4,1} = \frac12(c_6+c_8)+c_7$ & $g_4 = \frac12 (c_6-c_8)$ \\
		& $c_8$ & $g_{4,2} = -\frac12 (c_7+c_8)$ & $g_4' = c_7+c_8$
	\end{tabular}
	\caption{Conversion table for the EFT coefficients.}
	\label{tab:conversion}
\end{table}

\pagebreak

\bibliographystyle{JHEP}
\bibliography{References}
	
\end{document}